\documentclass[
 aip,
 jap,%
 reprint,%
]{revtex4-1}
\usepackage{textcomp}
\usepackage{amsmath,amssymb,amsfonts,amsthm}
\usepackage{algorithm}
\usepackage{algorithmic}
\usepackage{graphicx}
\usepackage{subcaption}
\usepackage{array}
\usepackage{pstricks}
\usepackage{comment}
\usepackage{txfonts}
\usepackage{placeins}

\newcolumntype{L}{>{\raggedright\arraybackslash}p}
\newcolumntype{C}{>{\centering\arraybackslash}m}

\theoremstyle{plain}

\theoremstyle{definition}
\newtheorem{theorem}{Property}

\newtheorem{defn}{Definition} 
\newtheorem{cor}{Corollary}

\usepackage{tikz}
\usepackage{circuitikz}

\newcounter{sarrow}

\usepackage[final]{changes}
\definechangesauthor[name={Sameh}, color=red]{SA}
\definechangesauthor[name={Greg}, color=blue]{GR}
\usepackage[colorinlistoftodos]{todonotes}
\usepackage{tikz-3dplot}

\usepackage{pgfplots}
\pgfplotsset{compat=1.14}

\begin{document}
%
\title{Properties of Translation Operator and the Solution of the Eigenvalue and Boundary Value Problems of Arbitrary Space-time Periodic Circuits}
%
%
%

\author{Sameh Y. Elnaggar}
\email{samehelnaggar@gmail.com }
\affiliation{Department of Electrical and Computer Enginering, Royal Military College of Canada, Kingston, ON, Canada.}
\author{Gregory N.~Milford}
\email{ g.milford@adfa.edu.au}
\affiliation{School of Engineering and Information Technology, University of New South Wales, Canberra}
\date{\today}

\begin{abstract}
The time periodic circuit theory is exploited to introduce an appropriate translation operator that is invariant under the change of the spatial unit cell. Additionally, some useful properties of the operator of arbitrary space time modulated structures are derived. By casting the problem in an eigenvalue problem form, the equivalency between solutions at different positions along the structure is demonstrated. It is shown that the underlying mathematical machinery is identical to the one used in the analysis of linear time invariant periodic structures, where a two step eigen-decompositions is performed. The first decomposition is in the temporal eigen-functions basis, which is followed by the decomposition of the translation operator in the spatial domain. The two step process results in the well-known dispersion relation.  We also prove that all points in the ($\beta$,$\omega$) plane parallel to the modulation velocity $\nu_m$ are equivalent in the sense that the eigenvectors are related by a harmonic shift operator. Additionally, the wave propagation inside the space time periodic circuit as well as the terminal characteristics are rigorously determined via the expansion of the total solution in terms of the eigenmodes, and after imposing the suitable boundary conditions. To validate and demonstrate the usefulness of the developed framework, two examples are provided. In the first, a space time modulated composite right left handed transmission line is studied and results are compared with time domain simulation. The second example is concerned with the characterization of the non-reciprocal behaviour observed on a nonlinear transmission line that was manufactured in our lab. Circuit parameters, extracted from measurements, are used to predict the wave behaviour along the transmission line and its effect on the terminal properties. Using the developed machinery it is shown that the passive interaction between different harmonics results in an observed giant non-reciprocity, where the difference between the forward and backward transmission coefficients can be greater than 30 dB. The frequencies at which non-reciprocity occurs and its strength agree with time domain simulation and measurements.
\end{abstract}

\maketitle

%

\section{Introduction}
%
%
%
%
Recently, there has been a surge of interest in studying the properties of space-time modulated systems that arise from the intrinsic asymmetric interaction of the space-time harmonics \cite{Trainiti2016,ElnaggarMilford2018}. Such asymmetry breaks the principle of reciprocity, hence enabling the design of novel non-reciprocal devices such as nonreciprocal antenna  \cite{Hadad2016, Taravati2017mixer,RamacciaALu2018}, magnet-less circulators  \cite{AluCaloz, Alu2016magnetless, KordAlu2018}, one-way beam splitters  \cite{TaravatiKishkarXiv}, isolators  \cite{Lira2012,Winn1999}, and space time modulated metasurfaces  \cite{Alu_PRB_2015,mazor2019}. On top of that, systems that possess time and/or space-time periodic elements may not be constrained by the physical limitations of linear time invariant (LTI) systems. For instance, a time modulated reactance does not necessarily result in a total reflection of an incident wave impinging the structure, hence enables the accumulation of energy  \cite{mirmoosa2019}. Additionally, a time switched transmission line was demonstrated to have a broadband matching capability not limited by the Bode-Fano criteria that impose a return loss/bandwidth trade-off  \cite{fano1950,beyondbf}. Ref. \onlinecite{taravati2019} provides an excellent review on the developments, applications and methods of analysis of space-time media.

The interest in space-time modulated media dates back to the mid of last century in the context of understanding the properties of distributed parametric amplifiers \cite{Tien1958,Cullen1958,Simon1960,Oliner1961,Cassedy1963,Cassedy1967}. The salient properties of such media emerge from Bloch-Floquet theory, which states that the eigensolutions are the sum of infinite space time harmonics. Unlike LTI systems, space time modulation leads to an asymmetric dispersion relation, where at a given frequency $\omega$, the forward and backward wavenumbers are not necessarily equal \cite{Trainiti2016, ElnaggarMilford2018}.

In general, space-time modulated systems are built from nonlinear lumped elements. The interaction of the nonlinearity and a strong pump results in a spatio-temporal modulation of one or more system parameter \cite{Qin2014,Taravati2018}. In a conventional treatment, it is generally assumed that the system is distributed such that the modulation wavelength is much longer than the length of the unit cell. Such assumption permits describing the system via partial differential equations (PDEs). Upon the application of the Bloch-Floquet condition, the PDEs reduce to an infinite system of homogeneous equations, where the eigenmodes can be calculated from its non-trivial solutions. Very recently, we have developed a circuit based framework that extends the theory of linear time periodic (LTP) circuits and systems, developed in Refs. \onlinecite{kurth1977,wereley1991,hindawi}, to space-time structures \cite{ElnaggarMilford2020}. The dispersion relation emerges from the application of the Bloch condition to a unit cell that links its input and output time harmonics. Therefore, it is valid for both electrically long and electrically short systems. Furthermore the framework enables the exploration of various structures such as Composite Right Left Handed (CRLH) transmission lines (TLs) and nonsinusoidal periodic modulation \cite{ElnaggarMilford2020}. The governing equations reduce to  generalized telegraphist's equations when the unit cell is infinitesimally small.

In the current manuscript, we exploit the circuit based approach developed in Ref. \onlinecite{ElnaggarMilford2020} to explore the translational properties of the unit cell operator, and derive equivalency relations between different eigenvalues and eigenvectors. We show that the approach is, in principle, equivalent to how periodic structures are described via the diagonalization of the translation operator, where the eigenmodes are nothing but the Bloch waves. Additionally for a generic unit cell and an arbitrary periodic modulation, the boundary value problem is solved via the expansion of the solution inside the structure in terms of the eigenvectors (eigenmodes).

Section \ref{sect:translational} starts with a brief review of how the immittance matrix, a generalization of the immittance circuit parameter in LTI systems, emerges from Bloch-Floquet theorem. We then proceed by showing how elements in cascade combine and how the ABCD parameters change between unit cells. In Section \ref{sect:DispersionEVP}, we focus on the eigenvalue problem that describes the system modal behaviour. The invariance of eigenvalues and eigenvectors resulting from the transformation of the system translation operator is discussed and a complete mathematical proof is provided in the appendix. We also show that for a generic space-time circuit, the eigensolutions along the modulation line are equivalent. Furthermore the section discusses how the LTI and LTP systems are formally equivalent in the sense of eigen-decompositions in time and spatial domains. Section \ref{sect:bvp} demonstrates how the driven modal solution is expanded in terms of the system eigenmodes. Additionally, expressions of the transmission and reflection coefficients are derived. In Section \ref{sect:results}, two systems are studied. In Subsection \ref{subsect:crlh} a CRLH TL is fully described using the developed machinery and results are compared to time domain simulation. Dispersion relations, eigenvalues, eigenvectors, waveforms, and transmission coefficient are computed for both the right hand (RH) and left hand (LH) regimes. Subsection \ref{subsect:nlrhtl} applies the framework to a nonlinear RH TL that has been fabricated in our lab. The modulation is achieved via a strong pump and hence, the TL operates in the sonic regime \cite{Cassedy1963}. Dispersion relations, eigenvalues and waveforms are computed and compared to simulation. Furthermore, the S parameters are calculated and compared to measurements.

\section{Translational Properties of Immittance and ABCD matrices}
\label{sect:translational}
In LTP circuits, Bloch Floquet theorem allows the voltage and current harmonics to be related via immittance matrices  \cite{kurth1977,ElnaggarMilford2020}. An immittance matrix is best visualized as a generalization of the concept of immittance, a scalar complex quantity, in LTI systems. Before proceeding with the detailed description, it is worth noting that we represent the $(m,n)$ element in matrix $\mathbf{A}$, using the notation $A_m^{n}$, i.e, the subscript (superscript) represents the row (column).

Without loss of generality, consider a time modulated capacitance $\tilde{C}(t)$. The instanteneous current $i(t)$ is given by
\begin{equation}
i(t)=\frac{d\tilde{C}(t)v(t)}{dt},
\end{equation}
where $v(t)$ is the capacitance voltage. Since $\tilde{C}$ is time periodic with a perioid $T$, it can be expanded in its Fourier components. Furthermore, the eigensolutions of LTP systems are in the form of $p(t)\exp(i\omega t),~p(t+T)=p(t)$  \cite{wereley1990,ElnaggarMilford2020}, hence the above relation can be re-written as
\begin{align}
\sum_{r=-\infty}^{+\infty} I_re^{i\left(\omega+r\omega_m\right)t}&=\frac{d}{dt}\sum_{q=-\infty}^{+\infty}\sum_{l=-\infty}^{+\infty }C_qV_le^{i\left(\omega+\left[q+l\right]\omega_m\right)t}\\
&=\sum_{q,l=-\infty}^{+\infty}i\left(\omega+\left[q+l\right]\omega_m\right)C_qV_le^{i\left(\omega+[q+l]\omega_m\right)\omega_mt}.
\end{align}
Matching the $\omega+r\omega_m$ frequency, one gets
\begin{equation}
I_r=\sum_{l=-\infty}^{+\infty}\underbrace{i\left(\omega+r\omega_m\right)C_{r-l}}_{\tilde{Y}_r^l}V_l,
\end{equation}
where $\tilde{Y}_r^l$ is the$(r,l)$ element of the admittance matrix $\tilde{\mathbf{Y}}$ and $r,l=-\infty,\cdots,-2,-1,0,1,2,\cdots,\infty$. Hence the voltage-current relation can vw described compactly by the matrix equation
\begin{equation}
\mathbf{I}=\tilde{\mathbf{Y}}\mathbf{V}.
\end{equation}
The entries of an arbitrary $k^\textnormal{th}$ row can be determined from the zeroth row, since
\begin{equation}
\label{eq:Yproperty}
\tilde{Y}_k^{k+l}\left(\omega\right)=\tilde{Y}_0^l\left(\tilde{\omega}_k\right),
\end{equation}
where $\tilde{\omega}_k\triangleq \omega+k\omega_m$.

Consider now a structure where the above capacitance is modulated via a travelling wave with speed $\nu_m$, i.e,
$$\tilde{C}(t,x)=\tilde{C}(t-x/\nu_m)$$
and $x$ is a multiple of the underlying spatial lattice distance $p$ (i.e, $x=np$, $n=-\infty,\cdots,-2,-1,0,1,2,\cdots,\infty$. Again, expanding $\tilde{C}$ in its Fourier components, and noting that the modulation frequency $\omega_m$ and wave-number $\beta_m$ are related by $\omega_m=\nu_m\beta_m$, give
\begin{equation}
\tilde{C}(t-x/\nu)=\sum_{r=-\infty}^{+\infty}C_re^{-ir\beta_m x}e^{i\omega_mt}.
\end{equation}
This implies that
\begin{equation}
\label{eq:immittance_transformation}
\tilde{Y}_{\color{blue}{q}}^{\color{red}{p}}(x)=\tilde{Y}_{\color{blue}{q}}^{\color{red}{p}}(0) e^{-i\left[{\color{blue}{q}}-{\color{red}{p}}\right]\beta_mx}.
\end{equation}
This means that the elements in a given row of an admittance matrix $\tilde{\mathbf{Y}}(x)$  are those in the same row of $\tilde{\mathbf{Y}}(0)$, but multipled by a phasor that rotates in the counter clockwise direction as we go from left to right. Additionally for a fixed column, the elements from top to bottom are multipled by a clockwise rotating phasor.

When a structure is constructed from different cascaded LTP elements, the dispersion properties are uniquely determined by the \emph{net} ABCD parameters of the unit cell. These in turn result from the multiplication of LTP impedance and admittance matrices. Therefore, it is crucial to understand the properties of the product of LTP matrices. Consider for instance the series $\mathbf{Z}$ and shunt $\mathbf{Y}$ of a lumped right handed transmission line. The $(r,r-s)$ element of $(\mathbf{Z}\mathbf{Y})_r^{r-s}$ is
\begin{align*}
(\mathbf{Z}\mathbf{Y})_r^{r-s}(x)= &\sum_{l=-\infty}^{+\infty} Z_r^{r-l}(0)e^{-il\beta_mx} Y_{r-l}^{r-s}(0)e^{-i\left[s-l\right]\beta_mx}\\
=&\sum_{l=-\infty}^{+\infty}Z_r^{r-l}(0)Y_{r-l}^{r-s}(0)e^{-is\beta_mx}\\
=&e^{-is\beta_mx}\sum_{l=-\infty}^{+\infty}Z_r^{r-l}(0)Y_{r-l}^{r-s}(0)\\
=&(\mathbf{Z}\mathbf{Y})_r^{r-s}(0)e^{-is\beta_mx}.
\end{align*}
Therefore, we have the following important property:

\theorem{
\label{prop1}
For a space-time periodic structure consisting of a cascade of space-time periodic unit cells, the ABCD parameters $\mathbf{X}=\mathbf{A},~\mathbf{B},~\mathbf{C}$ and $\mathbf{D}$ for a unit cell $x$ away from the origin are related to the ones at the origin by

\begin{equation}
\label{eq:paramtranslated}
\mathbf{X}_q^p(x)=\mathbf{X}_q^p(0)e^{-i[q-p]\beta_mx}=\mathbf{X}_q^p(0)\Gamma_{q-p}^{x/p},
\end{equation}
where $\Gamma_{q-p}\triangleq \exp{(-i[q-p]\beta_mp)}$.}
Hence as $x$ changes, the ABCD parameters follow the same transformation of immittance matrices (Eq. \ref{eq:immittance_transformation}). Note that if the modulation wavelength $\lambda_m\triangleq 2\pi/\beta_m$ is a multiple of $p$, and when $x$ is a multiple of $\lambda_m$, $\mathbf{\Gamma}$ becomes the identity matrix and $\mathbf{X}_q^p(x)=\mathbf{X}_q^p(0)$ as expected.

\section{Eigenvalue Problem and Dispersion Relation}
\label{sect:DispersionEVP}
The harmonics at the terminals of the $n^\textnormal{th}$ unit cell are related by the ABCD transfer matrix
$$
\begin{pmatrix}
\mathcal{V}[n]
\\
\mathcal{I}[n]
\end{pmatrix}
=
\begin{pmatrix}
\mathbf{A} & \mathbf{B}\\ \mathbf{C} &\mathbf{D}
\end{pmatrix}_n
\begin{pmatrix}
\mathcal{V}[n+1]
\\
\mathcal{I}[n+1]
\end{pmatrix},
$$

or in the more convenient form
\begin{equation}
\label{eq:BFC}
\mathbf{\Psi}_n =\mathbf{P}_n\mathbf{\Psi}_{n+1},
\end{equation}
where $\mathbf{\Psi}_r\triangleq (\mathcal{V}[r],\mathcal{I}[r])^t$ is an inifnite dimensional vector that stores the amplitude of all time harmonics at $x=rp$; and $\mathbf{P}_n$ is the ABCD matrix at the $n^\textnormal{th}$ unit cell.

We seek solutions of the form 
\begin{equation}
\label{eq:BFSol}
\mathbf{\Psi}_{n+1}=e^{-i\beta p}\mathbf{\Lambda}\mathbf{\Psi}_n,
\end{equation}
where

 $$\mathbf{\Lambda}=
\begin{pmatrix}
\mathbf{\Gamma} & \mathbf{0}
\\
\mathbf{0} &\mathbf{\Gamma}
\end{pmatrix} 
 $$
and $\Gamma_{rr}=\Gamma_r=\exp (-ir\beta_mp)$ and zero otherwise. The condition (\ref{eq:BFSol}) is equivalent to seeking a travelling wave solution of the form $\sum_{r=-\infty}^\infty \Psi_{0r}e^{i[\tilde{\omega}_rt-\tilde{\beta}_rnp]}$, where $\tilde{\beta}_r\triangleq \beta+r\beta_m$. Therefore, (\ref{eq:BFC}) and (\ref{eq:BFSol}) can be combined to give the eigenvalue problem (EVP)

\begin{equation}
\label{eq:evb}
\mathbf{T}_n\mathbf{\Psi}_n=\mathbf{P}_n\mathbf{\Lambda}\mathbf{\Psi}_n=e^{i\beta p}\mathbf{\Psi}_n. 
\end{equation}
Note that $\mathbf{T}_n\triangleq\mathbf{P}_n\mathbf{\Lambda}$ acts as a translation operator, where $\exp(i\beta p)$ and $\mathbf{\Psi}_n$ are its eigenvalue and eigenvector, respectively. Furthermore, $\mathbf{T}_n$ is a function of the operating frequency $\omega$. If $\mathbf{T}_n$ is a translation operator, we should expect that the eigen-solutions should be independent of the unit cell. The next property shows that $\mathbf{T}_n$ is indeed independent of the unit cell (i.e, independent of $n$).

\theorem{
\label{prop2}
Consider the EVP (\ref{eq:evb}) at which $\beta$ and $\mathbf{\Psi}_n$ is a solution. Then  

\begin{equation}
\label{eq:betadash}
\beta'=\beta
\end{equation}
 and the eigenvector
 \begin{equation}
 \label{eq:n+1}
 \underbrace{V_{k}'}_{{\color{blue}{\textnormal{at }n+1}}}=\underbrace{V_{k}}_{{\color{red}{\textnormal{at }n}}}\Gamma_{k},~\underbrace{I_{k}'}_{{\color{blue}{\textnormal{at }n+1}}}=\underbrace{I_{k}}_{{\color{red}{\textnormal{at }n}}}\Gamma_{k}
 \end{equation}
is a solution of 
 $$\mathbf{T}_{n+1}\mathbf{\Psi}_{n+1}=e^{i\beta' p}\mathbf{\Psi}_{n+1}.$$ 
 
The proof of property (\ref{prop2}) is presented in Appendix \ref{append:translationevp}. Equations (\ref{eq:betadash}) and (\ref{eq:n+1}) imply that regardless of the unit cell used, the EVP will always result in a unique propagation constant $\beta$. Additionally, the k\textsuperscript{th} component of the eigenvector changes in a way that is equivalent to the phase delay of the k\textsuperscript{th} harmonic along a unit cell, which is equal to $k\beta_mp$. Therefore, the solution of the EVP (\ref{eq:evb}) is independent of the unit cell.

The above result can be generalized to
\cor{The solution of (\ref{eq:evb}) using $\mathbf{T}_{n+q}$ is
$$\beta'=\beta$$
and
$$V_k'=V_k\Gamma_k^q,~~I_k'=I_k\Gamma_k^q.$$

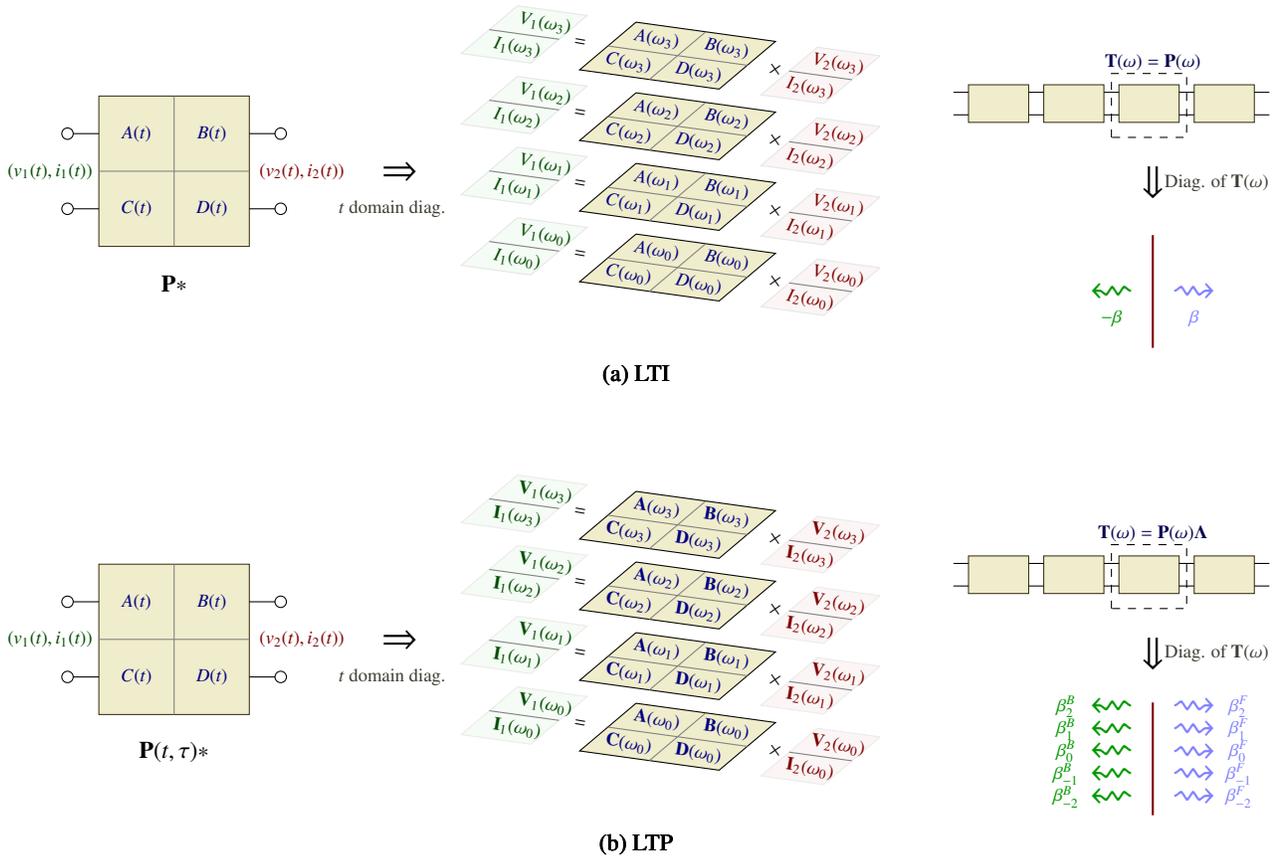
\begin{figure*}
\centering

\begin{subfigure}{0.32\linewidth}
\begin{tikzpicture}[scale=1]
\draw [color=green!20!orange!20!black!90,fill=green!30!orange!20] (0,0) rectangle+(2,2);
\draw [gray] (1,0)--(1,2);
\draw [gray] (0,1)--(2,1);
\node [blue!50!black] at (0.5,0.5){\scriptsize $C(t)$};
\node [blue!50!black] at (0.5,1.5){\scriptsize $A(t)$};
\node [blue!50!black] at (1.5,0.5){\scriptsize $D(t)$};
\node [blue!50!black] at (1.5,1.5){\scriptsize $B(t)$};
\draw [o-] (-0.5,0.5)--(0.0,0.5);
\draw [o-] (-0.5,1.5)--(0.0,1.5);
\draw [-o] (2,0.5)--(2.5,0.5);
\draw [-o] (2,1.5)--(2.5,1.5);
\node [green!30!black] at (-0.65,1){\scriptsize $(v_1(t),i_1(t))$};
\node [red!50!black] at (2.7,1){\scriptsize $(v_2(t),i_2(t))$};
\node [black] at (4.0,1){\Large $\Rightarrow$};
\node [olive!20!black] at (3.9,0.5) {\scriptsize \emph{t} domain diag.};
\node at (1,-0.5) {$\mathbf{P}\ast$};
\end{tikzpicture}
\end{subfigure}
\tdplotsetmaincoords{70}{112}
\begin{subfigure}{0.32\linewidth}
\begin{tikzpicture}[tdplot_main_coords,scale=2]
\def\s{0.5};
\foreach \z in {0,1,2,3}
{
\draw [fill=green!30!orange!20] (0,0,\s*\z)--(1,0,\s*\z)--(1,1,\s*\z)--(0,1,\s*\z)--(0,0,\s*\z);
\draw [gray] (0,0.5,\s*\z)--(1,0.5,\s*\z);
\draw [gray] (0.5,0,\s*\z)--(0.5,1,\s*\z);
\node [blue!50!black,rotate=-10] at (0.25,0.25,\s*\z){\scriptsize $A(\omega_\z)$};
\node [blue!50!black,rotate=-10] at (0.75,0.25,\s*\z){\scriptsize $C(\omega_\z)$};
\node [blue!50!black,rotate=-10] at (0.25,0.75,\s*\z){\scriptsize $B(\omega_\z)$};
\node [blue!50!black,rotate=-10] at (0.75,0.75,\s*\z){\scriptsize $D(\omega_\z)$};
\draw [fill=gray!50!green!50,opacity=0.1](0,-0.85,\s*\z)--(1,-0.85,\s*\z)--(1,-0.35,\s*\z)--(0,-0.35,\s*\z)--(0,-0.85,\s*\z);
\node [rotate=-10] at (0.5,-0.2,\s*\z){\scriptsize $=$};
\draw [gray](0.5,-0.85,\s*\z)--(0.5,-0.35,\s*\z);
\draw [fill=gray!50!red!50,opacity=0.1](0,1.75,\s*\z)--(1,1.75,\s*\z)--(1,1.3,\s*\z)--(0,1.3,\s*\z)--(0,1.75,\s*\z);
\node [rotate=-10] at (0.5,1.2,\s*\z){\scriptsize $\times$};
\draw [gray](0.5,1.75,\s*\z)--(0.5,1.3,\s*\z);
\node [green!30!black,rotate=-10] at (0.25,-0.55,\s*\z){\scriptsize $V_1(\omega_\z)$};
\node [green!30!black,rotate=-10] at (0.75,-0.55,\s*\z){\scriptsize $I_1(\omega_\z)$};
\node [red!50!black,rotate=-10] at (0.25,1.55,\s*\z){\scriptsize $V_2(\omega_\z)$};
\node [red!50!black,rotate=-10] at (0.75,1.55,\s*\z){\scriptsize $I_2(\omega_\z)$};
\node at (0,0,-1) {(a) LTI};
}
\end{tikzpicture}
\end{subfigure}
\begin{subfigure}{0.33\linewidth}
\begin{tikzpicture}
\def\x{1};
\foreach \i in{0,1,2,3}
{
\draw [green!20!orange!20!black!90,fill=green!30!orange!20] (\i,0) rectangle+(0.8,0.5);
\draw (\i-0.2,0.1)--(\i,0.1);
\draw (\i-0.2,0.4)--(\i,0.4);
}
\draw (4-0.2,0.1)--(4,0.1);
\draw (4-0.2,0.4)--(4,0.4);
\draw [dashed] (2-0.1,-0.2) rectangle+(1,0.85);
\node [blue!30!black] at (2.45,0.8){\scriptsize $\mathbf{T}(\omega)=\mathbf{P}(\omega)$};
\node [black] at (2.45,-0.8) {\Large $\Downarrow$};
\node [olive!20!black] at (3.3,-0.8) {\scriptsize Diag. of $\mathbf{T}(\omega)$};
\draw [red!50!black,thick] (2.45,-1.5)--(2.45,-3);
\node [blue!50!white] at (3,-2.25){\Large $\rightsquigarrow$};
\node [blue!50!white] at (3,-2.6){\scriptsize $\beta$};
\node [green!60!black] at (1.9,-2.25){\Large $\leftsquigarrow$};
\node [green!60!black] at (1.9,-2.6){\scriptsize $-\beta$};
\end{tikzpicture}
\end{subfigure}

\vspace{1cm}
\begin{subfigure}{0.32\linewidth}
\begin{tikzpicture}[scale=1]
\draw [green!20!orange!20!black!90,fill=green!30!orange!20] (0,0) rectangle+(2,2);
\draw [gray] (1,0)--(1,2);
\draw [gray] (0,1)--(2,1);
\node [blue!50!black] at (0.5,0.5){\scriptsize $C(t)$};
\node [blue!50!black] at (0.5,1.5){\scriptsize $A(t)$};
\node [blue!50!black] at (1.5,0.5){\scriptsize $D(t)$};
\node [blue!50!black] at (1.5,1.5){\scriptsize $B(t)$};
\draw [o-] (-0.5,0.5)--(0.0,0.5);
\draw [o-] (-0.5,1.5)--(0.0,1.5);
\draw [-o] (2,0.5)--(2.5,0.5);
\draw [-o] (2,1.5)--(2.5,1.5);
\node [green!30!black] at (-0.65,1){\scriptsize $(v_1(t),i_1(t))$};
\node [red!50!black] at (2.7,1){\scriptsize $(v_2(t),i_2(t))$};
\node [black] at (4.0,1){\Large $\Rightarrow$};
\node [olive!20!black] at (3.9,0.5) {\scriptsize \emph{t} domain diag.};
\node at (1,-0.5) { $\mathbf{P}(t,\tau)\ast$};
\end{tikzpicture}
\end{subfigure}
\tdplotsetmaincoords{70}{112}
\begin{subfigure}{0.32\linewidth}
\begin{tikzpicture}[tdplot_main_coords,scale=2]
\def\s{0.5};
\foreach \z in {0,1,2,3}
{
\draw [fill=green!30!orange!20] (0,0,\s*\z)--(1,0,\s*\z)--(1,1,\s*\z)--(0,1,\s*\z)--(0,0,\s*\z);
\draw [gray] (0,0.5,\s*\z)--(1,0.5,\s*\z);
\draw [gray] (0.5,0,\s*\z)--(0.5,1,\s*\z);
\node [blue!50!black,rotate=-10] at (0.25,0.25,\s*\z){\scriptsize $\mathbf{A}(\omega_\z)$};
\node [blue!50!black,rotate=-10] at (0.75,0.25,\s*\z){\scriptsize $\mathbf{C}(\omega_\z)$};
\node [blue!50!black,rotate=-10] at (0.25,0.75,\s*\z){\scriptsize $\mathbf{B}(\omega_\z)$};
\node [blue!50!black,rotate=-10] at (0.75,0.75,\s*\z){\scriptsize $\mathbf{D}(\omega_\z)$};
\draw [green!20!orange!20!black!90,fill=gray!50!green!50,opacity=0.1](0,-0.85,\s*\z)--(1,-0.85,\s*\z)--(1,-0.35,\s*\z)--(0,-0.35,\s*\z)--(0,-0.85,\s*\z);
\node [rotate=-10] at (0.5,-0.2,\s*\z){\scriptsize $=$};
\draw [gray](0.5,-0.85,\s*\z)--(0.5,-0.35,\s*\z);
\draw [green!20!orange!20!black!90,fill=gray!50!red!50,opacity=0.1](0,1.75,\s*\z)--(1,1.75,\s*\z)--(1,1.3,\s*\z)--(0,1.3,\s*\z)--(0,1.75,\s*\z);
\node [rotate=-10] at (0.5,1.2,\s*\z){\scriptsize $\times$};
\draw [gray](0.5,1.75,\s*\z)--(0.5,1.3,\s*\z);
\node [green!30!black,rotate=-10] at (0.25,-0.55,\s*\z){\scriptsize $\mathbf{V}_1(\omega_\z)$};
\node [green!30!black,rotate=-10] at (0.75,-0.55,\s*\z){\scriptsize $\mathbf{I}_1(\omega_\z)$};
\node [red!50!black,rotate=-10] at (0.25,1.55,\s*\z){\scriptsize $\mathbf{V}_2(\omega_\z)$};
\node [red!50!black,rotate=-10] at (0.75,1.55,\s*\z){\scriptsize $\mathbf{I}_2(\omega_\z)$};
\node at (0,0,-1) {(b) LTP};
}
\end{tikzpicture}
\end{subfigure}
\begin{subfigure}{0.33\linewidth}
\begin{tikzpicture}

\foreach \i in{0,1,2,3}
{
\draw [green!20!orange!20!black!90,fill=green!30!orange!20] (\i,0) rectangle+(0.8,0.5);
\draw (\i-0.2,0.1)--(\i,0.1);
\draw (\i-0.2,0.4)--(\i,0.4);
}
\draw (4-0.2,0.1)--(4,0.1);
\draw (4-0.2,0.4)--(4,0.4);
\draw [dashed] (2-0.1,-0.2) rectangle+(1,0.85);
\node [blue!30!black] at (2.45,0.8){\scriptsize $\mathbf{T}(\omega)=\mathbf{P}(\omega)\mathbf{\Lambda}$};
\node [black] at (2.45,-0.8) {\Large $\Downarrow$};
\node [olive!20!black] at (3.3,-0.8) {\scriptsize Diag. of $\mathbf{T}(\omega)$};
\draw [red!50!black,thick] (2.45,-1.45)--(2.45,-2.95);
\foreach \shift in {-2,-1,0,1,2}
{
\node [blue!50!white] at (3,-2.1+0.3*\shift){\Large $\rightsquigarrow$};
\node [blue!50!white] at (3.6,-2.1+0.3*\shift){\scriptsize $\beta^F_{\shift}$};
\node [green!60!black] at (1.9,-2.1+0.3*\shift){\Large $\leftsquigarrow$};
\node [green!60!black] at (1.3,-2.1+0.3*\shift){\scriptsize $\beta^B_{\shift}$};
}
\end{tikzpicture}
\end{subfigure}
\caption{Eigen-decomposition of (a) LTI and (b) LTP systems in the frequency and subsequently in the spatial domains.}
\label{fig:mathematicalstructure}
\end{figure*}

It is worth to digress here and discuss how the above formulation has the same mathematical foundations used to describe LTI systems. The discussion provides a deeper insight into the \emph{mathematical} structure of the framework and highlights similarities as well as differences between LTI and LTP systems. We refer to Fig. \ref{fig:mathematicalstructure} that depicts a conceptual diagram of the applied flow. Fig. \ref{fig:mathematicalstructure}(a) illustrates the rather implicit flow one usually uses to describe LTI circuits. The basic relations of a given circuit are given via differential and algebraic equations, leading to a LTI state space representation of the system. This in turn allows the  output parameters to be related to the input ones via the convolution with the $2\times 2$ impulse response matrix, shown in the left side of the Fig. by the $A(t)$, $B(t)$, $C(t)$  and $D(t)$ functions. For the time periodic (or in general time varying) circuits, the output and input parameters can be related via the aid of the state transition matrix, rendering the analysis of such general systems challenging. Fortunately for LTI systems, the convolution operation with the impulse response can be easily diagonalized, where the set of complex functions $\exp(i\omega t)$ are the eigen-functions of the convolution operator $\mathbf{P}\ast$. Hence, it allows the system to be represented by simple matrix multiplications at each frequency $\omega$ as shown in the middle panel of Fig.\ref{fig:mathematicalstructure}(a). In another words, the frequency domain representation is merely an expansion of the input and output parameters in the eigen-functions basis of $\mathbf{P}\ast$. For LTP systems, Floquet theorem tells us that the eigen-functions are in the form $\sum Q_r\exp(i\tilde{\omega}_r t)$. Therefore, the input and output voltages and currents become infinitely long complex vectors that store the coefficients $Q_r$ and are connected via the multiplication with LTP ABCD matrix. In another words, the LTP ABCD matrix is the operator that connects the output and input vectors expanded in the LTP eigen-functions. 

When a two port LTI network is cloned to form a periodic structure, the network ABCD matrix does not change as one moves from one unit cell to the other. Such ABCD matrix maps the input applied to one of its terminals to the output, hence it represents the translation operator $\mathbf{T}$ (i.e, $\mathbf{T}=\mathbf{P}$). $\mathbf{T}$ is a linear operator in the two dimensional complex space $\mathcal{C}^2$. Diagonalization of $\mathbf{T}$ is equivalent to finding its eigen-functions, which are nothing but the Bloch waves. The eigenvalues are conveniently represented as $\exp{(i\beta p)}$, where $p$ is the unit cell length. Therefore for each $\omega$, representing a temporal eigenvalue, there are two spatial eigenvalues $\pm\beta$ that diagonalize the spatial operator. The ordered pair $(\omega,\pm\beta)$ describes the propagation of waves in a periodic structure at $\omega$. On the other hand, property \ref{prop1} implies that the ABCD matrix $\mathbf{P}_n$ in a space-time modulated structure changes from one point to the other (i.e, depends on $n$). Nevertheless, property \ref{prop2} shows that $\mathbf{T}_n$ does indeed act as a translation operator. Since the space of time harmonics is infinite, the diagonalization of $\mathbf{T}_n$ (i.e, enforcing the Bloch condition) results in an infinite number of eigen-functions as the last panel of Fig. \ref{fig:mathematicalstructure}(b) illustrates. Note that if the modulation is removed $\mathbf{\Lambda}$ reduces to the identify matrix and $\mathbf{T}_n$ becomes the ABCD matrix $\mathbf{P}_n$. Although the  mathematical spaces for the LTI and LTP systems are different, the circuit based formalism suggests that space-time structures are formally equivalent to the well known periodic structures. The infinite dimension of the LTP space and its translation operator highlights the \emph{richeness} in the spectrum (i.e, the eigen-functions) of space time periodic structures, which eventually may lead to non-reciprocity.

There are an infinite number of eigenfunctions $\mathbf{\Psi}_n$ for any given frequency $\omega$. When the modulation strength goes to zero, the $\mathbf{A},\mathbf{B},\mathbf{C}$ and $\mathbf{D}$ matrices become diagonal and the eigenmodes become the solution of the LTI system at frequencies $\tilde{\omega}_r$. The propagation constants $\tilde{\beta}_r$ satsify the well known dispersion relation \cite{Collin2007,Pozar}
$$\tilde\beta_rp = \cos^{-1}\left(\frac{A_r^r(\omega)+D_r^r(\omega)}{2}\right).$$
Furthermore, the eigenvectors become the Bloch waves. Fig. \ref{fig:infiniteeigensolutions} depicts the dispersion relation of a RH TL for an infinitesimal modulation strength. The modes are labelled as shown for different values of $r$. Note that for each $r$, there are two branches representing the solutions $\pm\tilde\beta_r$.
 
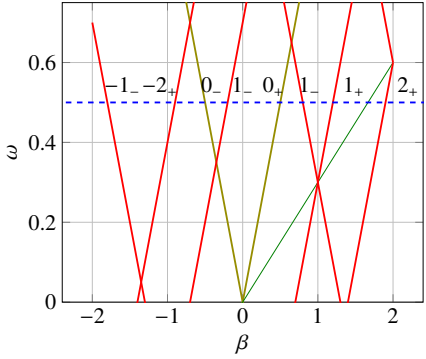
\begin{figure}[!htb]
\begin{tikzpicture}[scale=0.7]
\def\vm {0.3};
\def \w {0.5};
\begin{axis}[xlabel=\large $\beta$,ylabel=\large $\omega$,title=\large ,ymin=0,ymax=0.75,grid]
\pgfplotsset{every tick label/.append style={font=\large}}
\addplot[samples=51, domain=-2:2,smooth, line width=1pt,olive]{x};
\addplot[samples=51, domain=-2:2,smooth, line width=1pt,olive]{-x};
[dashed]\addplot[samples=2,domain=-2:2,smooth,line width=.5pt,color=green!50!black]{\vm*x};
\foreach \xshift in{-2,-1,1,2}
\addplot[samples=5,smooth, domain=-2:2, line width=1pt,color=red]{x-\xshift+\vm*\xshift}
;
\foreach \xshift in{-2,-1,1,2}
\addplot[samples=5,smooth, domain=-2:2, line width=1pt,color=red]{-(x-\xshift)+\vm*\xshift}
;

\node at(\w-0.09,\w+0.04){\large $0_+$};
\node at(\w+0.39,\w+0.04){\large $1_-$};
\node at(1.48,\w+0.04){\large $1_+$};
\node at(2.2,\w+0.04){\large $2_+$};
\node at(0,\w+0.04){\large $1_-$};
\node at(-\w+0.09,\w+0.04){\large $0_-$};
\node at(-1.1,\w+0.04){\large $-2_+$};
\node at(-1.6,\w+0.04){\large $-1_-$};
\draw [dashed,blue,line width=1pt] (-2.5,\w)--(2.5,\w);
\end{axis}
\end{tikzpicture}
\caption{A typical dispersion relation for a right handed medium, when the modulation is very small (i.e, the bandgaps approach zero width). The main branches are distinguished by the olive lines (online). At a given frequency, shown by the blue dashed line, different modes are possibly excited.}
\label{fig:infiniteeigensolutions}
\end{figure} 

For the subsequent discussion of the properties of $\mathbf{T}_n$, it is useful to introduce the harmonic shift operator $\mathcal{S_U}$.

\defn{$\mathcal{S_U}$ is a linear operator on $\mathbf{\Psi}_n=[\cdots, \psi_{k-1},\psi_k,\psi_{k+1},\cdots]_n^t$ that has the following effect
$$\left(\mathcal{S_U}\Psi_n\right)_k=\left(\Psi_n\right)_{k+1}.$$
i.e, $\mathcal{S_U}$ shifts the vector $\mathbf{\Psi}$ up by one position. Similarly $\mathcal{S_D}\triangleq\mathcal{S_U}^{-1}$ shifts the vector down by one position.
}

If $(\omega,\beta)$ is a solution of (\ref{eq:evb}), then the following is a general property of arbitrary space time modulated structures (Proof in Appendix \ref{append:relationbetomega}).

\theorem{
\label{theorem:solfromsol}
Let $(\beta,\omega)$ be a solution to the eigenvalue problem (\ref{eq:evb}), then $(\beta+l\beta_m,\omega+l\omega_m)$, where $l\in\mathbb{Z}$ is also a solution.
Moreover if $\mathbf{\Psi}_n$ is the eigenvector at $(\beta,\omega)$ then $\mathcal{S}_\mathcal{U}^{l}\mathbf{\Psi}_n$ is an eigenvector at $(\beta+l\beta_m,\omega+l\omega_m)$.
}

\cor{ All points $(\beta+l\beta_m,\omega+l\omega_m)$ along the line $\omega'=\omega +\nu_m(\beta'-\beta)$ are equivalent in the sense that the eigenvectors are all related by the harmonic shift operator $\mathcal{S}$. 
}

The previous property should not be surprising. The change $\omega\rightarrow \omega+l\omega_m$ and $\beta\rightarrow \beta+l\beta_m$ is equivalent to re-numbering the harmonics. Such property can be exploited to understand how the different modes at a given frequency behave by restricting the analysis to the zero order branches only. Consider for example the dispersion relation shown in Fig. \ref{fig:dispersionmodes}. Suppose one is interested in the modal behaviour at $\omega=1.5\textnormal{ a.u}$. Three modes are highlighted in the Fig. Property \ref{theorem:solfromsol} shows that the (2) eigenvector is a shifted copy of the (2') one. The (2') mode is at the intersection point of the +0 and -1 branches. As is already established when the modulation strength is relatively large, this point corresponds to the center of a bandgap, where there is a strong interaction with the -1 harmonic {\cite{Oliner1961,Cassedy1963}. Therefore, the voltage of this mode will contain non-zero entries only in the -1\textsuperscript{th} and 0\textsuperscript{th} locations. Property \ref{theorem:solfromsol} implies that at $\omega=1.5\textnormal{ a.u}$, the (2) mode has a zero entry in the 0\textsuperscript{th} location and hence does not couple to the excitation at $\omega$.

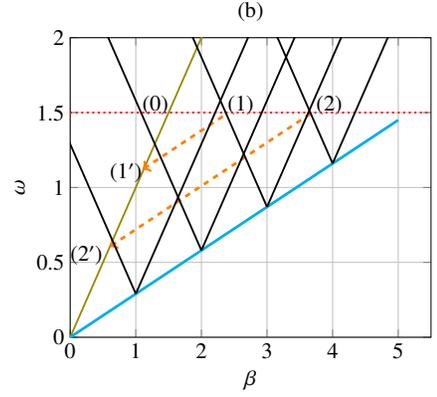
\begin{figure}
\begin{tikzpicture}[scale=0.70]
\def\vm {0.29};
\def \w {0.5};
\begin{axis}[xlabel=\large $\beta$,ylabel=\large $\omega$,title=\large (b),ymin=0,ymax=2.0,xmin=0,grid,cycle list name=exotic ]
\pgfplotsset{every tick label/.append style={font=\large}}
\addplot[samples=51, domain=-5:5,smooth, line width=1pt,olive]{x};
\addplot[samples=51, domain=-5:5,smooth, line width=1pt,olive]{-x};
\addplot[samples=5,domain=-5:5,smooth,line width=1.5pt,color=cyan]{\vm*x};
\addplot[samples=5,domain=1.1:2.4,smooth,line width=1.5pt,color=orange,dashed,<-]{\vm*x+0.8};
\addplot[samples=5,domain=0.6:3.70,smooth,line width=1.5pt,color=orange,dashed,<-]{\vm*x+0.43};
\node at (1.3,1.55) {\large $(0)$};
\node at (2.6,1.55) {\large $(1)$};
\node at (3.95,1.55) {\large $(2)$};
\node at (0.85,1.1) {\large $(1')$};
\node at (0.25,0.55) {\large $(2')$};
\draw [dotted,red,line width=1pt](0,1.5)--(5.5,1.5);
\foreach \xshift in{1,2,3,4}
\addplot[samples=5,smooth, domain=\xshift:2\xshift, line width=1pt]{x-\xshift+\vm*\xshift}
;
\foreach \xshift in{1,2,3,4}
\addplot[samples=5,smooth, domain=-0.5\xshift:\xshift, line width=1pt]{-(x-\xshift)+\vm*\xshift}
;
\end{axis}
\end{tikzpicture}
\caption{The dispersion relation of a RH TL when the modulation strength is infinitesimally small. Different modes at some frequency $\omega=1.5$ are highlighted showing their corresponding counterparts on the main branch, albiet at different frequencies.}
\label{fig:dispersionmodes}
\end{figure}

For the subsequent analysis the $^\textnormal{th}$ eigenvalue and the corresponding eigenvector will be identified by the $(k)$ superscript. In general the mode voltage $\mathcal{V}^{(k)}$ and mode current $\mathcal{I}^{(k)}$ are related via the \emph{Bloch Admittance} matrix $$\bar{\bar{\mathbf{Y}}}=\left[e^{i\beta^{k)} p}\mathbf{e}-\mathbf{D\Gamma}\right]^{-1}\mathbf{C\Gamma}.$$

The total voltage $v^{(k)}[n]$ of the the $k^\textnormal{th}$ mode at any unit cell $n^\textnormal{th}$ is
\begin{equation}
\label{eq:vtotalonline}
v^{(k)}[n]=\sum_{r=-\infty}^{+\infty}\mathcal{V}_r^{(k)}e^{i[\tilde{\omega}_rt-\tilde{\beta}_r^{(k)}np]}+c.c.
\end{equation}
Additionally, the current of the $k^\textnormal{th}$ mode becomes
\begin{align}
{i}^{(k)}[n]&=\sum_{r=-\infty}^{+\infty}\mathcal{I}_r^{(k)}e^{i[\tilde{\omega}_rt-\tilde{\beta}_r^{(k)}np]}+c.c.
\end{align}
The general solution is the superposition of all modes
\begin{align}
v[n]&=\sum_{k=1}^{\infty} a_kv^{(k)}[n]+c.c. \nonumber \\ 
	&=\sum_{k=1}^\infty\sum_{r=-\infty}^\infty a_k\mathcal{V}_r^{(k)}e^{i[\tilde{\omega}_rt-\tilde{\beta}_r^{(k)}np]}+c.c. \label{eq:vexpansion}
\end{align}
and
\begin{align}
i[n]&=\sum_{k=1}^{\infty} a_ki^{(k)}[n]+c.c. \nonumber \\
	&=\sum_{k=1}^\infty\sum_{r=-\infty}^\infty a_k\mathcal{I}_r^{(k)}e^{i[\tilde{\omega}_rt-\tilde{\beta}_r^{(k)}np]}+c.c. \label{eq:iexpansion}
\end{align}

\section{Boundary Value Problem}
\label{sect:bvp}
\begin{figure*}
\centering
\begin{circuitikz}[american voltages,scale=1.25]

\foreach \x in{0,1.2,2.4}
	\draw 
		(\x,0) rectangle (\x+1,1)
		(\x+1,0.1) -- (\x+1.2,0.1)
		(0+\x+1,0.9) -- (0+\x+1.2,0.9);
\draw (0.5,0.5) node{\scriptsize$n=1 $};		
\draw 
	[dotted] (3.6,0.1)--(4,0.1)
	[dotted] (3.6,0.9)--(4,0.9);
\foreach \x in{4,5.2,6.4}
	\draw 
		(\x,0) rectangle (\x+1,1)
		(\x+1,0.1) -- (\x+1.2,0.1)
		(0+\x+1,0.9) -- (0+\x+1.2,0.9);
\draw (6.9,0.5) node{\scriptsize$n=N$};		
\draw 
	(7.6,0.1)--(9,0.1)
	(7.6,0.9)  to [short, i=\scriptsize$i_L(t)$] (9,0.9)
	(9,0.9) to [european resistor,/tikz/circuitikz/bipoles/length=0.6cm, l^=\scriptsize$Z_L$] (9,0.1)
	(8.8,0.9) to [open,v_=\scriptsize$v^{(L)}(t)$] (8.8,0.1);	
\draw 
	(0,0.9)--(-0.3,0.9)
	(-0.3,0.9) to [european resistor,/tikz/circuitikz/bipoles/length=0.6cm, l_=\scriptsize$Z_0$](-1.4,0.9)
	to [sV,l_=\scriptsize$v_s(t)$,/tikz/circuitikz/bipoles/length=0.6cm](-1.4,0.1)--(0,0.1);
\draw 
  (-0.3,0.7) node{\Large$\rightsquigarrow$}
  (-0.35,0.5) node{\scriptsize$V^{(inc)}$}
  (-0.3, 0.2)  node{\Large$\leftsquigarrow$}
  (-0.35,-0.1) node{\scriptsize$V^{(ref)}$};
\draw [dashed,thick,blue](-0.6,1.2)--(-0.6,-0.15);
\draw [dashed,thick,blue](8,1.2)--(8,-0.15);
\node [blue]at (-0.4,1.2){\scriptsize P1};
\node [blue]at (7.8,1.2){\scriptsize P2};
\draw [shade,opacity=0.15,rotate around= {-90:(0.58,0.4)}] (0,-0.2) rectangle(0.97,7.2);  
\end{circuitikz}
\caption{$N$ unit cells of a space time periodic structure connected to an input voltage source with a reference impedance $Z_0$ and a load of impedance $Z_L$.}
\label{fig:bvp}
\end{figure*}
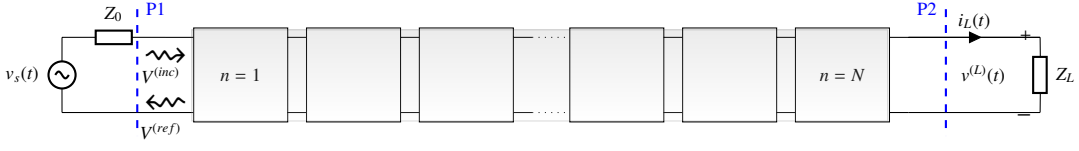
Consider the structure in Fig. \ref{fig:bvp} that represents a generic space-time modulated structure that is connected to a source and load. The incident $(v^{(inc)},i^{(inc)})$ and reflected $(v^{(ref)},i^{(ref)})$ waves appear on a transmission line of characteristic impedance $Z_0$ (usually a 50 $\Omega$ microstrip or coaxial TL) that connects the structure to a voltage source $v_s(t)$. At the input side,
\begin{align*}
v_s(t)&=v^{(inc)}(t)+v^{(ref)}(t)+Z_0\left[i^{(inc)}(t)+i^{(ref)}(t)\right]\\
&=2v^{(inc)}(t),
\end{align*}
where the last equality stems from the fact that $i^{(inc)}(t)=v^{(inc)}/Z_0$ and $i^{(ref)}=-v^{(ref)}/Z_0$.

At the input of the first unit cell ($x=0$), the boundary conditions are imposed by requiring the voltage and current to be continuous i.e, 
\begin{align*}
v(x=0^-)&=v(x=0^+) \textnormal{ and }\\ i(x=0^-)&=i(x=0^+).
\end{align*}

We consider sinusoidal excitation, i.e, $v_s(t)=V_s\cos(\omega t+\phi)$. Therefore,
\begin{align}
\frac{V_s}{2}e^{i(\omega t+\phi)}+v^{(ref)}+c.c.	 &=\sum_{k=1}^\infty\sum_{r=-\infty}^\infty a_k\mathcal{V}_r^{(k)}e^{i\tilde{\omega}_rt}+c.c. \label{eq:Vat0}
\end{align}
and
\begin{align}
\frac{V_s}{2}e^{i(\omega t+\phi)}-v^{(ref)}+c.c.&=Z_0\sum_{k=1}^\infty\sum_{r=-\infty}^\infty a_k\mathcal{I}_r^{(k)}e^{i\tilde{\omega}_rt}+c.c. \label{eq:Iat0}
\end{align}
At the load side $v(x=Np^-)=v(x=Np^+)$. Therefore,
\begin{align}
\sum_{k=1}^\infty\sum_{r=-\infty}^\infty a_k\mathcal{V}_r^ke^{i[\tilde{\omega}_rt-\tilde{\beta}_rNp]}+c.c&=
v^{(L)}(t) \label{eq:VatN}
\end{align}
and
\begin{align}
Z_L\sum_{k=1}^\infty\sum_{r=-\infty}^\infty a_k\mathcal{I}_r^ke^{i[\tilde{\omega}_rt-\tilde{\beta}_rNp]}+c.c&=
v^{(L)}. \label{eq:IatN}
\end{align}
In general $v^{(ref)}$ and $v^{(L)}$ can be written as
\begin{equation}
v^{(ref)}=\sum_{r=-\infty}^\infty \tilde{V}^{(ref)}_re^{i\tilde{\omega}_rt}
\end{equation}
and
\begin{equation}
v^{(L)}=\sum_{r=-\infty}^\infty \tilde{V}_r^{(L)}e^{i\tilde{\omega}_rt}.
\end{equation}

Adding (\ref{eq:Vat0}) and (\ref{eq:Iat0})
\begin{equation}
\sum_{k=1}^\infty\sum_{r=-\infty}^\infty a_k\left(\mathcal{V}_r^{(k)}+Z_0\mathcal{I}_r^{(k)}\right)e^{\tilde{\omega}_rt}+c.c=V_se^{i(\omega t+\phi)}+c.c.
\end{equation}
Equating the $\exp(i\tilde{\omega}_rt)$ coefficients:
\begin{align}
\sum_{k=1}^\infty a_k\underbrace{\left(\mathcal{V}_r^{(k)}+Z_0\mathcal{I}_r^{(k)}\right)}_{{\color{blue}{2\mathcal{V}^{inc1,k}_r}}}=V_se^{i\phi}\delta_r^0,~r=\cdots,-2,-1,0,1,2,\cdots,
\end{align}
where $\mathcal{V}^{inc1,k}_r$ is the contribution of the $k^\textnormal{th}$ mode to the wave incident on P1. Therefore, the above set of equations can be written as
\begin{equation}
\label{eq:bvpinfinite}
\sum_{k=1}^\infty \mathcal{V}_r^{inc1,k}a_k=V^{inc}\delta_r^0,~r=\cdots,-2,-1,0,1,2,\cdots,
\end{equation}
where $V^{inc}=V_se^{i\phi}/2$. Equation (\ref{eq:bvpinfinite}) shows that the the coefficients $a_k$ are such that the net effect of the branches is balanced with the excitation at frequency $\omega$ and they \emph{distructively} interfere at any other harmonics.

Subtracting (\ref{eq:IatN}) from (\ref{eq:VatN})  and matching the $\tilde{\omega}_r$ harmonic:

\begin{equation}
\label{eq:bvpinfiniteatload}
\sum_{k=1}^{\infty}a_k\underbrace{\left(\mathcal{V}_r^{(k)}-Z_L\mathcal{I}_r^{(k)}\right)}_{{\color{red}2\mathcal{V}^{inc2,k}_{r}}}e^{-i\tilde{\beta}_r^{(k)}Np}=0.
\end{equation}
The term in bracket represents the wave reflected from the load $Z_L$ when the output port is terminated in $Z_L$. For the remaining of the manuscript, it is assumed that $Z_L=Z_0$ (i.e., the structure is terminated in the reference impedance $Z_0$). Therefore, (\ref{eq:bvpinfiniteatload}) implies that the $a_k$ coefficients are the ones that result in a null reflection from the load at all harmonics.

In practical applications, only a limited number $N_\textnormal{H}$ of harmonics are significant. For convenience, we consider $N_\textnormal{H}$ to be an odd number $2\mathcal{N}_s+1$, where $\mathcal{N}_s=0,1,2,\cdots$. This selection allows the symmetric inclusion of harmonics from $-\mathcal{N}_s$ to $\mathcal{N}_s$. Furthermore, We consider the number of branches to be $2N_\textnormal{H}$ to account for forward and backward waves. For each harmonic, the truncated versions of (\ref{eq:bvpinfinite}) and (\ref{eq:bvpinfiniteatload}) provide two equations in the $2N_\textnormal{H}$ coefficients $a_k$. Taking all $N_\textnormal{H}$ harmonics into account, we end up with a system of $2N_\textnormal{H}$ equations in $2N_\textnormal{H}$ unknowns that can be written as
\begin{equation}
\label{eq:bvp}
\begin{pmatrix}
\mathcal{V}_{-\mathcal{N}_s}^{inc1,1} &\mathcal{V}_{-\mathcal{N}_s}^{inc1,2} &\cdots &\mathcal{V}_{-\mathcal{N}_s}^{inc1,2N_\textnormal{H}}\\

\vdots &\vdots &\vdots &\vdots\\
\mathcal{V}_{0}^{inc1,1} &\mathcal{V}_{0}^{inc1,2} &\cdots &\mathcal{V}_{0}^{inc1,2N_\textnormal{H}}\\
\vdots &\vdots &\vdots &\vdots\\
\mathcal{V}_{\mathcal{N}_s}^{inc1,1} &\mathcal{V}_{\mathcal{N}_s}^{inc1,2} &\cdots &\mathcal{V}_{\mathcal{N}_s}^{inc1,2N_\textnormal{H}}\\

\mathcal{V}_{-\mathcal{N}_s}^{inc2,1} &\mathcal{V}_{-\mathcal{N}_s}^{inc2,2} &\cdots &\mathcal{V}_{-\mathcal{N}_s}^{inc2,2N_\textnormal{H}}\\

\vdots &\vdots &\vdots &\vdots\\
\mathcal{V}_{0}^{inc2,1} &\mathcal{V}_{0}^{inc2,2} &\cdots &\mathcal{V}_{0}^{inc2,2N_\textnormal{H}}\\
\vdots &\vdots &\vdots &\vdots\\
\mathcal{V}_{\mathcal{N}_s}^{inc2,1} &\mathcal{V}_{\mathcal{N}_s}^{inc2,2} &\cdots &\mathcal{V}_{\mathcal{N}_s}^{inc2,2N_\textnormal{H}}
\end{pmatrix}
\begin{pmatrix}
a_1\\ \vdots \\ a_{\mathcal{N}_s+1} \\ \vdots \\ a_{N_\textnormal{H}}\\a_{N_\textnormal{H}+1}\\ \vdots \\ \vdots \\ \vdots \\ a_{2N_\textnormal{H}}
\end{pmatrix}
=
\begin{pmatrix}
0\\
\vdots\\
V^{(inc)}\\
\vdots\\
0\\
\vdots\\
\vdots\\
\vdots\\
\vdots\\
0
\end{pmatrix}
\end{equation}

Furthermore, the output contains the different harmonics $\tilde{\omega}_r$. Therefore, the transmisson coefficient of the $r^\textnormal{th}$ harmonic $S_{21}^{(r,0)}$ is defined to be

\begin{equation}
\label{eq:s21}
S_{21}^{(r,0)}=\frac{\tilde{V}_r^{(L)}}{V^{(inc)}},
\end{equation}
where from (\ref{eq:VatN})
$$\tilde{V}_r^{(L)}=\sum_{k=1}^{\infty}a_k\mathcal{V}_r^{(k)}e^{-i\tilde{\beta}_r^{(k)}Np}.$$
Similarly the reflection coefficient in the $r^\textnormal{th}$ harmonic is
$$S_{11}^{(r,0)}=\frac{\tilde{V}_r^{(ref)}}{V^{(inc)}},$$
where
$$\tilde{V}_r^{(ref)}=-V^{(inc)}\delta_0^r+\sum_{k=1}^\infty a_k\mathcal{V}_r^{(k)}.$$

\section{Results and Discussion}
\label{sect:results}
To verify and demonstrate the use of the machinery developed in Sections \ref{sect:DispersionEVP} and \ref{sect:bvp}), we will apply the framework to analyze two main structures. The first is a space-time periodic CRLH TL. Such idealistic model allows a thorough analysis of the propagation behaviour that can be compared with state space time domain simulations. Next, we use the framework to reproduce and give insight into the nonreciprocial behaviour observed on a nonlinear right handed transmission line (NL RH TL) that has been manufactured in our lab. Although the analysis carried out below is not meant to be exhaustive, it provides a useful and a systematic procedure to describe complex space-time periodic structures.
\subsection{Composite Right-Left Handed Space-time modulated TL}
\label{subsect:crlh}
The CRLH consists of $N=40$ unit cells as one shown in Fig. \ref{fig:CRLHstructure}, where the right handed capacitance $C_R$ is space-time modulated. The first unit cell is connected to a source of impedance $50~\Omega$. The load is also assumed to be $50~\Omega$. KCL and KVL along with the current and voltage relations in the time domain are used to derive a state space model (SSM) of the circuit that can be written as 
$$\dot{\mathbf{x}}=\mathbf{A}(t)\mathbf{x}+\mathbf{B}(t)u,$$
where $\mathbf{x}$ is an $N\times 1$ vector that stores the state variables (current in inductors and voltages across capacitors), $\mathbf{A}$ is a $N\times N$ matrix, $\mathbf{B}$ is a $N\times 1$ vector that connects the input excitation to the states.
\begin{figure}[!htb]
\centering
\ctikzset{/tikz/circuitikz/bipoles/length=0.8cm}
\begin{tikzpicture}[scale=0.7]
\draw (0,0) to [L,l_=\scriptsize $L_R$] (2,0)
      to [C,l^=\scriptsize $C_L$] (4,0);
\draw (4,0) to [L, l_=\scriptsize $L_L$] (4,-2);
\draw (5,0) to [C,l^=\scriptsize $C_R$] (5,-2);
\draw [-o](4,0)--(5.9,0);
\draw [o-o](-0.4,-2) --(5.9,-2);
\draw [o-] (-0.4,0) -- (0,0);
\draw [shade, yellow,opacity=0.3] ((4.4,-2.2) rectangle (5.5,0.2);
\end{tikzpicture}
\caption{Unit cell of a space-time modulated CRLH TL. The Right handed Capacitance $C_R$ is modulated as a travelling wave $C_R=C_{R0}\left[1+M\cos(\omega_mt-\beta_mn)\right]$.}
\label{fig:CRLHstructure}
\end{figure}
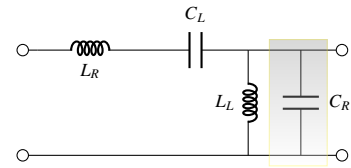
\begin{figure}[!t]
\centering
\begin{subfigure}[b]{1\linewidth}
\includegraphics[width=\linewidth]{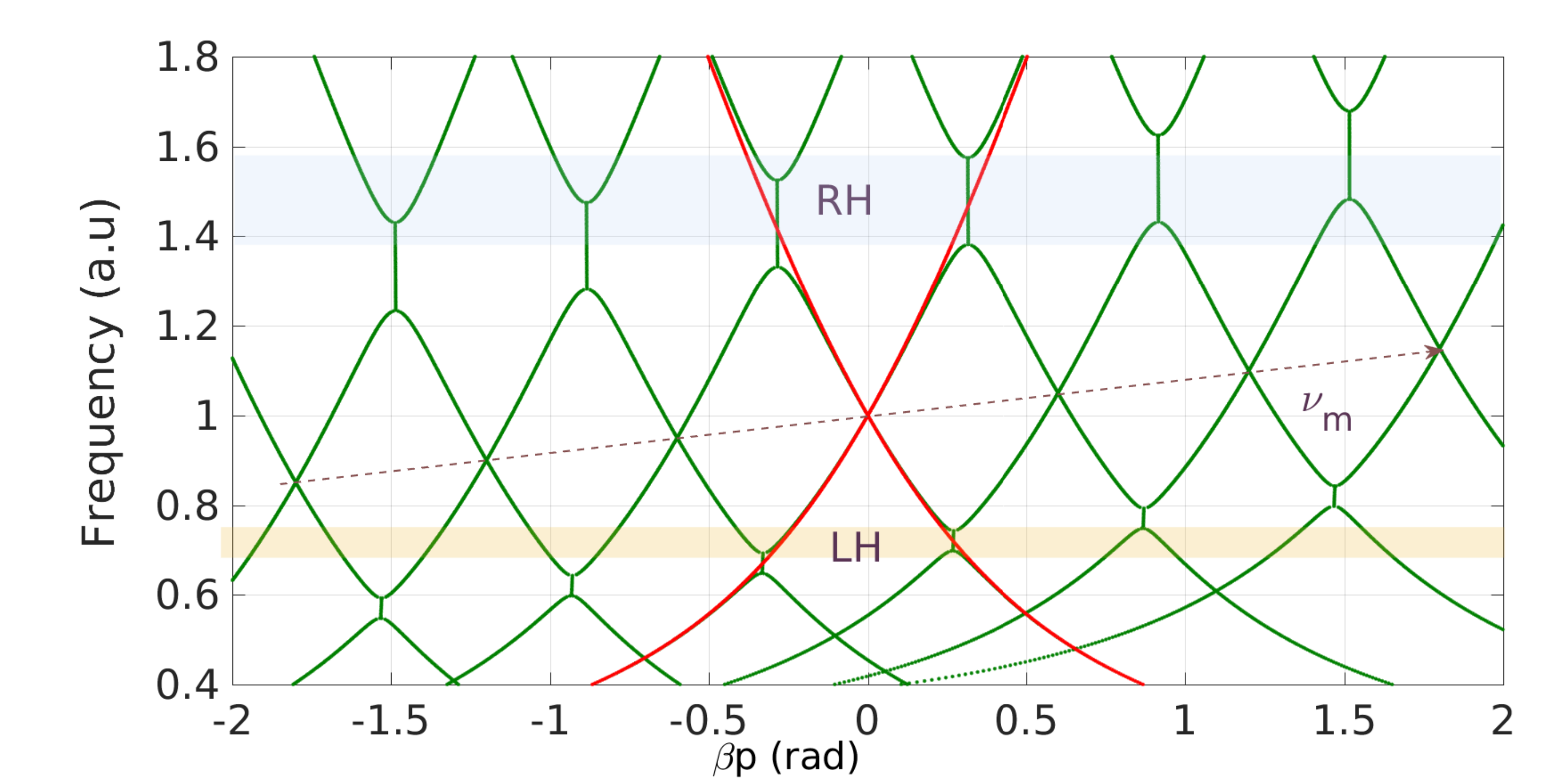}
\caption{Dispersion Relation of a CRLH TL modulated by a forward travelling wave.}
\end{subfigure}

\begin{subfigure}[b]{0.49\linewidth}
\includegraphics[width=\linewidth]{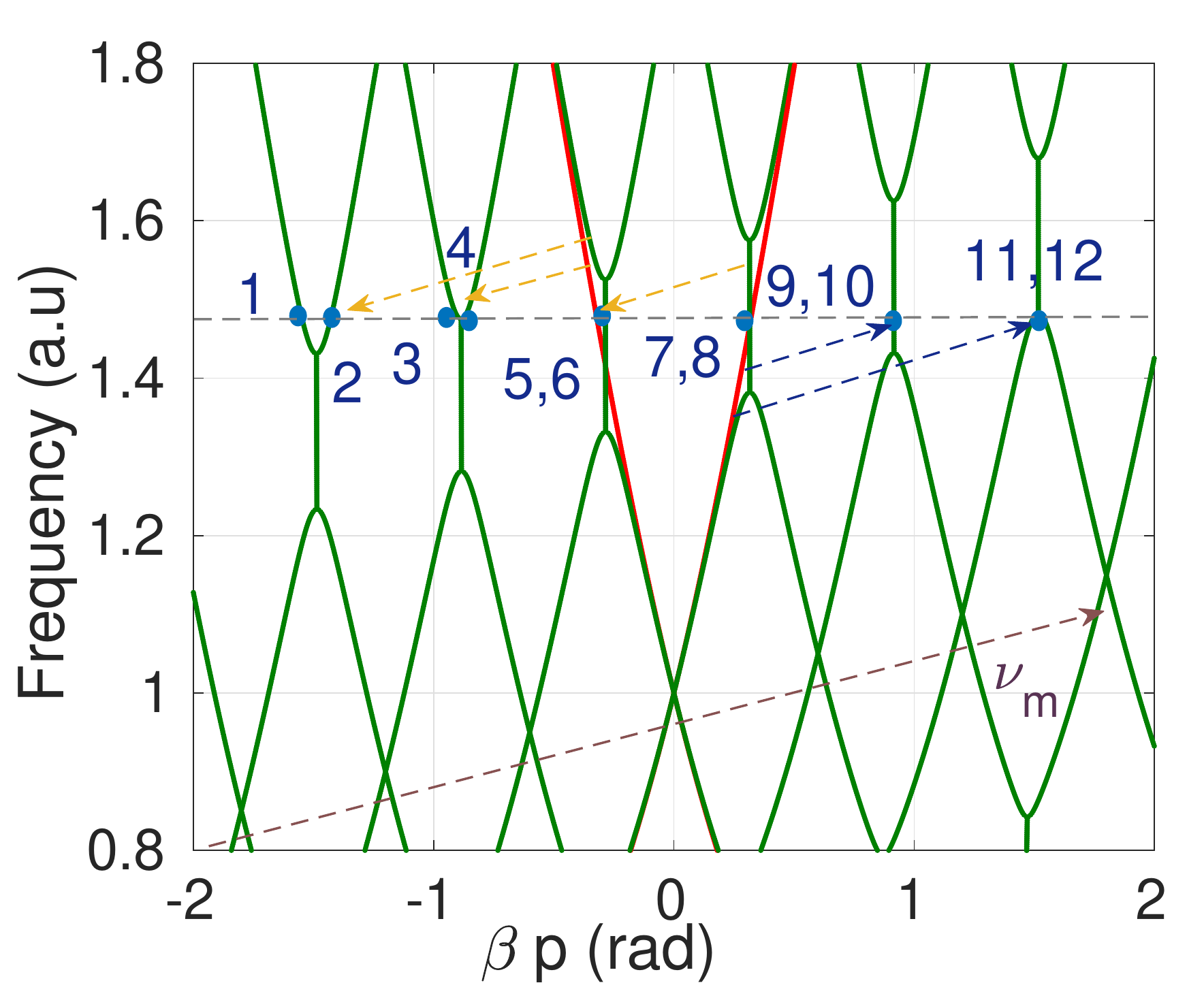}
\caption{Zoomed view of the RH region.}
\end{subfigure}
\begin{subfigure}[b]{0.49\linewidth}
\includegraphics[width=\linewidth]{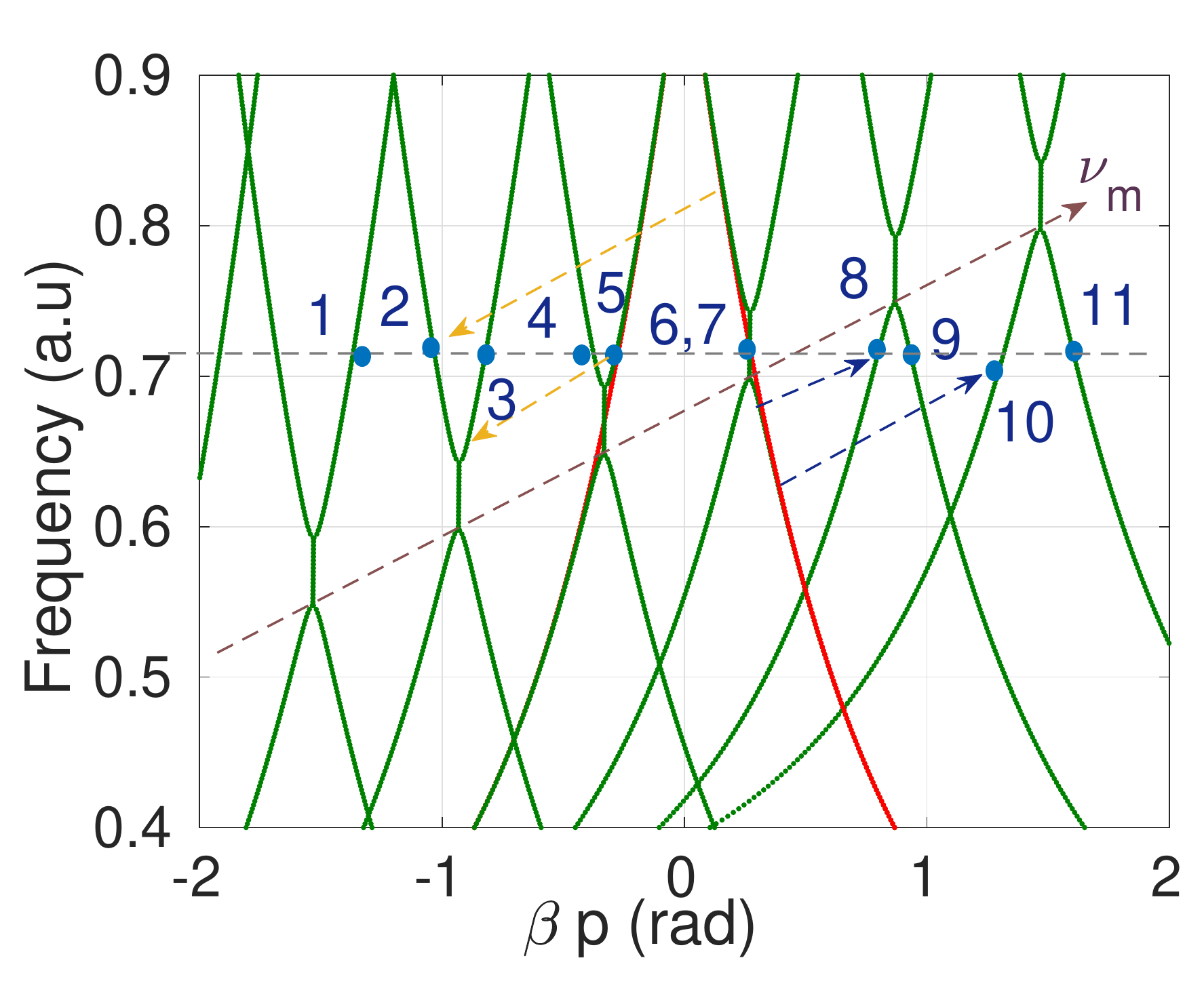}
\caption{Zoomed view of the LH region.}
\end{subfigure}
\caption{The Dispersion Relation of a CRLH TL modulated by a forward travelling wave. $M=0.4, \omega_{se}=\omega_{sh}=1~\textnormal{ a.u.},\omega_{RH}=2.5~\textnormal{a.u.}$.}
\label{fig:DispersionCRLH}
\end{figure}
The unit cell shown in Fig. \ref{fig:CRLHstructure} can be divided into three sub-units: (1) the LTI series impedance $Z_{se}\triangleq i\omega L_R-i/\omega C_L$, (2) shunt admittance $1/\omega L_L$, and (3) the LTP admittance $\tilde{\mathbf{Y}}_R$. Hence, the ABCD matrix of the unit cell can be formed by cascading its three sub-units. Therefore, the different eigenvalues $e^{i\beta^{(k)}p}$ and the corresponding eigenvectors $(\mathcal{V}^{(k)},\mathcal{I}^{(k)})^t$ are determined through the use of (\ref{eq:evb}). Subsequently, when the TL is excited by a sinusoidal source of frequency $\omega$, the boundary value problem (\ref{eq:bvp}) is solved and the modes coefficients $a_k$ are computed.

The LTP dispersion relation when $C_R$ is sinusoidally modulated, i.e, $C_R=C_{R0}\left[1+M\cos(\omega_mt-\beta_m np)\right]$ is obtained from the eigenvalues as depicted in Fig. \ref{fig:DispersionCRLH} (a). The LTI dispersion relation (when $M=0$) is superimposed to highlight the right hand(RH) and left hand (LH) regions. The LH region is in the low frequency range, frequencies below 1 a.u., where the phase and group velocities are opposite \cite{eleftheriades2005,CalozItoh}. For a balanced operation, the series and shunt resonances were both set to unity \cite{CalozItoh}. Figs. \ref{fig:DispersionCRLH} (b) and (c) show a close up view of the dispersion relation in the RH and LH regions, respectively. Different modes are highlighted and labelled. It is worth noting that whenever two branches intersect, degeneracy occurs and two of the eigenvalues form a complex conjugate pair. The point of intersection represents the center of a bandgap, where strong coupling with time harmonics may be significant. For instance consider Fig. \ref{fig:DispersionCRLH} (b), the points $7$ and $8$ represents two eigenvectors that have two complex conjugate eigenvalues. 

\begin{figure}[!t]
\centering
\begin{subfigure}[b]{0.49\linewidth}
\includegraphics[width=\linewidth]{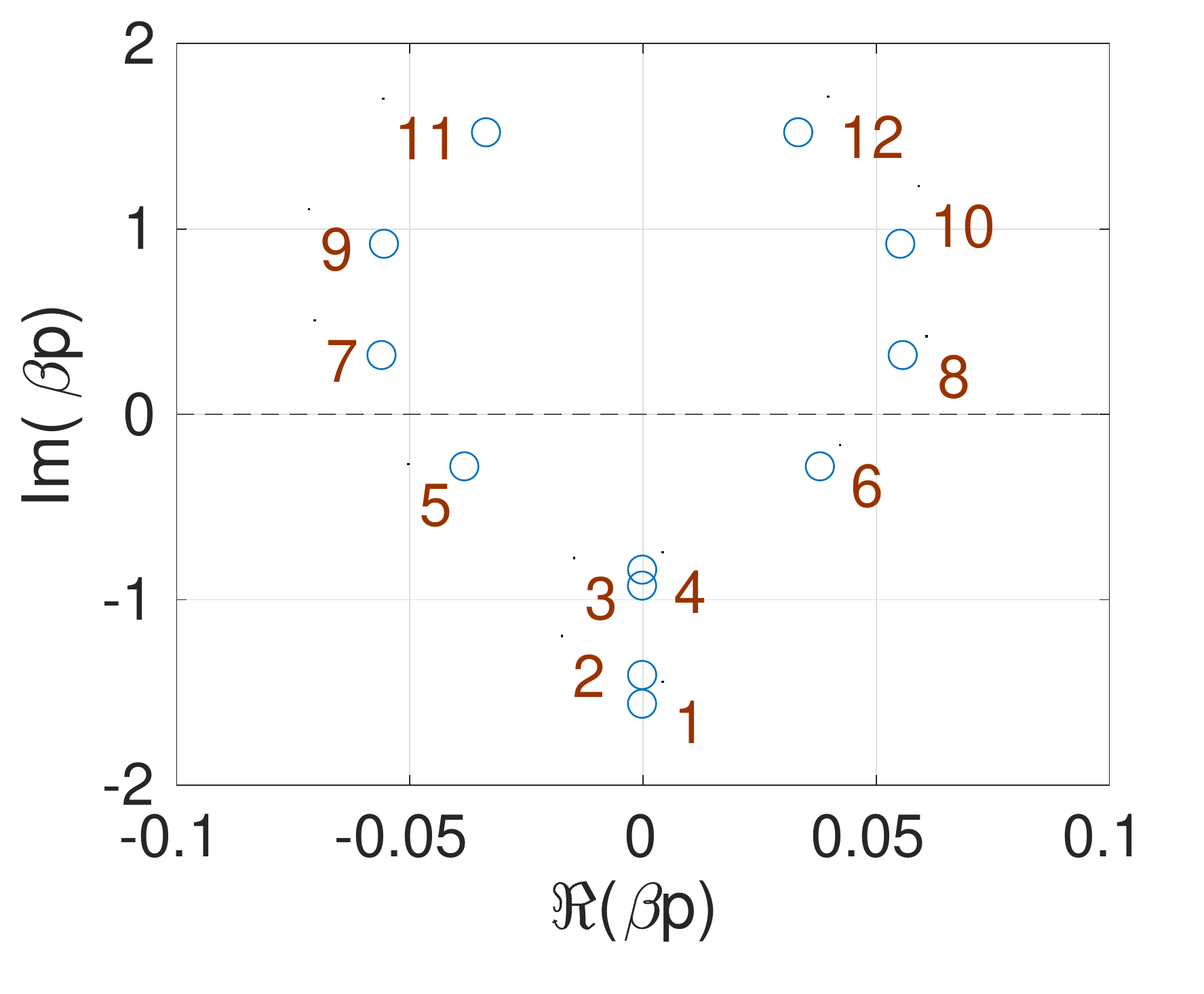}
\caption{}
\end{subfigure}
\begin{subfigure}[b]{0.49\linewidth}
\includegraphics[width=\linewidth]{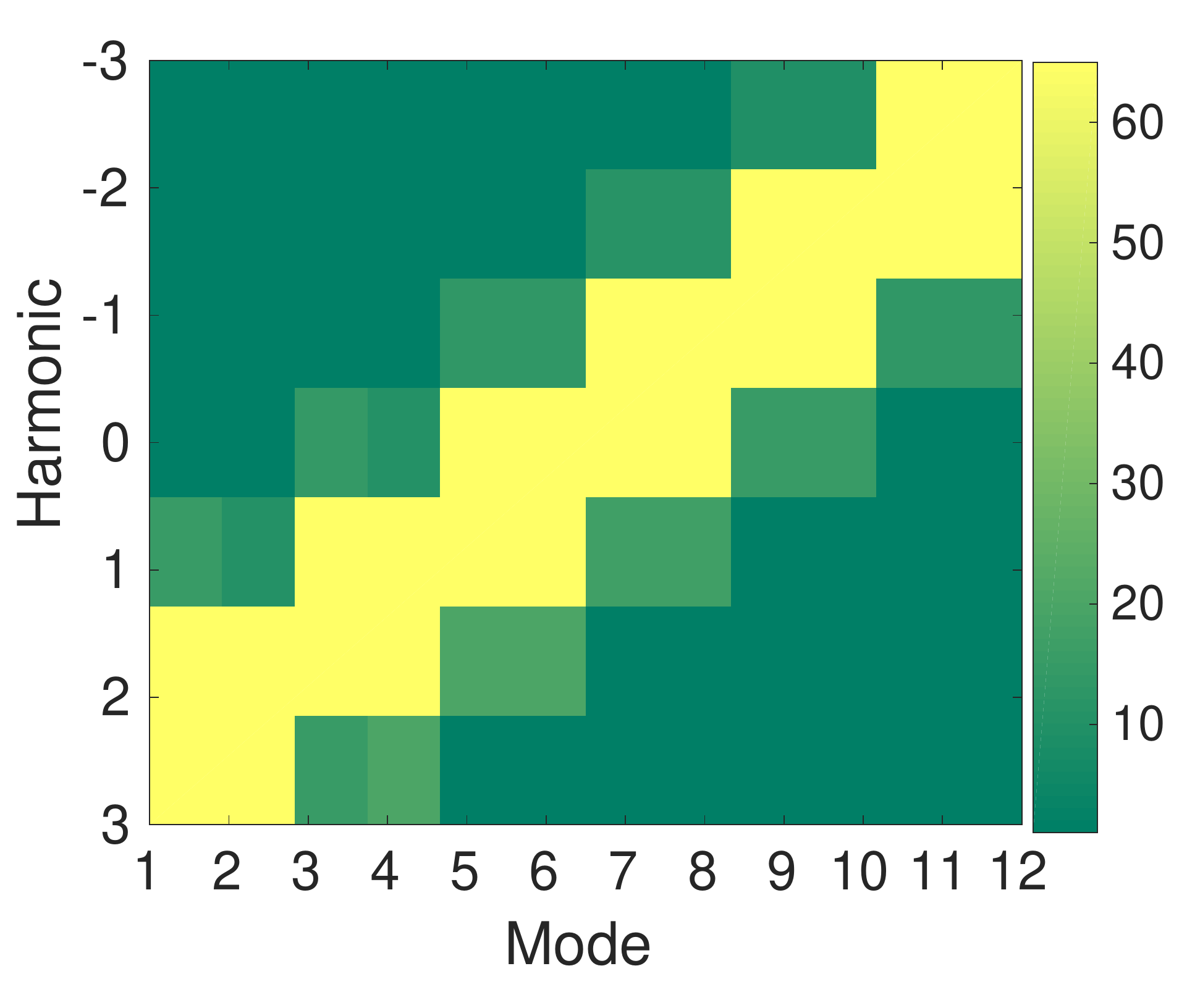}
\caption{}
\end{subfigure}

\begin{subfigure}[b]{0.49\linewidth}
\includegraphics[width=\linewidth]{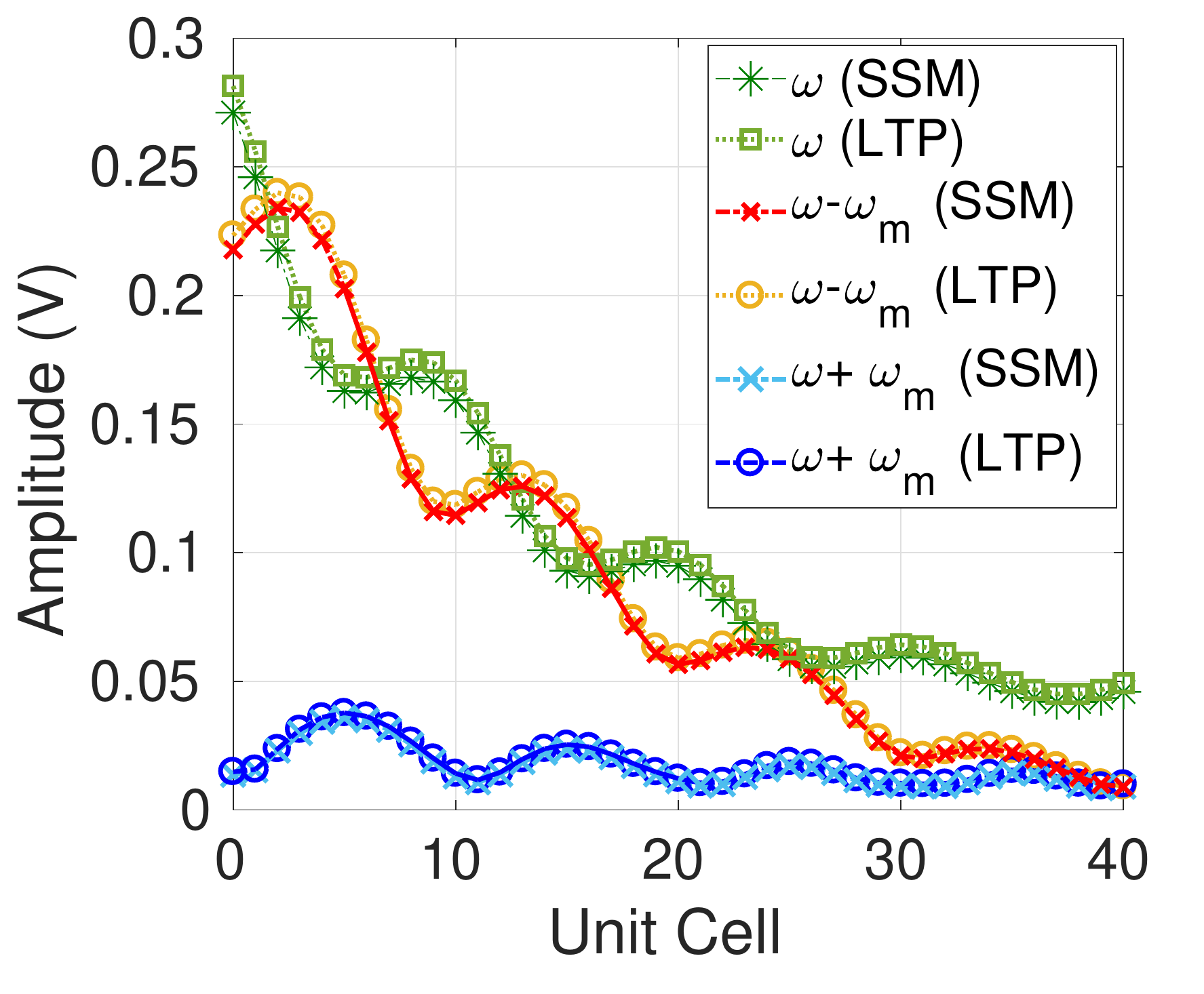}
\caption{}
\end{subfigure}
\begin{subfigure}[b]{0.49\linewidth}
\includegraphics[width=\linewidth]{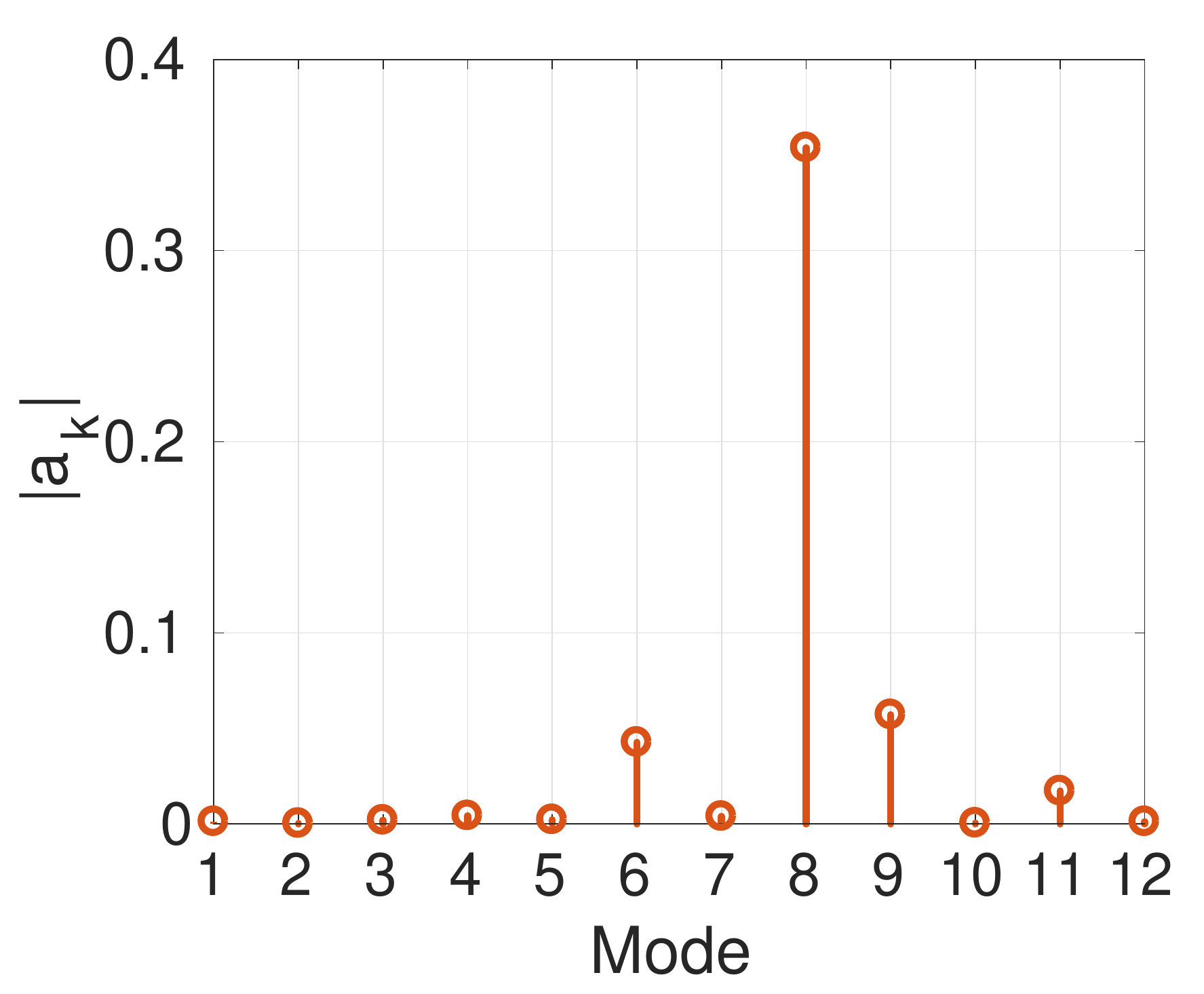}
\caption{}
\end{subfigure}
\caption{Computed eigenvalues, eigenvectors and waveforms at the center of the RH BG of the LTP CRLH TL, when $M=0.4$. (a) Real and imaginary of $\beta p$ for different modes. (b)  Magnitude of different components of eigenvectors $\mathcal{V}^{(k)}$. (c) Amplitude of waves at signal frequency $\omega$, the $\omega-\omega_m$ and $\omega+\omega$ harmonics. (d) Magnitude of $a_k$.}
\label{fig:waveformsrhcrlh}
\end{figure}

The eigenvalues, when the frequency is at the center of the RH BG ($f=1.5\textnormal{ a.u}$), are depicted in Fig. \ref{fig:waveformsrhcrlh}(a). Unlike the first four, the higher eigenvalues appear as complex conjugate pairs. This is not surprising since they correspond to points inside BGs as Fig. \ref{fig:DispersionCRLH}(b) shows. For any given eigenvalue, the magnitude of the components of the corresponding eigenvector is plotted in Fig. \ref{fig:waveformsrhcrlh}(b). For a given eigenvector (mode), the y-axis represents the strength of the $r^\textnormal{th}$ harmonic. According to (\ref{eq:vexpansion}), the waveform inside the space-time periodic structure is the linear superposition of the different eigenvectors. Figure. \ref{fig:waveformsrhcrlh}(d) plots the magnitude of the expansion coefficient $a_k$. Clearly, the wave behaviour is dominated by the $8^\textnormal{th}$ eigenvector, which corresponds to one of the modes inside the BG of the main branch as illustrated in Fig. \ref{fig:DispersionCRLH}(b). Fig. \ref{fig:waveformsrhcrlh}(b) shows that the $0^\textnormal{th}$ and $-1^\textnormal{th}$ harmonics are dominant and of the same order of magnitude, inferring a strong an interaction between the fundamental and its $-1^\textnormal{th}$ harmonic. Additionally, there is a small contribution coming from the 6\textsuperscript{th} and 9\textsuperscript{th} modes. The Fig. also shows that the $6^\textnormal{th}$ and $9^\textnormal{th}$ modes have significant components at the 0,+1, and 0,-1 enteries, respectively. Note that the behaviour of the two modes can be also deduced from property \ref{theorem:solfromsol}. Indeed, the equivalent mode of the $6^\textnormal{th}$ ($9^\textnormal{th}$) one on the main branch is inside the first BG. Hence the equivalent mode has a non-zero -1, and 0 entries. Since the eigenvector of the $6^\textnormal{th}$ ($9^\textnormal{th}$) mode is a down (up) shifted copy, it has non-zero values at the +1,0 (-1,0) entries, in agreement with Fig. \ref{fig:waveformsrhcrlh}(b).

\begin{figure}[!t]
\centering
\begin{subfigure}[b]{0.49\linewidth}
\includegraphics[width=\linewidth]{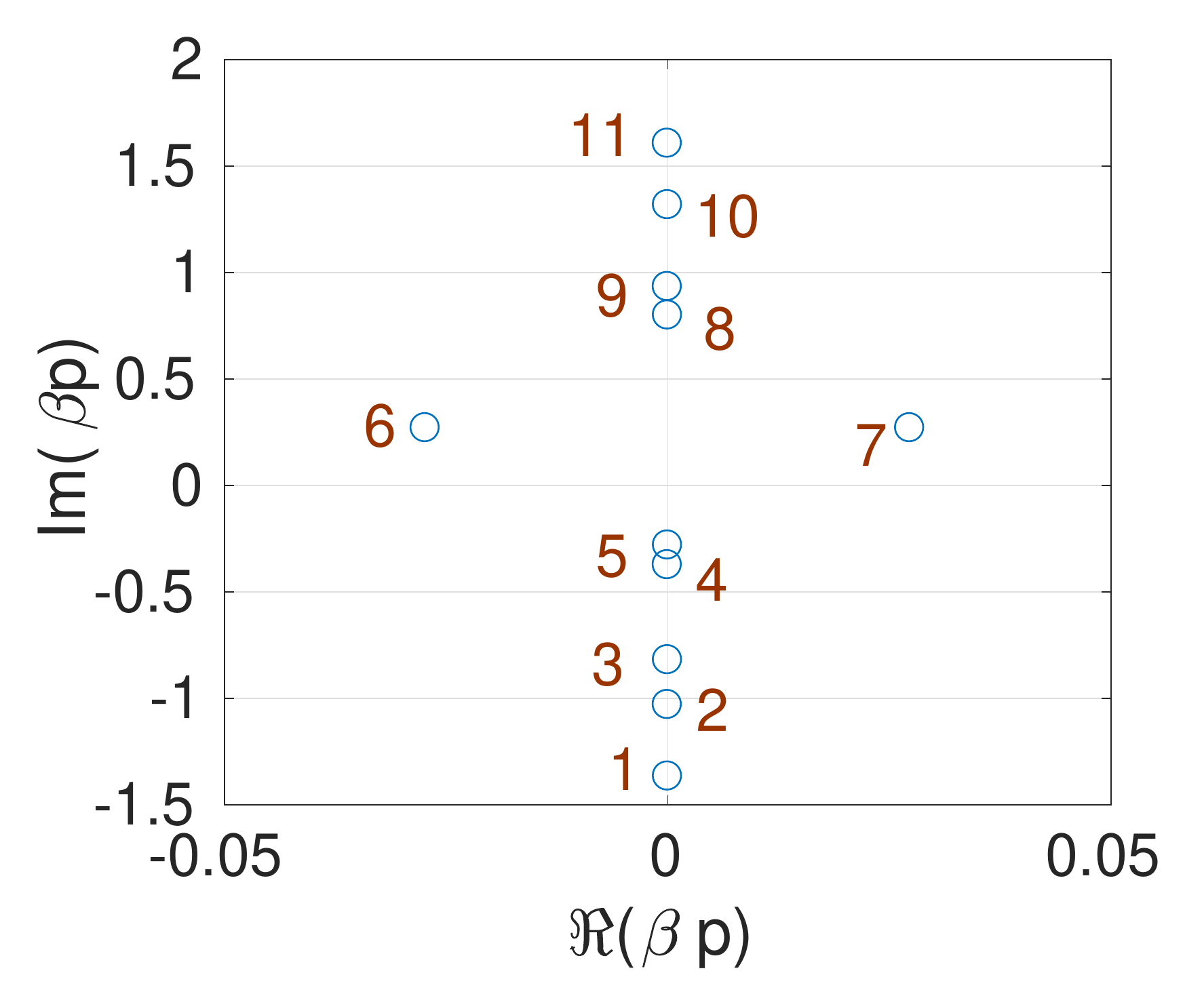}
\caption{}
\end{subfigure}
\begin{subfigure}[b]{0.49\linewidth}
\includegraphics[width=\linewidth]{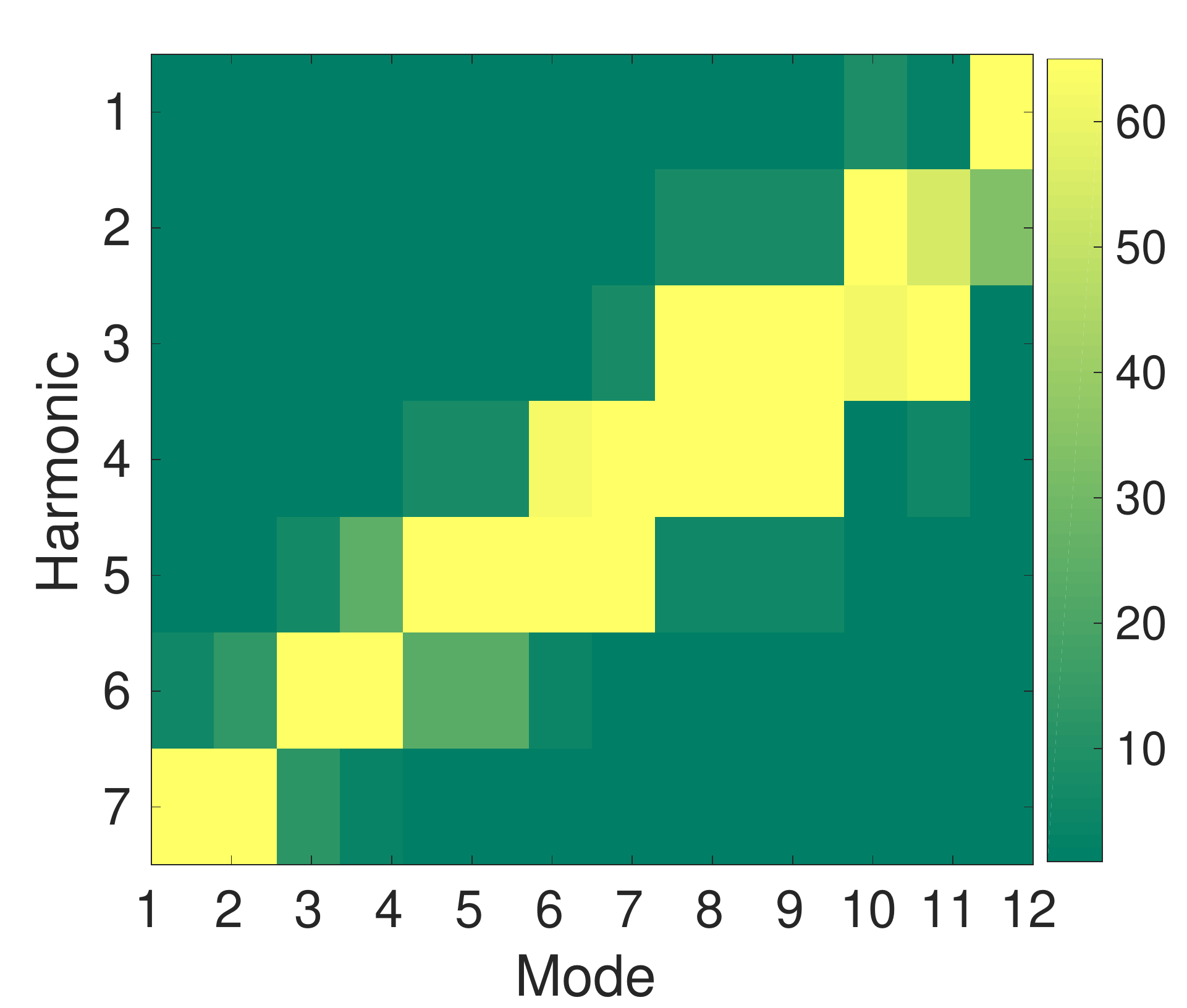}
\caption{}
\end{subfigure}

\begin{subfigure}[b]{0.49\linewidth}
\includegraphics[width=\linewidth]{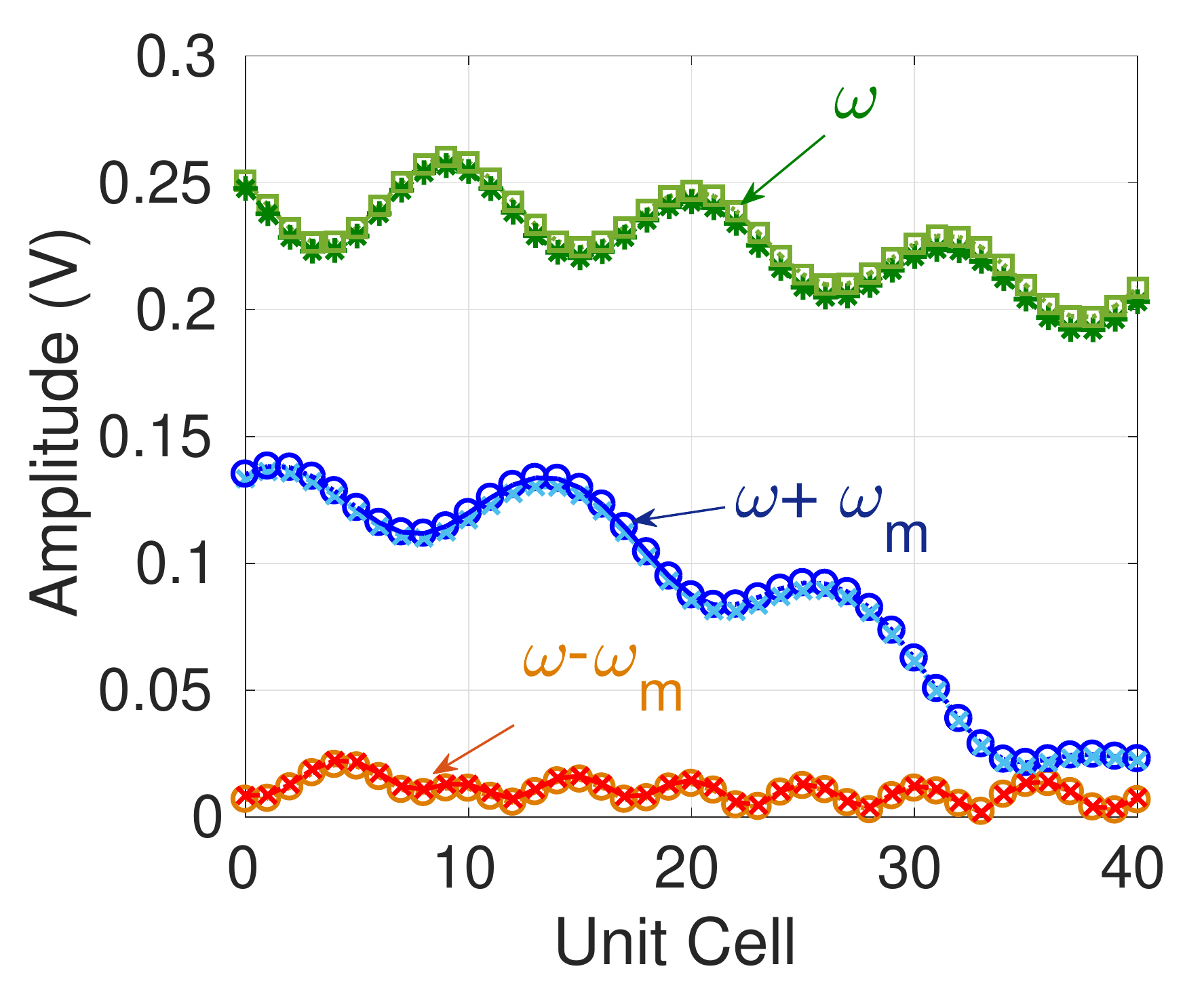}
\caption{}
\end{subfigure}
\begin{subfigure}[b]{0.49\linewidth}
\includegraphics[width=\linewidth]{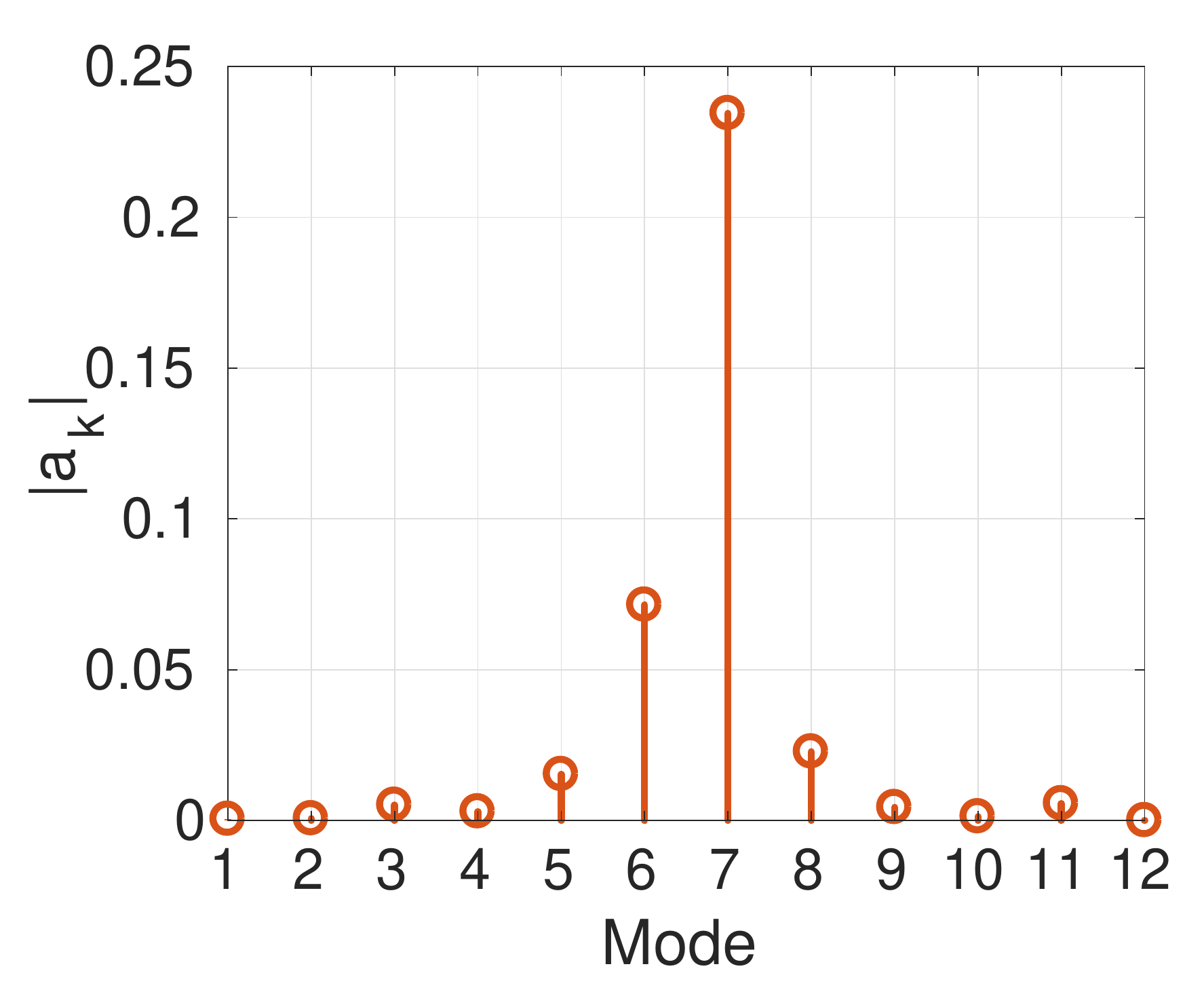}
\caption{}
\end{subfigure}
\caption{Computed eigenvalues, eigenvectors and waveforms at the center of the LH BG of the LTP CRLH TL, when $M=0.4$. (a) Real and imaginary of $\beta p$ for different modes. (b)  Magnitude of different components of eigenvectors $\mathcal{V}^{(k)}$. (c) Amplitude of waves at signal frequency $\omega$, the $\omega-\omega_m$ and $\omega+\omega$ harmonics. (d) Magnitude of $a_k$.}
\label{fig:waveformslhcrlh}
\end{figure}

To assess how accurate the LTP approach can predict the wave behaviour inside the structure, the waveform at the middle of the RH BG, at the three frequencies $\omega$, $\omega-\omega_m$ and $\omega+\omega_m$ are calculated using (\ref{eq:vexpansion}) and compared with the solution of the SSM. The time domain data obtained from the SSM simlation is transformed to the frequency domain, where the frequencies of interest are isolated. Figure \ref{fig:waveformsrhcrlh}(c) reports the amplitude of the three harmonics. As shown, there is an excellent agreement between LTP and SSM. Additionally, the amplitude of the main component at $\omega$ rapidly decreases inside as the wave penetrates into the structure, where it is scattered (mainly) in the -1 harmonic back to the source. Furthermore, there is a non vanishing contribution, coming from the +1 harmonic, as a result of the excitation of the 6\textsuperscript{th} mode.

The same procedure is repeated but for $\omega$ at the center of the LH BG (Fig. \ref{fig:DispersionCRLH}(c)). Unlike the RH BG, the incident and modulating waves are contra-directional. This is due to the left handedness of the CRLH in this regime. Therefore, the incident wave scatters in the $\omega+\omega_m$ (blue shifted) as Fig. \ref{fig:waveformslhcrlh}(c) highlights. The scattering, however, is not as strong as in the RH BG case. This is due to the smaller magnitude of the real part of the eigenvalue (Fig. \ref{fig:waveformslhcrlh}(a)) and witnessd by the slight reduction of the amplitude of the fundamental component (Fig. \ref{fig:waveformslhcrlh}(c)). It is expected that a modulation of $L_L$ or $C_L$ will result in a wider LH BG.

Finally, the transmission coefficent is calculated via (\ref{eq:s21}) and from the voltage and current time series obtained from the SSM computation over a wide frequency range that includes both the RH and LH BGs. Fig. \ref{fig:S21CRLH}(a) and (b) present the results, when the incident and modulation waves are co and contra-directional, respectively. The Figs. show that LTP based calculations are in a very good agreement with SSM. Furthermore, space time modulation has the effect of attenuating the transmitted signal (-15 dB for the LH BG and -30 dB for the RH BG), compared to almost 0 dB when modulation is absent.

\begin{figure}[!t]
\centering
\begin{subfigure}[b]{0.6\linewidth}
\includegraphics[width=\linewidth]{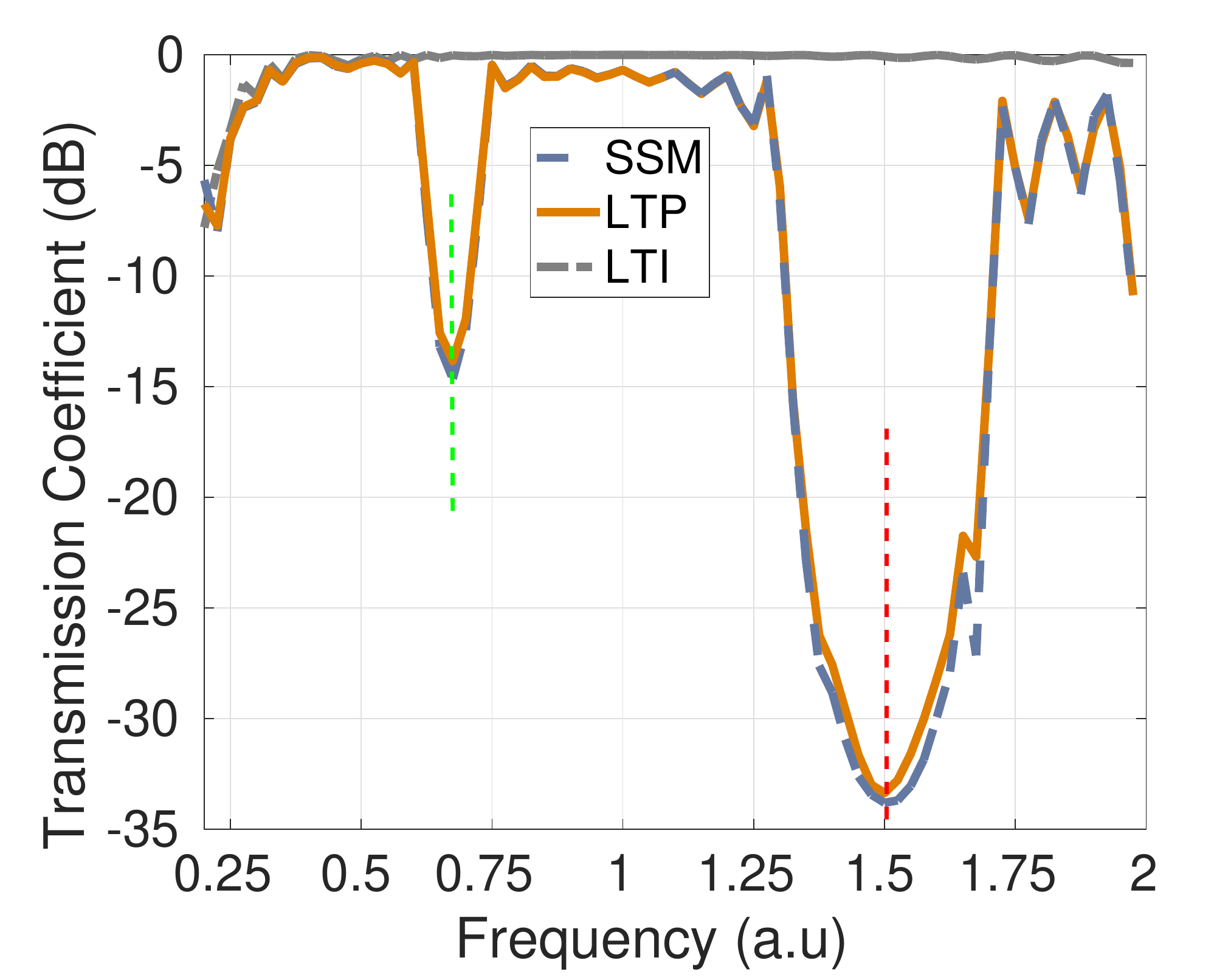}
\caption{Forward modulation (i.e, $\beta_m>0$).}
\end{subfigure}

\begin{subfigure}[b]{0.6\linewidth}
\includegraphics[width=\linewidth]{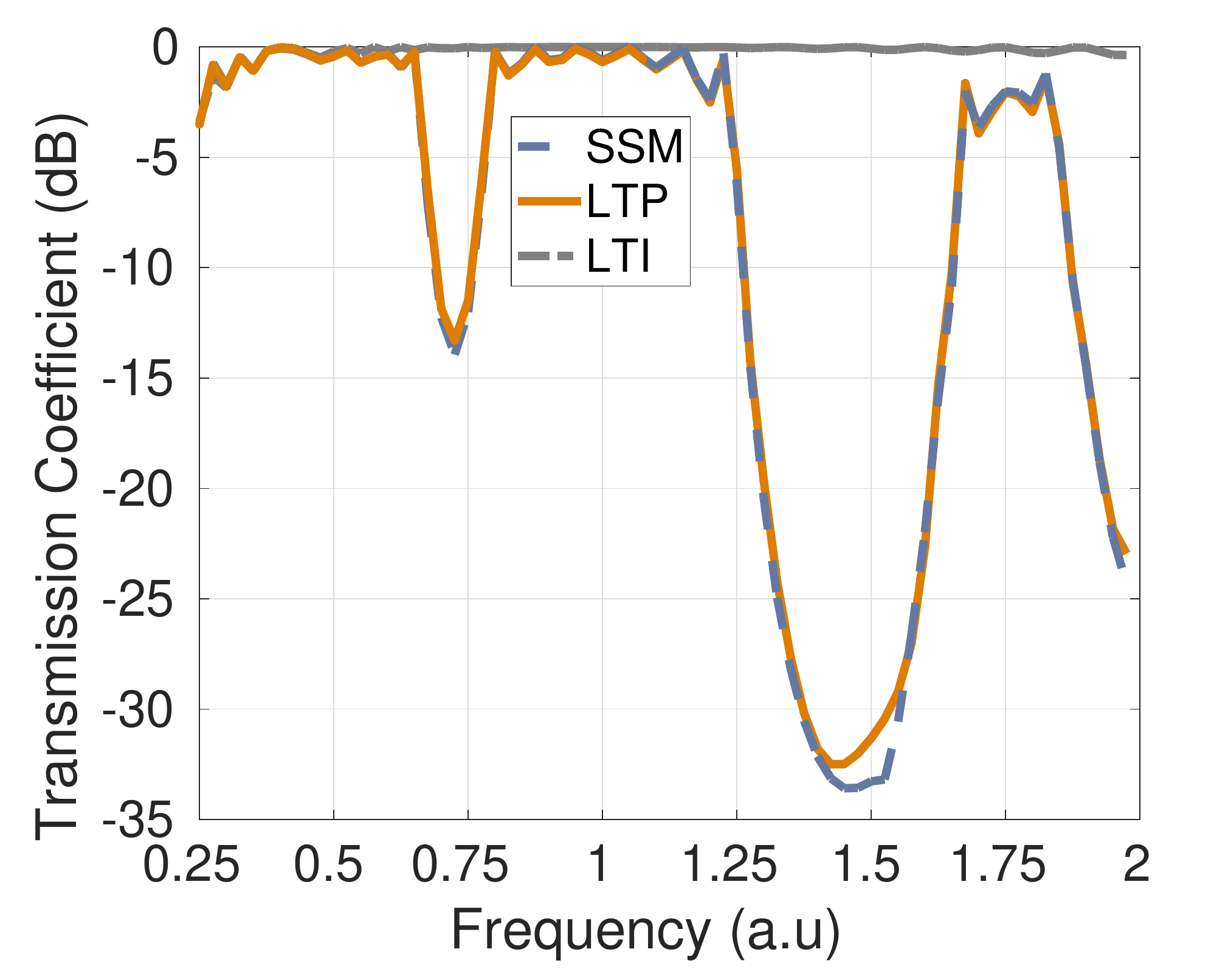}
\caption{Backward modulation (i.e, $\beta_m<0$).}
\end{subfigure}
\caption{Transmission Coefficient calculated for a space time modulated CRLH, with a modulation depth $M=0.8$ using the LTP formalism and brute force time domain computation.}
\label{fig:S21CRLH}
\end{figure}

\subsection{Analysis of a nonlinear right handed transmission line (NLRHTL)}
\label{subsect:nlrhtl}

\begin{figure*}[!t]
\centering
\ctikzset{/tikz/circuitikz/bipoles/length=0.8cm}
\begin {tikzpicture}[scale=0.85]
\draw (-0.38,0) node{\Large$\rightsquigarrow$};
\draw (-0.3,-0.3) node{\scriptsize$V_{0m}\cos\omega_mt$};
\draw (0,0) --(0.5,0);
\draw [fill=green!10](0.5,-1.2) rectangle+(1.5,1.4);
\draw [->, thick] (0.65,0) --+(1.2,0);
\draw [->,thick,blue,dashed] (0.65,-1.1) .. controls (0.7,-0.1) .. (1.85,-0.14);
\draw (2,0)--+(0.5,0);
\draw [dashed, blue] (0.65,-1.5) node[rotate=90]{\Large$\rightsquigarrow$};
\node[blue] at (1.6,-1.5) {\scriptsize$V_s\cos(\omega t+\phi)$};
\draw (2.5,-0.2) rectangle+(1.5,0.4);
\draw (2.5+1.5,0)--(2.5+2,0);
\draw (2.5+2,-1.1) to [empty varcap]+(0,1.1);
\draw (4.5,0)--+(0.5,0);
\draw (5,-0.2) rectangle+(1.5,0.4);
\draw (6.5,0) --+(0.5,0);
\draw [dotted] (7,0)--(8,0);
\draw [dashed, fill=yellow!15] (8,-1.3) rectangle+ (5,2.1);
\draw(8,0)--+(0.5,0);
\draw (8.5,-0.2) rectangle+(1.5,0.4);
\draw [<->,gray] (8.5,0.35) --+(1.5,0);
\node at (9.25,0.5) {\scriptsize$p/2$};
\node at(9.25,0.0) {\scriptsize$Z_c,\tau_d$};
\draw (8.5+1.5,0)--(8.5+2,0);
\draw (8.5+2,-1.1) to [empty varcap]+(0,1.1);
\node at (11.25,-0.6) {$C_n(v)$};
\draw (10.5,0)--+(0.5,0); 
\draw (11,-0.2) rectangle+(1.5,0.4);
\draw [<->,gray] (11,0.35) --+(1.5,0);
\node at (11.75,0.5) {\scriptsize$p/2$};
\node at(11.75,0.0) {\scriptsize$Z_c,\tau_d$};
\draw (12.5,0) --+(0.5,0);
\draw [dotted] (13,0)--(14,0);
\draw(14,0)--+(0.5,0);
\draw (14.5,-0.2) rectangle+(1.5,0.4);
\draw (14.5+1.5,0)--(14.5+2,0);
\draw (14.5+2,-1.1) to [empty varcap]+(0,1.1);
\draw (16.5,0)--+(0.5,0);
\draw (17,-0.2) rectangle+(1.5,0.4);
\draw (18.5,0) --+(0.5,0);
\node at (19,-0.6) {\Large$\rightsquigarrow$};
\node at (19.6,-0.6) {$Z_0$};
\draw (2,-1.1)--(19,-1.1);
\draw [shade,opacity=0.2,ultra thin](19,0.2) rectangle +(1,-1.6);
\node at (10.4,-1.85) {(a)};
\end{tikzpicture}
\begin{circuitikz}[scale=0.8]
\draw 
(0,-1) to [/tikz/circuitikz/bipoles/length=0.8cm,sV,l_=\scriptsize$V_{0m}\sin (\omega_mt)$](0,0) to [/tikz/circuitikz/bipoles/length=0.8cm,sV,l_=\scriptsize$V_{s0}\sin(\omega t+\phi)$] (0,1) to (0,1.5) to (0.5,1.5)
 to [/tikz/circuitikz/bipoles/length=0.8cm,european resistor,l^=\scriptsize$50~\Omega$] (1.5,1.5) to[/tikz/circuitikz/bipoles/length=0.8cm, C,l^=\scriptsize$v_\textnormal{dci}$] (2.5,1.5) to [/tikz/circuitikz/bipoles/length=0.8cm,L,l^=\scriptsize$L$] (4,1.5) to [/tikz/circuitikz/bipoles/length=0.8cm,C,l_=\scriptsize$C$] (4,-1) (5,-1) to [/tikz/circuitikz/bipoles/length=0.8cm,empty varcap] (5,1.5) to (4,1.5)
 (5,1.5) to [/tikz/circuitikz/bipoles/length=0.8cm,L,l^=\scriptsize$L$] (6.5,1.5) to [/tikz/circuitikz/bipoles/length=0.8cm,C,l_=\scriptsize$C$](6.5,-1);
 \draw [dotted] (7.5,0) -- (8,0);
\draw (8.5,1.5) to [/tikz/circuitikz/bipoles/length=0.8cm,L,l^=\scriptsize$L$] (10,1.5)to [/tikz/circuitikz/bipoles/length=0.8cm,C,l_=\scriptsize$C$] (10,-1) (11,-1) to [/tikz/circuitikz/bipoles/length=0.8cm,empty varcap,l_=\scriptsize$C_n(v)$] (11,1.5) to (10,1.5)
 (11,1.5) to [/tikz/circuitikz/bipoles/length=0.8cm,L,l^=\scriptsize$L$] (12.5,1.5) to [/tikz/circuitikz/bipoles/length=0.8cm,C,l^=\scriptsize$C$](12.5,-1);
  \draw [dotted] (13.5,0) -- (14,0);
  
  \draw (14,1.5) to [/tikz/circuitikz/bipoles/length=0.8cm,L,l^=\scriptsize$L$] (15.5,1.5)to [C,l_=\scriptsize$C$] (15.5,-1) (16.5,-1) to [/tikz/circuitikz/bipoles/length=0.8cm, empty varcap] (16.5,1.5) to (15.5,1.5)
 (16.5,1.5) to [/tikz/circuitikz/bipoles/length=0.8cm,L,l_=\scriptsize$L$] (18,1.5) to [/tikz/circuitikz/bipoles/length=0.8cm,C,l_=\scriptsize$C$](18,-1);
 \draw (18,1.5) to [/tikz/circuitikz/bipoles/length=0.8cm, C,l^=\scriptsize$v_\textnormal{dc0}$] (19,1.5) to [/tikz/circuitikz/bipoles/length=0.8cm,european resistor, l^=\scriptsize$50~\Omega$] (19,-1);
 \draw (0,-1) to (6.5,-1);
 \draw (8.5,-1) to (12.5,-1);
 \draw (14,-1) to (19,-1);
 \draw (18,1.5) to [/tikz/circuitikz/bipoles/length=0.8cm,L,l^=\scriptsize$L_\textnormal{RFC}$] (18,2.5) (18,3) node{\scriptsize$V_B$};
 \draw [-o] (18,2.5) --(18,2.75);
 \node at (10.5,-2) {(b)};
 \draw (9,-5.5) to [/tikz/circuitikz/bipoles/length=0.8cm,empty varcap] (9,-3.5) (9.75,-4.5) node{\Large$\Rightarrow$} (10.5,-3) to [/tikz/circuitikz/bipoles/length=0.8cm, european resistor, l=$R_s$] (10.5,-4) to [/tikz/circuitikz/bipoles/length=0.8cm,L, l=$L_s$] (10.5,-5) to [/tikz/circuitikz/bipoles/length=0.8cm, C,l=$C_n(v)$](10.5,-6);
 \draw [-o] (10.5,-6) -- (10.5,-6.25);
 \draw [-o] (10.5,-3) --(10.5,-2.75);
 \draw [-o] (9,-5.5)--(9,-5.75);
 \draw [-o] (9,-3.5) -- (9,-3.25);
 \node at (10.5,-6.8) {(c)};
 \draw [shade,opacity=0.2](12,-5.5) rectangle+(0.75,2.0);
 \draw [-o](12+0.5*0.75,-5.5) --(12+0.5*0.75,-6.2);
 \draw [-o] (12+0.5*0.75,-3.5)--((12+0.5*0.75,-2.8);
 \node at (13.25,-4.5) {$\tilde{\mathbf{Y}}_{sh}$};
\end{circuitikz}
\caption{(a) Schematics of a synthesized NLRHTL built from microstrip sections that are loaded by shunt varactors. The length of the unit cell $p$ is approximately 6.5 mm, the microstrip dimensions and substrate are such that $Z_c=64.26~\Omega,~\tau_d\approx 27.2~\textnormal{ps}$. (b) The microstrips are modelled by lumped LC sections. Additionally, DC blocking capacitors are included. Furthermore the bias voltage $V_B$ is applied via a ferrite bead that is modelled by a high inductance $L_\textnormal{RFC}$. (c) The varactor is modelled by a series RLC circuit. $R_s$ represents the ohmic losses, $L_s$ the parasitics due to bond-wires and soldering, and $C(u_{pn})$ is the nonlinear capacitance value.}
\label{fig:NLRHTL}
\end{figure*}
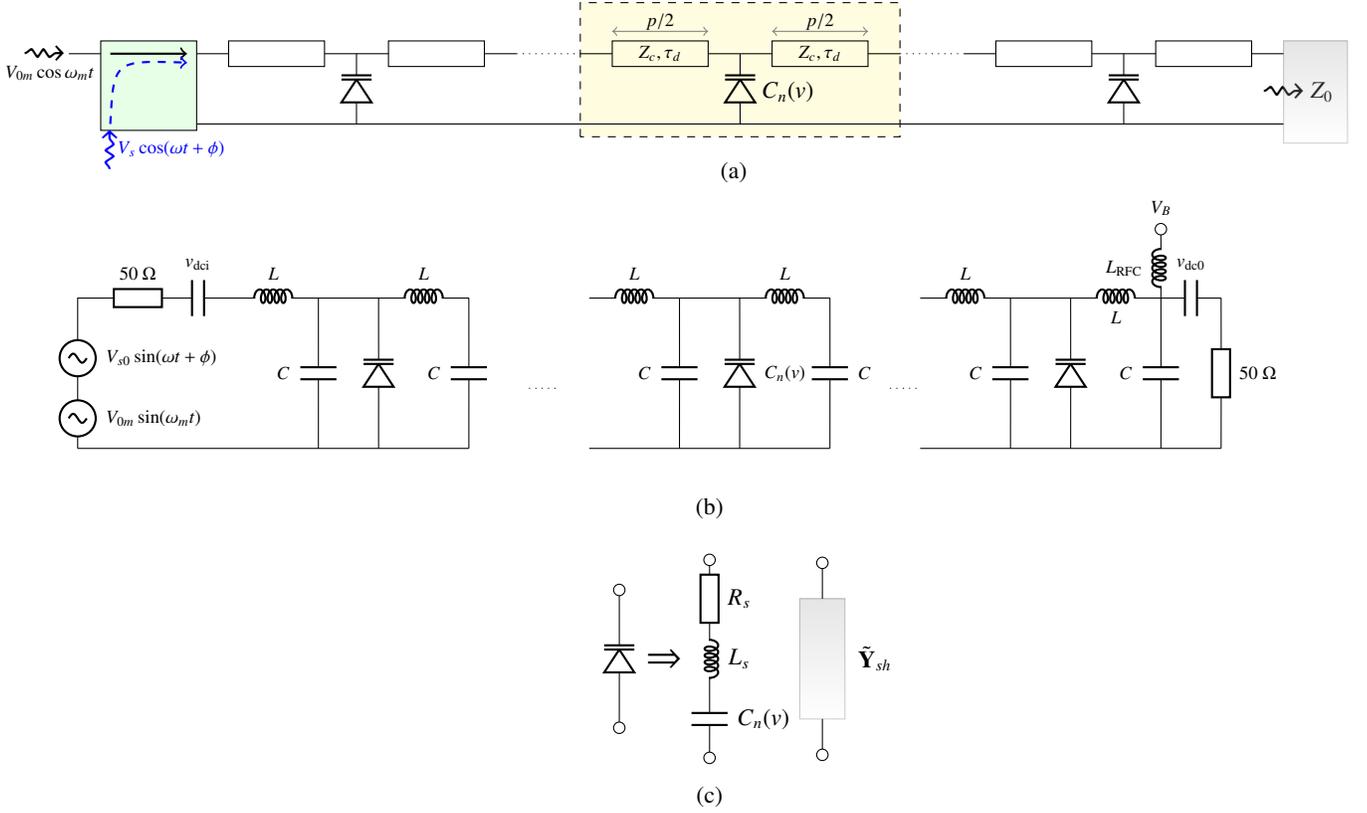
A modulating sinusoid $v_m(t)$ and an input signal $v_s(t)$ was applied to nonlinear right handed transmission line (NLRHTL) that is built from twenty unit cells, as shown in Fig. \ref{fig:NLRHTL}(a). The inputs $v_m$ and $v_s$ are combined using a directional coupler as highlighted. Each unit cell consists of a $p\approx 6.5\textnormal{ mm}$ long microstrip loaded at its center by a varactor (M/A-COM, MA46H120). The circuit is etched on a 25 mil thick Rogers RO3010 substrate. The varactors are bonded in place using H20E conductive epoxy.

The capacitance of the varactor $C_v$ depends on the voltage across its terminals $u(t)=u_m(t)+u_s(t)$, where $u_m(t)$ and $u_s(t)$ are the voltages due to the modulating input and signal, respectively. The current through the varactor is given by

$$i(t)=C_v(u)\frac{du}{dt}.$$
Since $C_v(u)=C_v(u_m+u_s)\approx C_v(u_m)+u_sdC_v/dt|_{u_m}$, it can be shown that the current $i_s(t)$ due to the signal excitation is 
$$i_s(t)=\frac{d}{dt}C_v(u_m)u_s(t),$$
where $C_v(u_m)$ is the capacitance evaluated at $u_m(t)$, which is periodic with frequency $\omega_m$. Hence, the system is linearized about the limit cycle steady state \cite{Sameh_JAP_NLD}. Fig. \ref{fig:NLRHTL}(c) shows the varactor's equivalent circuit. $R_s$ models the ohmic losses in the semiconductor bulk, contact and bondwires, $L_s$ accounts for the inductance of the bondwires, and $C$ represents the varactor capacitance. The different circuit parameters were extracted from measuring the S parameters at different bias voltage and fitting the response via the use of the Vector Fitting technique.  \cite{gustavsen1999rational}

At low frequency, the microstrip line can be described by lumped circuits as in Fig. \ref{fig:NLRHTL}(b). The $p/2$ microstrip line section is modelled as a lumped LC network, such that $L=\tau_dZ_{c}$ and $C=\tau_d/Zc$, where $\tau_d$ and $Z_c$ are the delay and characteristic impedance of the microstrip, respectively. The lumped circuit approximation allows the convenient representation of the NL RH TL in a SSM form. In this case, the biasing circuit and blocking capacitances can be included as in Fig. \ref{fig:NLRHTL}(b).

\subsubsection{Dispersion Relation}
As a first step, the LTI dispersion relation of the structure is extracted from measuring the small signal S parameters for different bias voltages and compared to the circuit models. Fig. \ref{fig:ltidispersion} shows the dispersion relation curves of four bias voltages. Clearly, both the lumped and distributed circuit models are in agreement with measurement; confirming the validity of the the lumped circuit model.

\begin{figure}[!htb]
\centering
\begin{subfigure}[b]{0.48\linewidth}
\includegraphics[width=\linewidth]{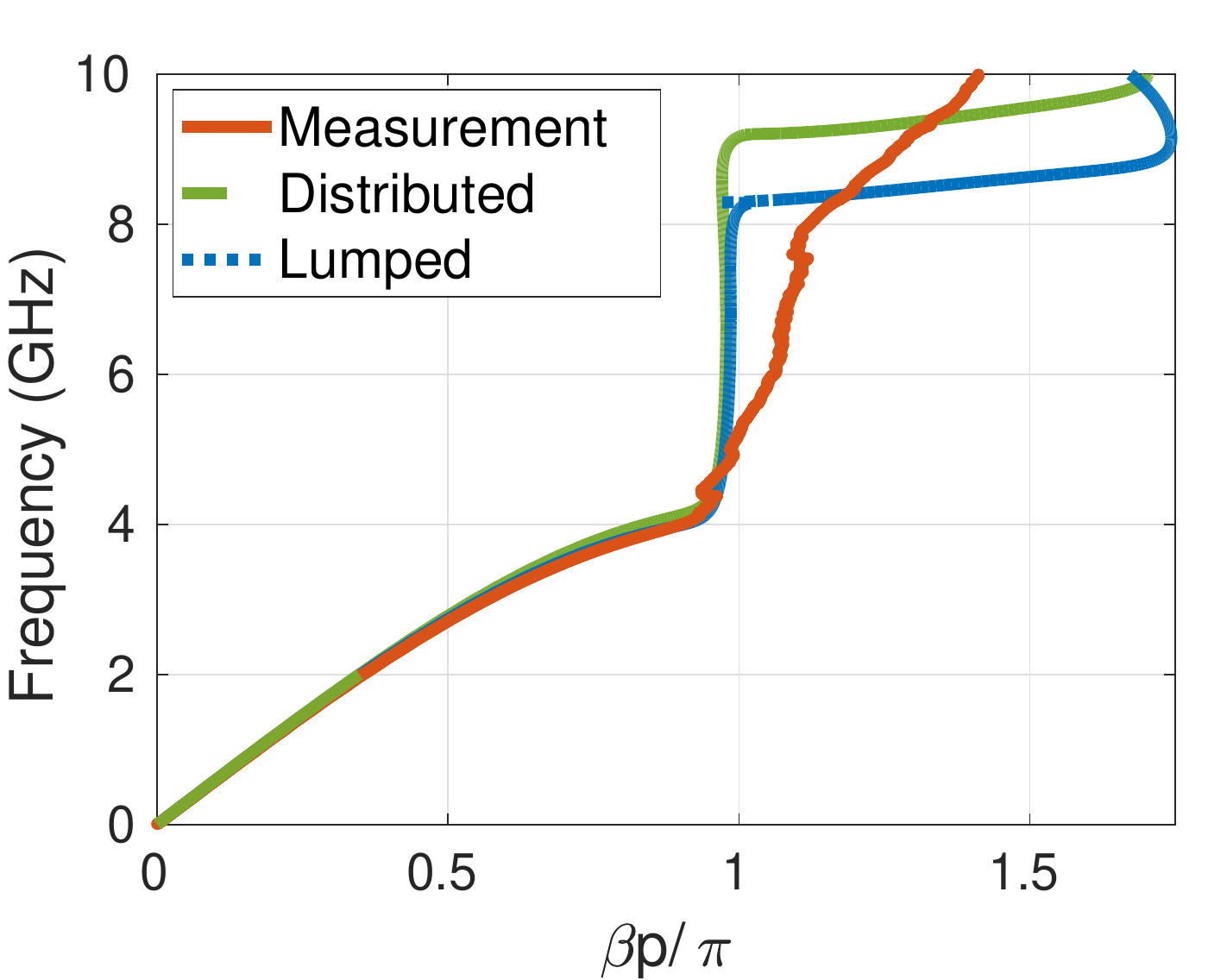}
\caption{\scriptsize $V_{B}=0~\textnormal{V}$}
\end{subfigure}
\centering
\begin{subfigure}[b]{0.48\linewidth}
\includegraphics[width=\linewidth]{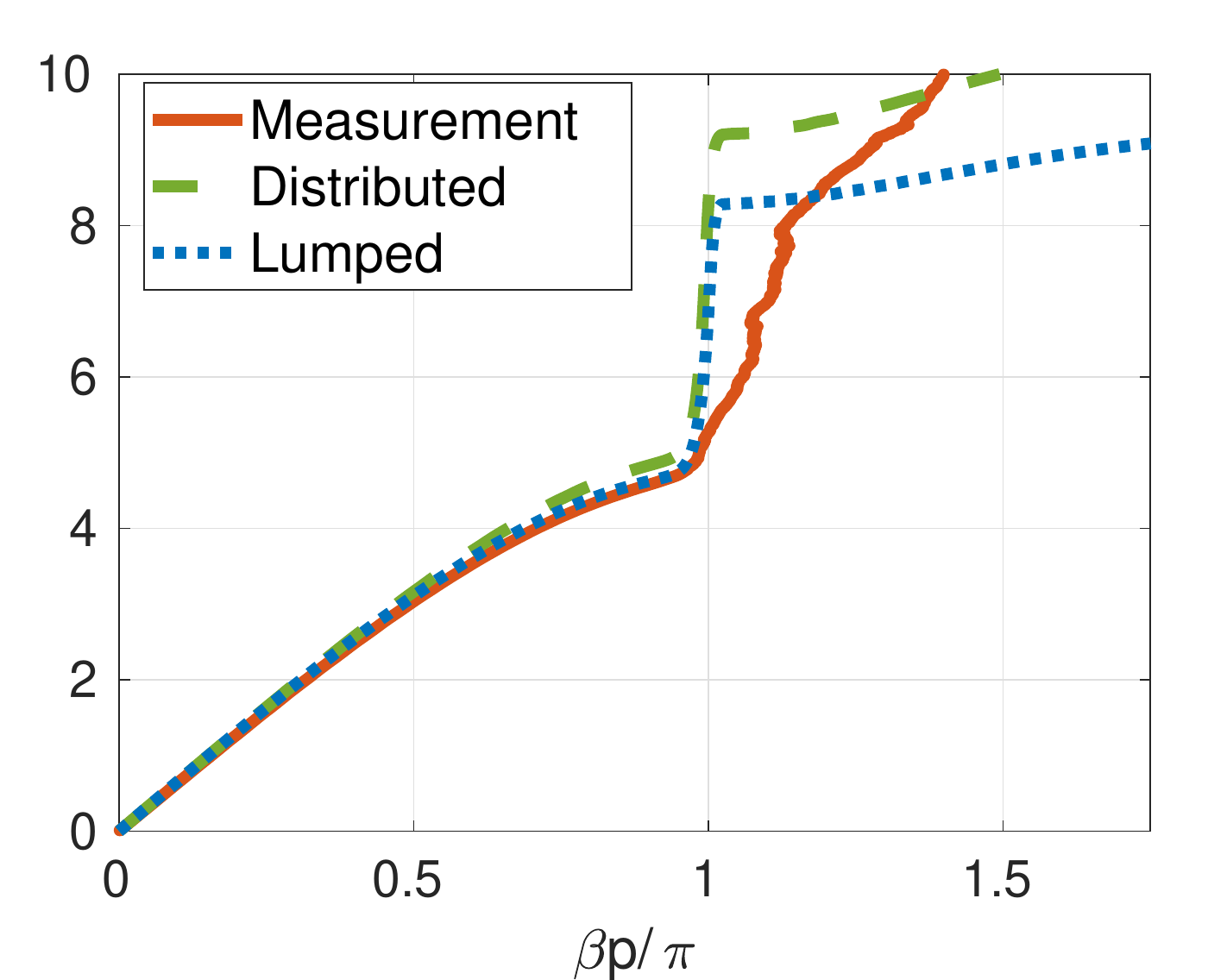}
\caption{\scriptsize $V_{B}=1~\textnormal{V}$}
\end{subfigure}

\begin{subfigure}[b]{0.48\linewidth}
\includegraphics[width=\linewidth]{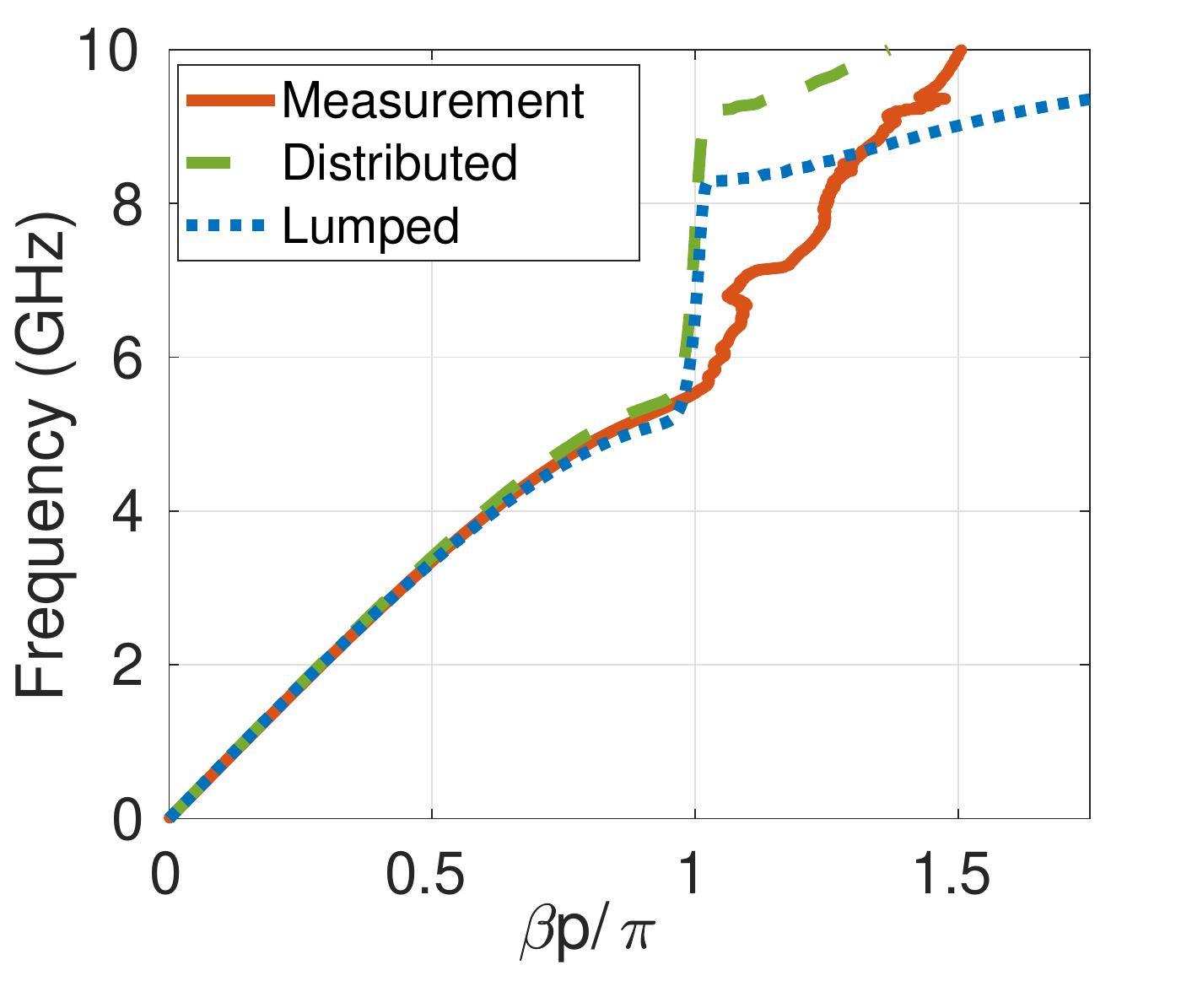}
\caption{\scriptsize $V_{B}=2~\textnormal{V}$}
\end{subfigure}
\begin{subfigure}[b]{0.48\linewidth}
\includegraphics[width=\linewidth]{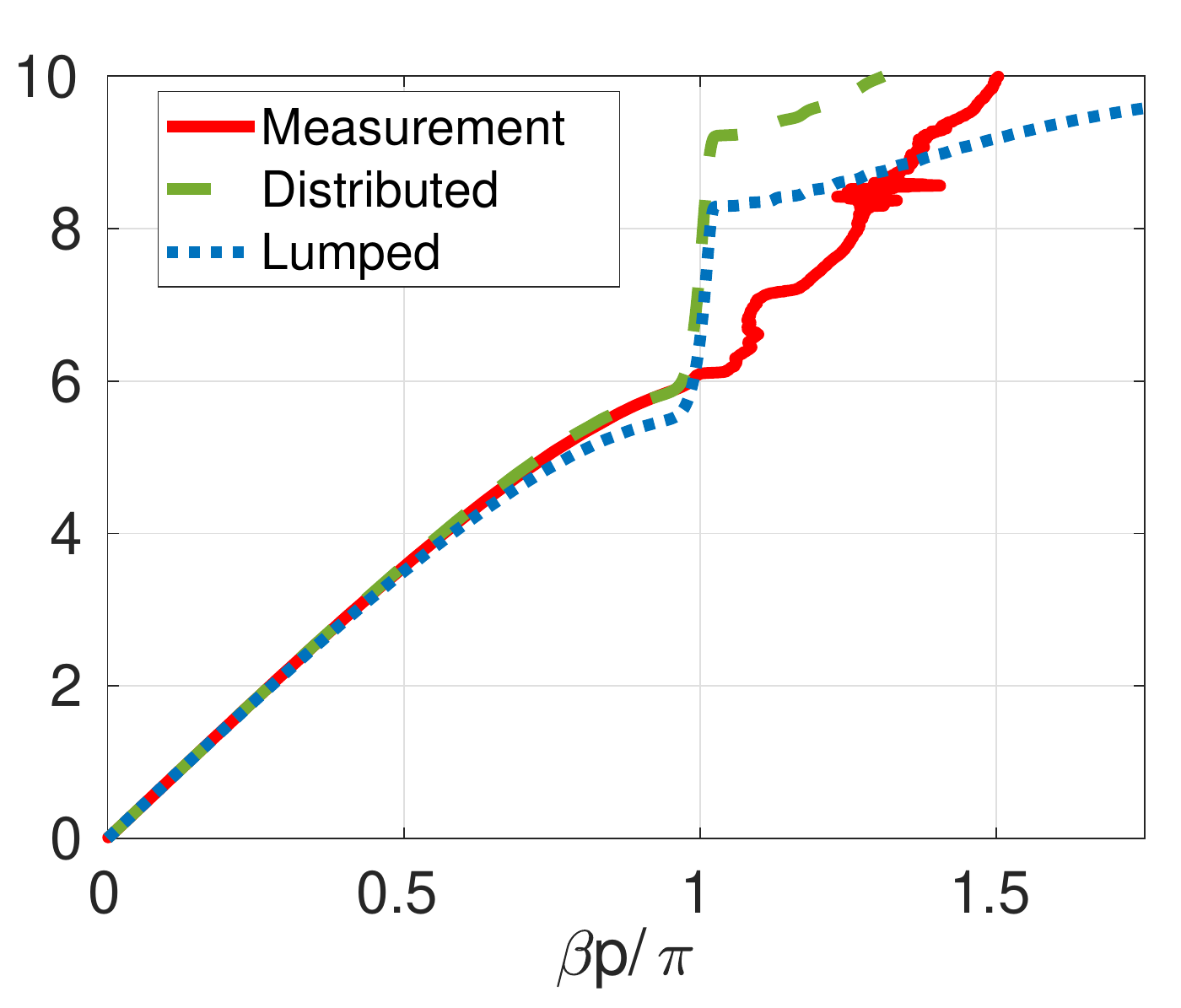}
\caption{\scriptsize $V_{B}=3~\textnormal{V}$}
\end{subfigure}
\caption{LTI dispersion relation of the NLRHTL for different bias voltages.}
\label{fig:ltidispersion}
\end{figure}
\begin{figure}[!htb]
\centering
\begin{subfigure}[b]{0.38\linewidth}
\includegraphics[width=\linewidth]{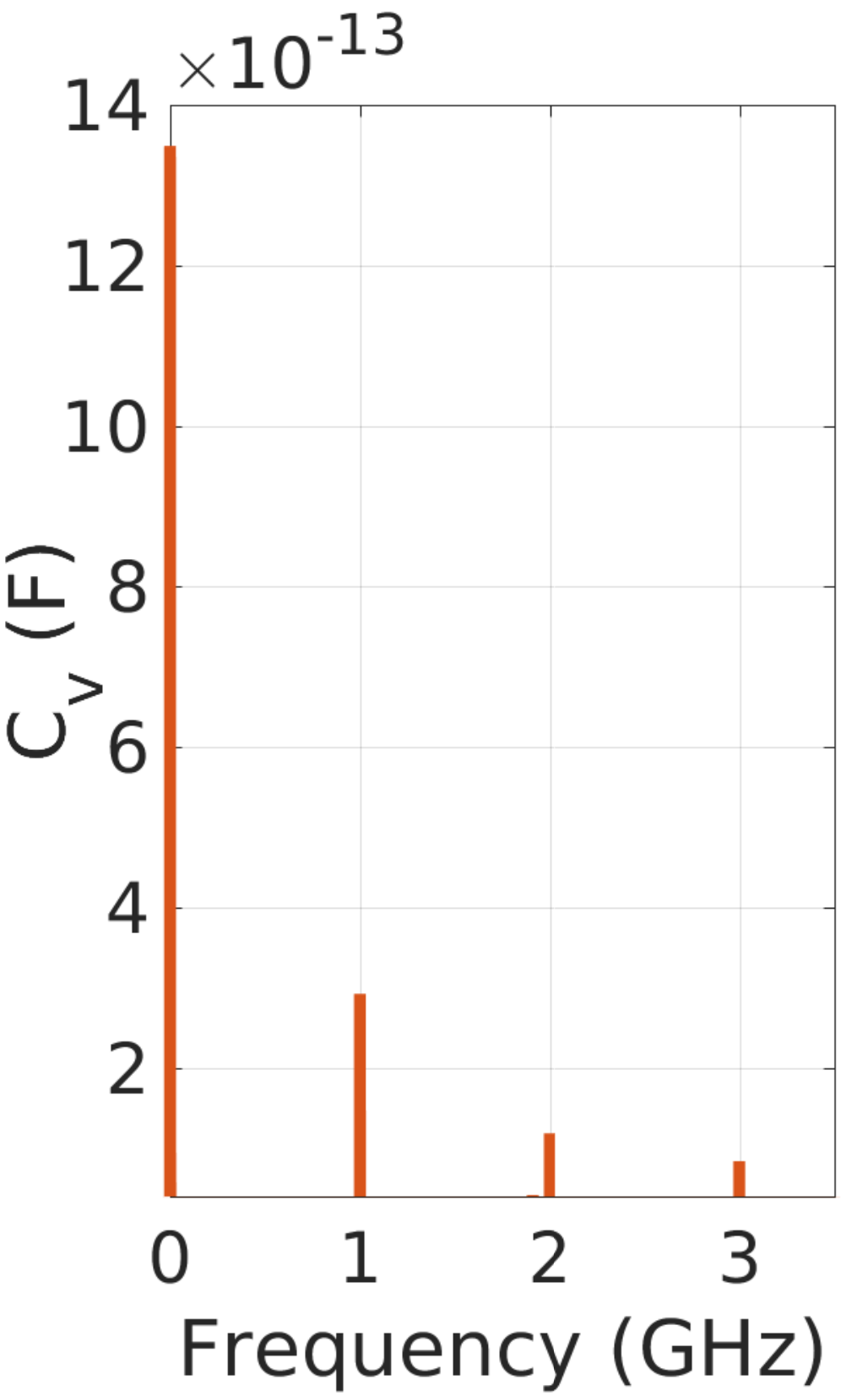}
\caption{ }
\end{subfigure}
\centering
\begin{subfigure}[b]{0.61\linewidth}
\includegraphics[width=\linewidth]{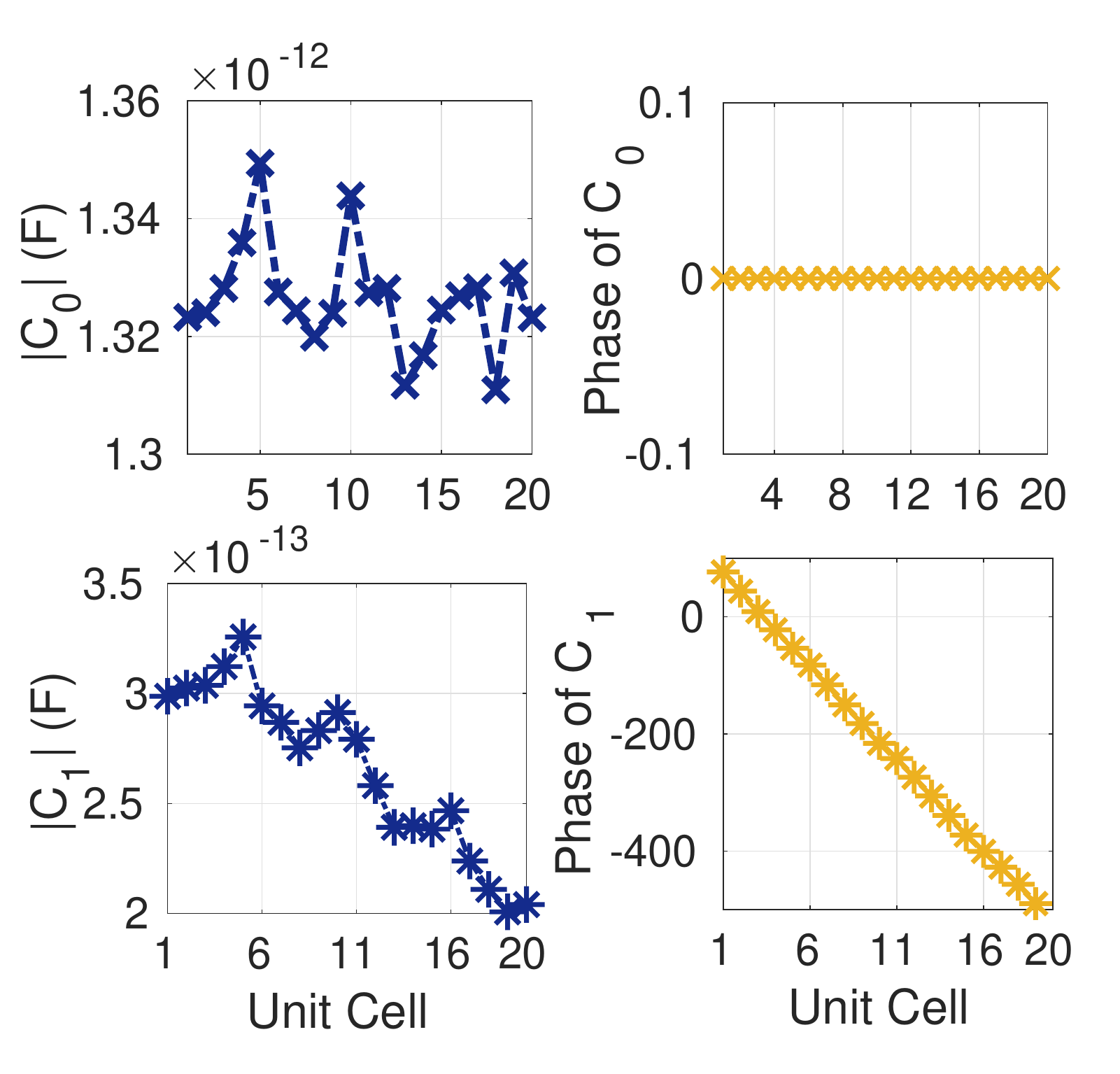}
\caption{ }
\end{subfigure}
\caption{Varactor Capacitance for the NL RH TL. (a) Spectrum of $C_v$ at the 10\textsuperscript{th} unit cell, when $f_m=1~\textnormal{GHz}$ and $V_m=15~\textnormal{dBm}$. (b) Capacitance of varactor at each unit cell, obtained from the state space model.}
\label{fig:cv}
\end{figure}
In the presence of the modulating signal with frequency $\omega_m$, the varactor capacitance $C_v$ becomes periodic with a period of $2\pi/\omega_m$. Therefore, it can be expanded in Fourier series
$$C_v(t)=\sum_{r=-\infty}^{+\infty}\tilde{C}_re^{ir\omega_mt}.$$
Since the amplitude of modulation is large ($u_m\gg u_s$), the DC capacitance $\tilde{C}_0$  may deviate from the small signal value. In the subsequent analysis the DC and first harmonic only ($\tilde{C}_0$ and $\tilde{C}_1$) will be considered. They are calculated from a time domain simulation of NLRHTL. The system differential equations are solved to compute the voltages $u_{m}$ across the different varactors. Consequently, $\tilde{C}_0$ and $\tilde{C}_1$ are calculated from the Fourier transform of $C(u_m)$. Fig. \ref{fig:cv}(a) shows the computed spectrum of $C_v$ across the 10\textsuperscript{th} varactor. The DC capacitance $\tilde{C}_0$ has increased from approximately 1.2 pF to 1.35 pF. The modulation strength $M\triangleq\tilde{C}_1/\tilde{C}_0\approx 0.2$ assuming an excitation of strength $\sim 10-15 \textnormal{ dBm}$. Fig. \ref{fig:cv}(b) demonstrates how $\tilde{C}_0$ and $\tilde{C}_1$ change from one unit cell to the other. Although not constant, we will assume that both $\tilde{C}_0$ and $\tilde{C}_1$ are constants and fixed to their average values. 

The unit cell can be represented by the block diagram in Fig. \ref{fig:UC}. Hence the ABCD matrix of the unit cell is $\mathbf{T}=\mathbf{T}_\textnormal{LTI}\mathbf{T}_\textnormal{LTP}\mathbf{T}_\textnormal{LTI}$. For the microstrip, the ABCD parameters are identical to the LTI counterpart, but calculated at each harmonic frequency $\tilde{\omega}_r$. 

The LTP block $\mathbf{T}_\textnormal{LTP}$ represents  the ABCD parameters of the shunt varactor, which is modelled by a shunt time periodic admittance $\tilde{\mathbf{Y}}_{sh}=\left(\mathbf{Z}_{se}+\tilde{\mathbf{Y}}^{-1}\right)^{-1}=\tilde{\mathbf{Y}}\left(\mathbf{e}+\mathbf{Z}_{se}\tilde{\mathbf{Y}}\right)^{-1}$. The second term in the last expression is the inverse of a tridiagonal matrix and can be computed using closed form expressions as in Ref. \onlinecite{tridiagonal}.
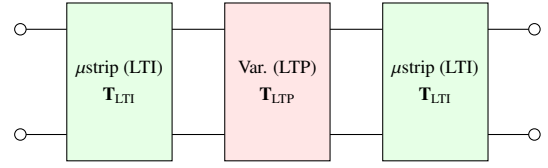
\begin{figure}[!htb]
\centering
\begin{tikzpicture}[scale=0.7]
\draw [o-] (0,1) -- (1,1);
\draw [o-] (0,-1) -- (1,-1);
\draw [fill=green!10] (1,-1.5) rectangle+(2,3);
\node at (2,0.25) {\scriptsize $\mu$strip (LTI)};
\node at (2,-0.25){\scriptsize $\mathbf{T}_\textnormal{LTI}$};
\draw [-] (3,1) -- (4,1);
\draw [-] (3,-1) -- (4,-1);
\draw [fill=red!10] (4,-1.5) rectangle+(2,3);
\node at (5,0.25) {\scriptsize Var. (LTP)};
\node at (5,-0.25){\scriptsize $\mathbf{T}_\textnormal{LTP}$};
\draw (6,-1)--(7,-1);
\draw (6,1)--(7,1);
\draw [fill=green!10] (7,-1.5) rectangle+(2,3);
\node at (8,0.25) {\scriptsize $\mu$strip (LTI)};
\node at (8,-0.25){\scriptsize $\mathbf{T}_\textnormal{LTI}$};
\draw [-o] (9,-1) -- (10,-1);
\draw [-o] (9,1) -- (10,1);
\end{tikzpicture}
\caption{The unit cell of the modulated TL modelled as the cascade of three sections: two LTI sections representing the microstrip lines and a LTP section describing the terminals behaviour of the varactor.} 
\label{fig:UC}
\end{figure}

The speed of modulation $\nu_m$ is determined from the phase $\phi$ of $\tilde{C}_1$, where 
$$\nu_m=2\pi f_m p\Bigr|\frac{\Delta n}{\Delta\phi}\Bigr|,$$
which is expected to \emph{slightly} deviate from the TL LTI speed. For a given modulation frequency $f_m$ and strength $M$, the dispersion relation can be determined from the solution of (\ref{eq:evb}). Fig.  \ref{fig:ltpdispersion} depicts the dispersion relation for $f_m=1~\textnormal{GHz}$, and when the modulation propagates in forward (Fig. \ref{fig:ltpdispersion}(a)) and backward (Fig. \ref{fig:ltpdispersion}(b)) directions. As shown, $\nu_m$ is very close to the LTI speed, suggesting that the LTP system is in the sonic regime \cite{Cassedy1962,Oliner1961}. 
\begin{figure}[!htb]
\centering
\begin{subfigure}{0.49\linewidth}
\includegraphics[width=\linewidth]{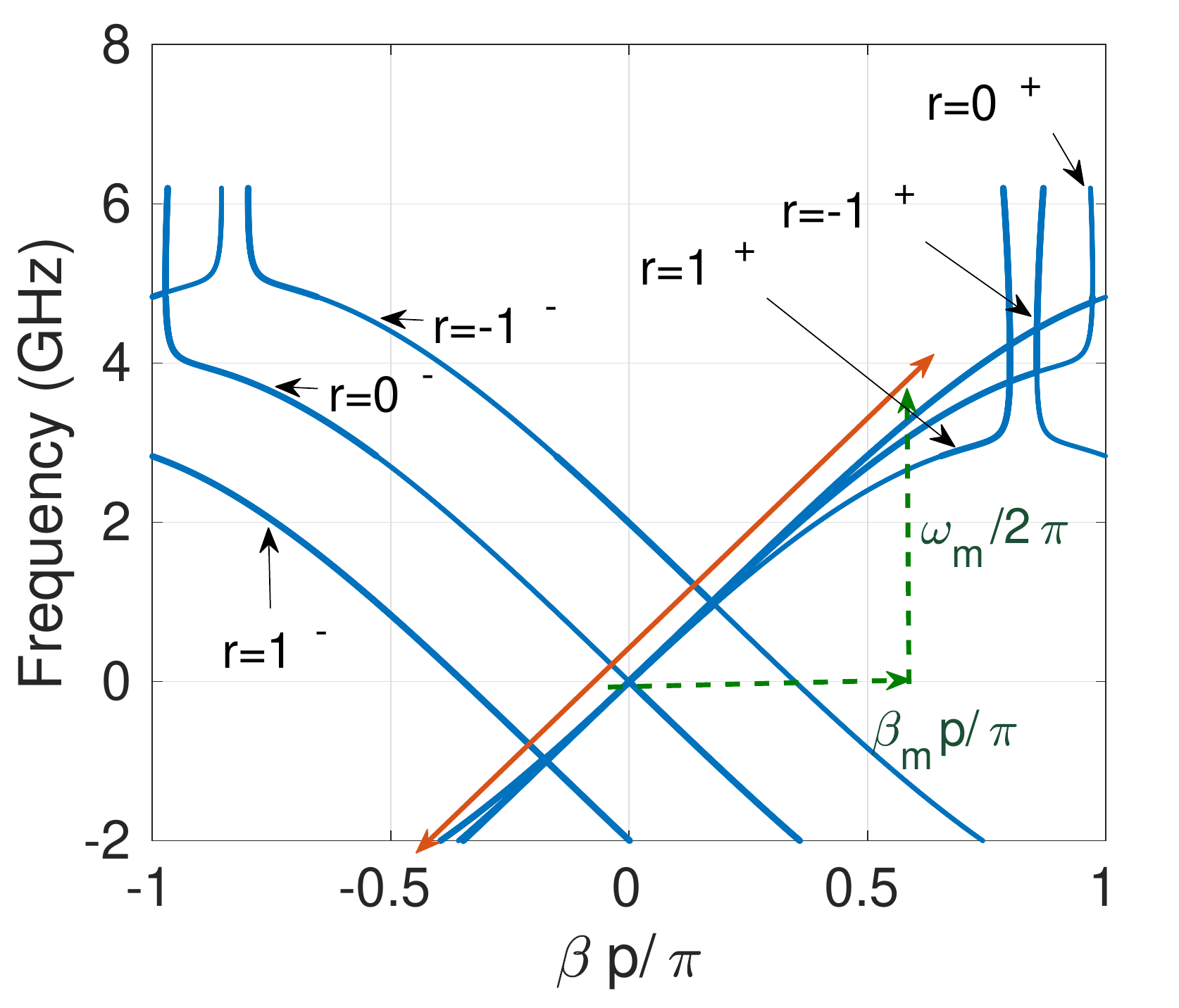}
\caption{}
\end{subfigure}
\begin{subfigure}{0.49\linewidth}
\includegraphics[width=\linewidth]{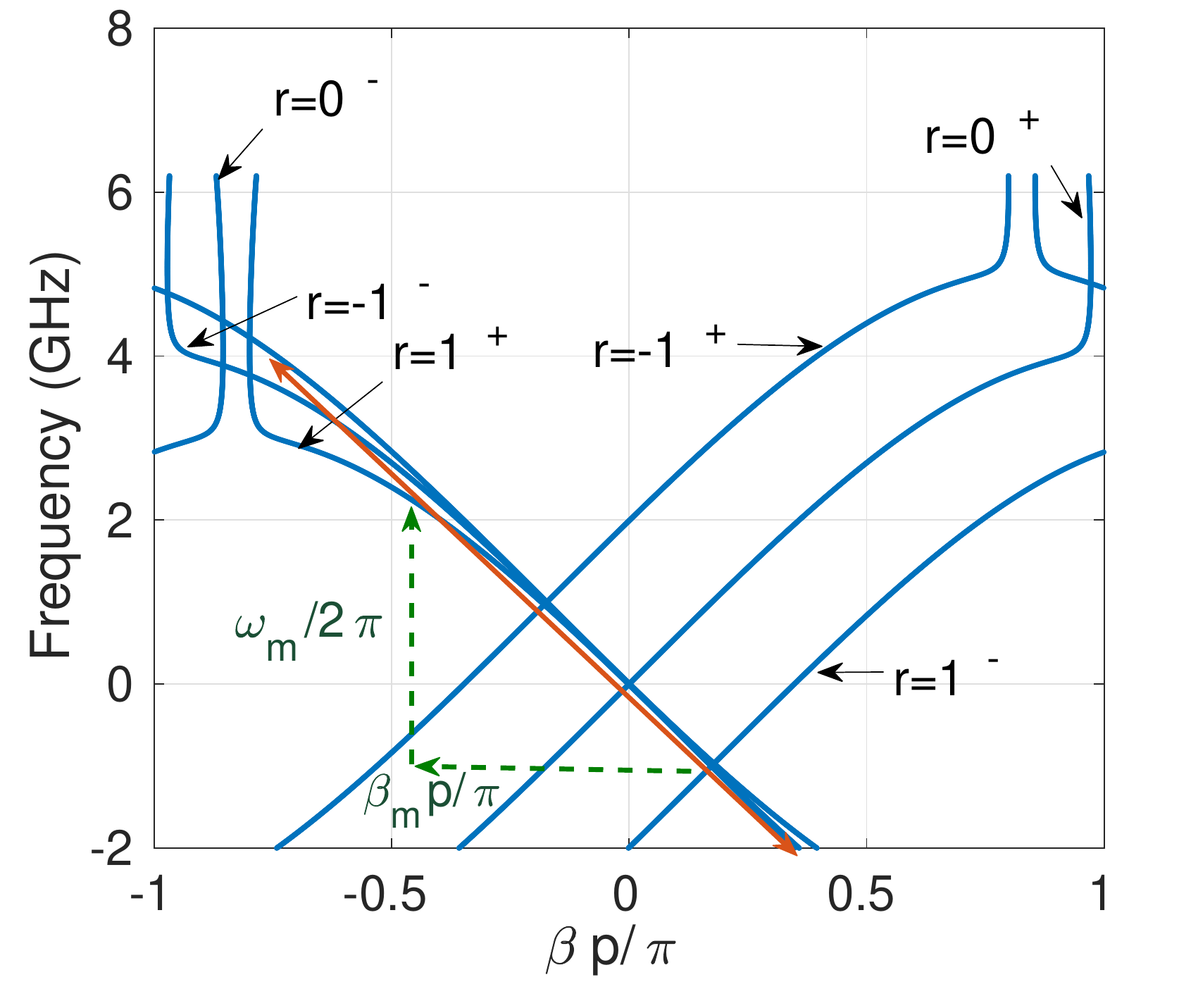}
\caption{}
\end{subfigure}
\caption{LTP dispersion relation of the NLRHTL, when $f_m=1~\textnormal{GHz}$. (a) Forward Modulation. (b) Backward Modulation.}
\label{fig:ltpdispersion}
\end{figure}

To explore the interaction between the different modes and how they contribute to the overall propagation, consider the situation where the modulation and signal are co-directional. Using 14 modes, the eigenvectors are calculated as reported in Fig. \ref{fig:waveform_evec_ak}(b). The TL is excited with a sinusoidal signal of frequency $f= 2.82~\textnormal{GHz}$. As will be shown later, at this frequency, maximum non-reciprocity is observed. The plot shows the magnitude of the components of each eigenvector normalized to its maximum value. Modes of interest are the ones that strongly couple with the input excitation; hence they have significant components at $\omega$ (or the 0\textsuperscript{th} harmonic as highlighted in Fig. \ref{fig:waveform_evec_ak}(b)) and can potentially be excited. Additionally, the BVP (\ref{eq:bvp}) is invoked to compute the different $a_k$ values that in turn determine the strength of the excited modes as Fig. \ref{fig:waveform_evec_ak}(d) shows. The waveforms at different frequencies are the superposition of the corresponding eigenvectors as presented in Fig. \ref{fig:waveform_evec_ak}(c) and confirmed with SSM  in Fig. \ref{fig:waveform_evec_ak}(a). Furthermore, Fig. \ref{fig:waveform_evec_ak}(e) demonstrates that the waveforms can be approximated by the dominant eigenmodes (i.e, the ones that couple with the input excitation such that their expansion coefficients $a_k$ are non-vanishing). The signal at $\omega$ is significantly reduced at the output due to the interaction with its harmonics, mainly the -1 harmonic. The absence of bandgaps in the dispersion plots in Fig. \ref{fig:ltpdispersion} suggests that this type of interaction is passive in nature (i.e, $\beta$ is imaginary) \cite{Pierce1954}. Such implication can be demonstrated by plotting $\beta$ in the complex plane as in Fig. \ref{fig:nltl_eigenvals_100}(a). Note that the excited modes, as witnessed by the values of $|a_k|$ in Fig. \ref{fig:waveform_evec_ak}(d), have imaginary propagation constants. Additionally, a SSM compuation of the same TL, but with an $N=100$ unit cells is performed and the results are reported in  Fig. \ref{fig:nltl_eigenvals_100}(b). Up to the 20\textsuperscript{th} cell, the wave behaviour and interaction between harmonics resemble that of an $N=20$ unit cells shown in Figs. \ref{fig:waveform_evec_ak}(a), (c) and (e), where energy is mainly transferred from the fundamental to its -1 harmonic. Nevertheless, for the subsequent stages, up to the 60\textsuperscript{th} cell, energy is pumped back to the fundamental and the amplitude of the fundamental harmonic increases. 

The strong interaction between the fundamental and its -1 harmonic is apparent from the measured output spectrum (Fig. \ref{fig:waveform_evec_ak}(f)). Here, the input port was fed by an RF source that was swept over a frequency range around 2.82 GHz and the output of the spectrum was measured by a spectrum analyzer.  The spectrum shows that once the modulation is turned on, the interaction is mainly with the -1 harmonic. Note that modulation and its higher harmonics (1, 2 and 3 GHz) appear as spikes in the measured spectrum.
\begin{figure}[!htb]
\centering
\begin{subfigure}{0.49\linewidth}
\includegraphics[width=\linewidth]{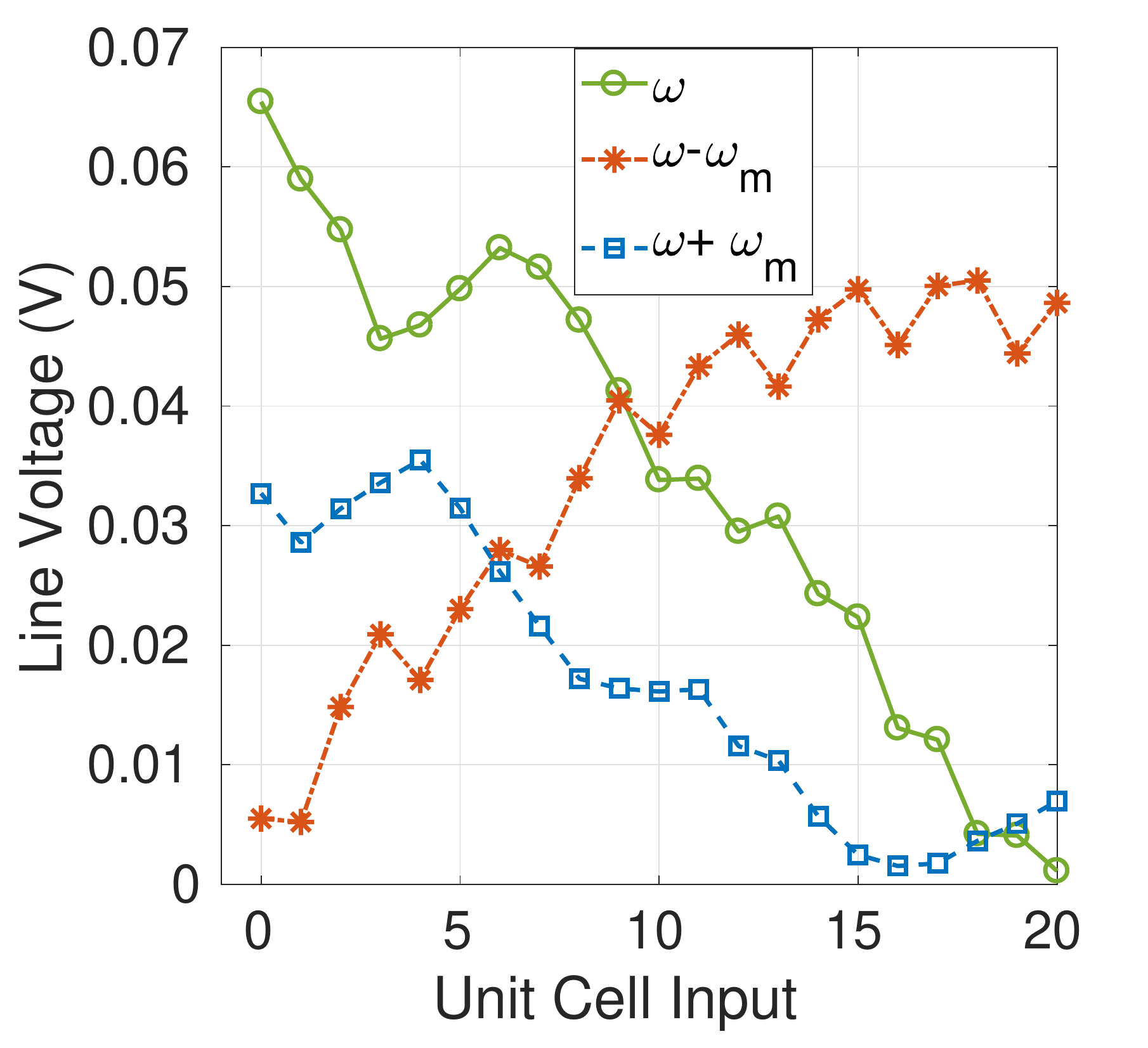}
\caption{}
\end{subfigure}
\begin{subfigure}{0.49\linewidth}
\includegraphics[width=\linewidth]{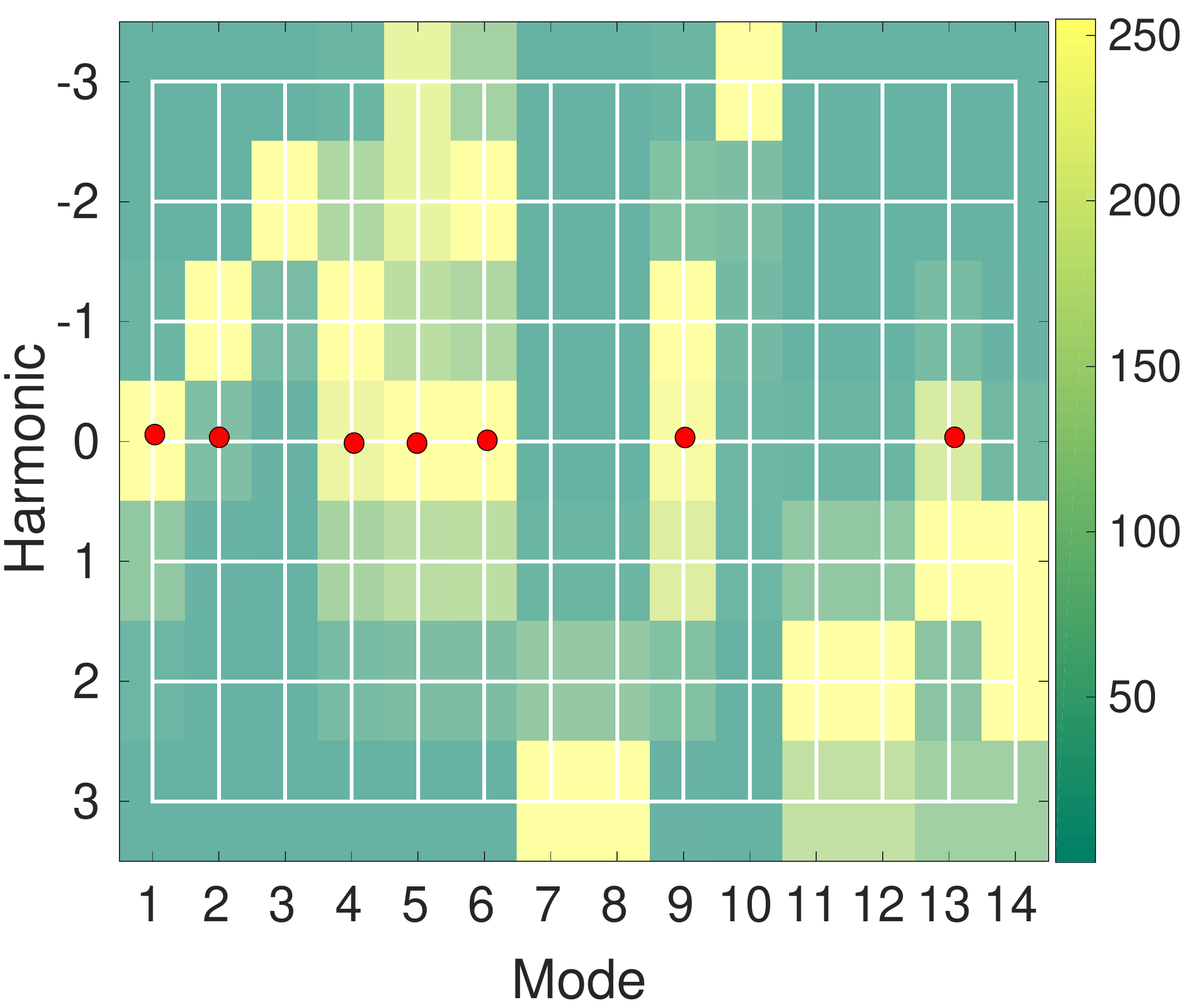}
\caption{}
\end{subfigure}
\begin{subfigure}{0.49\linewidth}
\includegraphics[width=\linewidth]{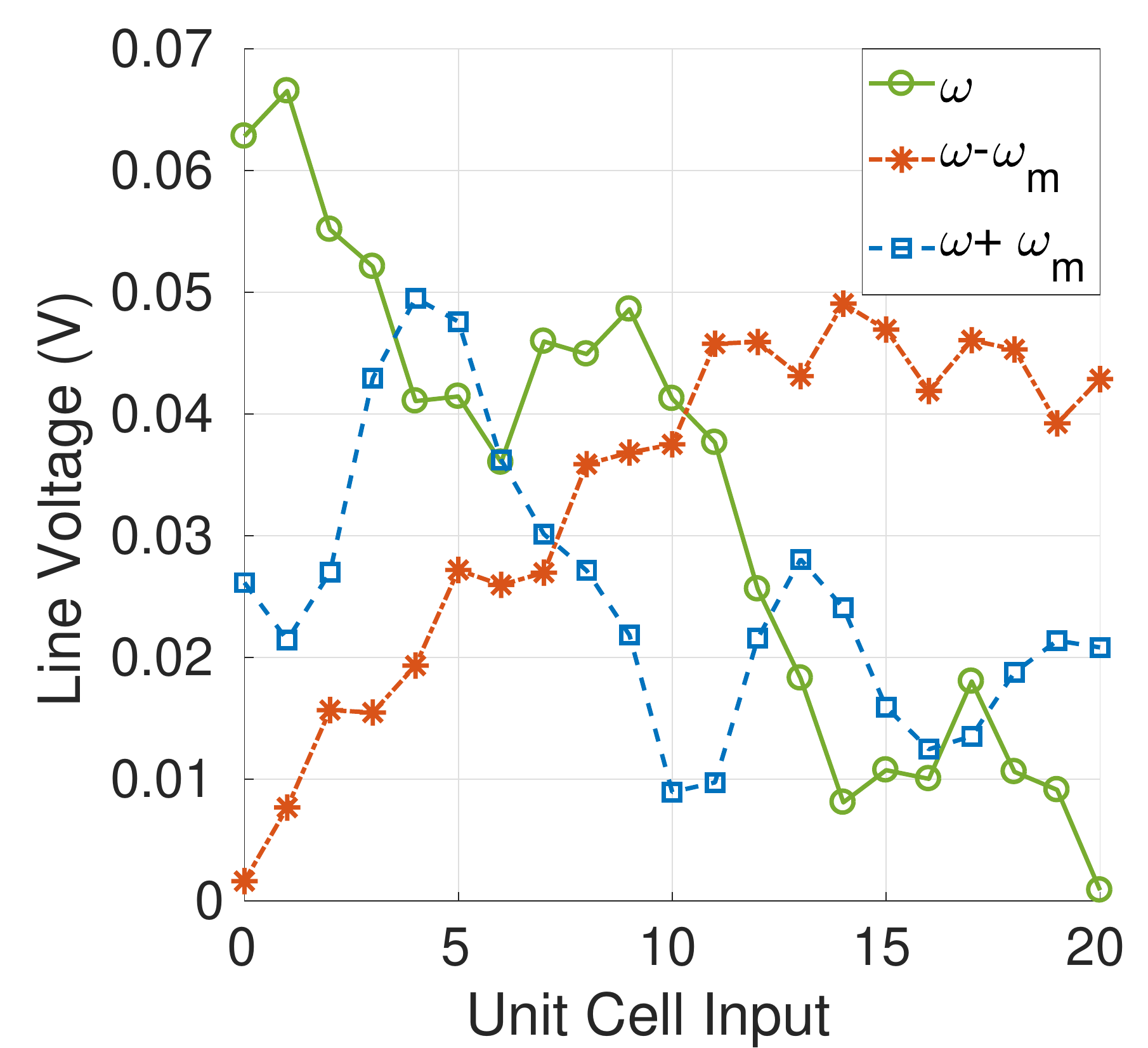}
\caption{}
\end{subfigure}
\begin{subfigure}{0.49\linewidth}
\includegraphics[width=\linewidth]{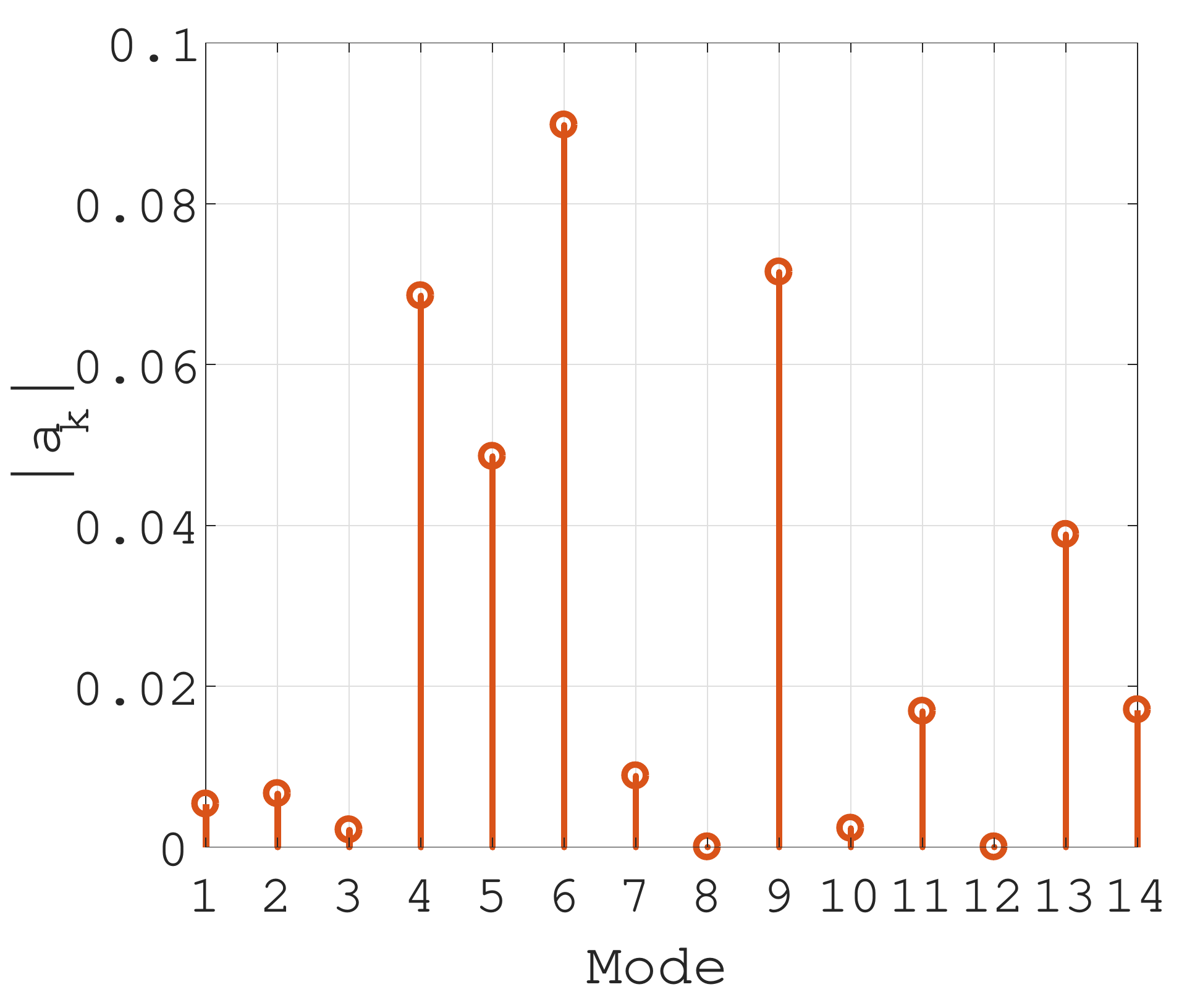}
\caption{}
\end{subfigure}
\begin{subfigure}{0.49\linewidth}
\includegraphics[width=\linewidth]{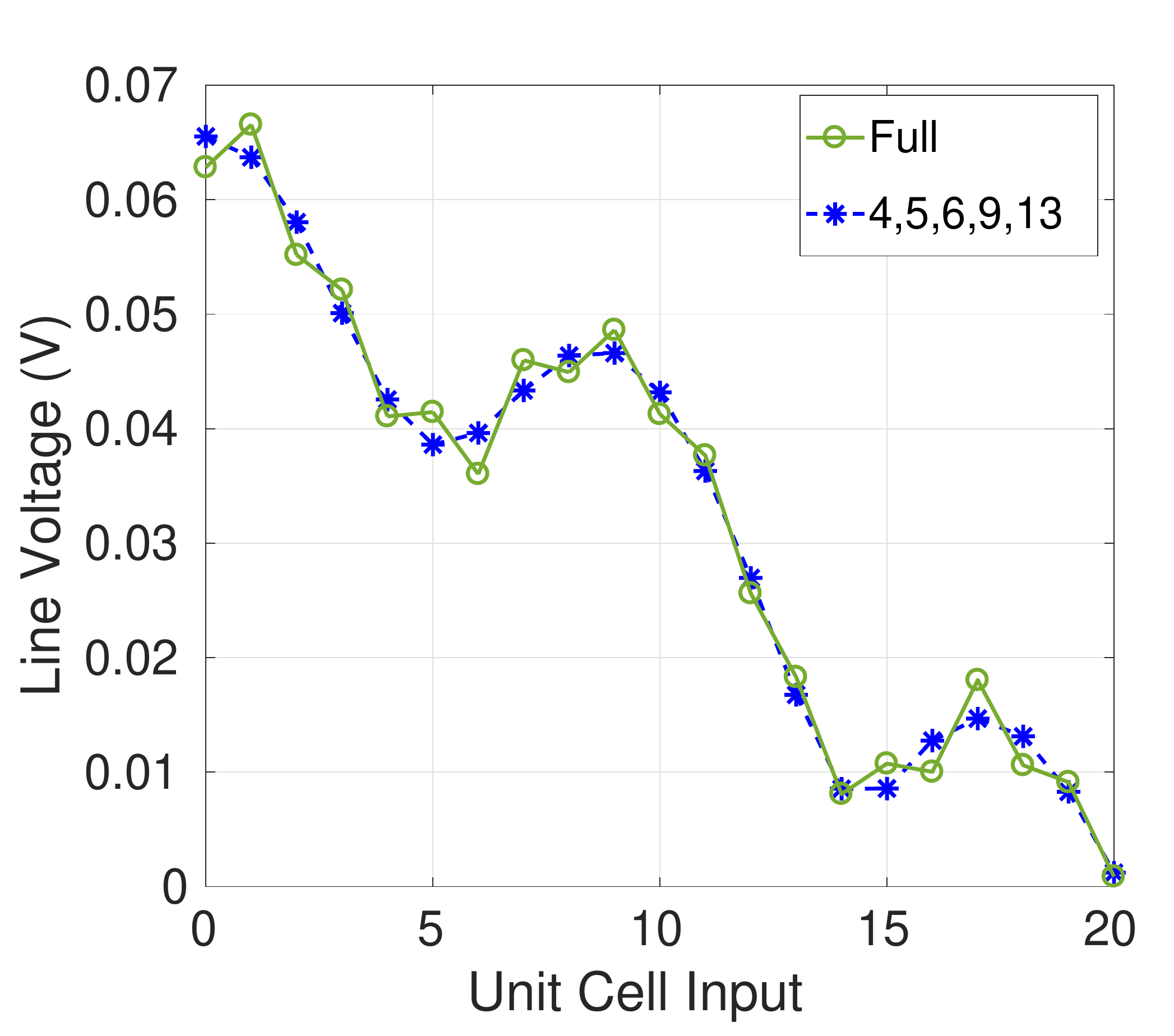}
\caption{}
\end{subfigure}
\begin{subfigure}{0.49\linewidth}
\includegraphics[width=\linewidth]{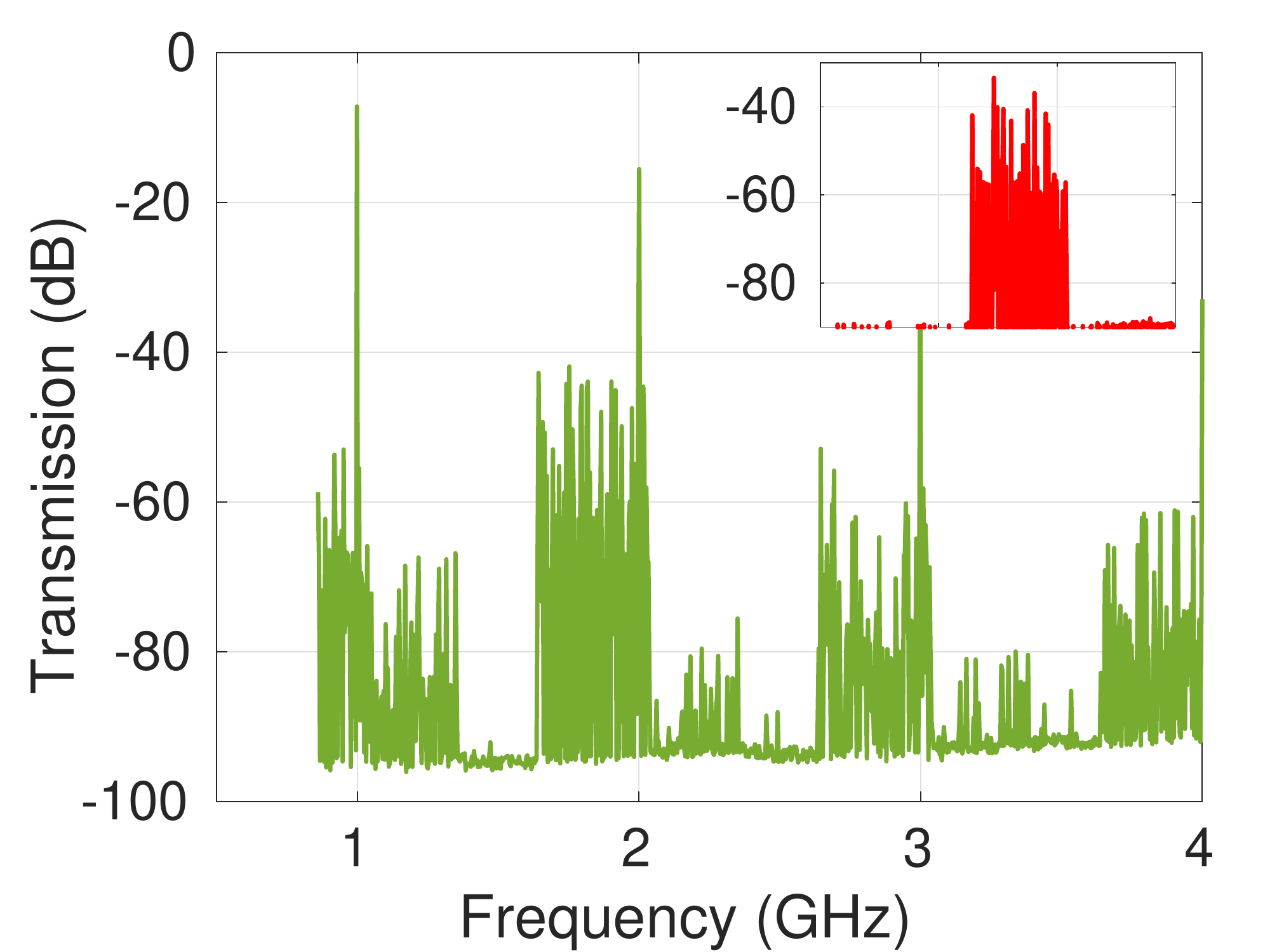}
\caption{}
\end{subfigure}
\caption{Amplitude of frequency components computed at $\omega$, $\omega-\omega_m$ and $\omega+\omega_m$ for the N=20 NL RH TL, using (a) SSM, (c) LTP. (b) Magnitude of the first 14 modes. The modes strongly couple to the excitation are highlighted with the (red: online) dots. (d) The magnitude of the expansion coefficient $a_k$ of the different modes. (e) The voltage at $\omega$ calculated using the full 14 modes and compared with the one calculated using the relevant modes only. (f) Measured spectrum of the NL RH TL, where $\omega$ was allowed to sweep slowly over a frequency range around the dip in $S_{21}$. The inset shows the spectrum when the modulation is removed (i.e, pump excitation turned off).}
\label{fig:waveform_evec_ak}
\end{figure}

\begin{figure}[!htb]
\centering
\begin{subfigure}{0.49\linewidth}
\includegraphics[width=\linewidth]{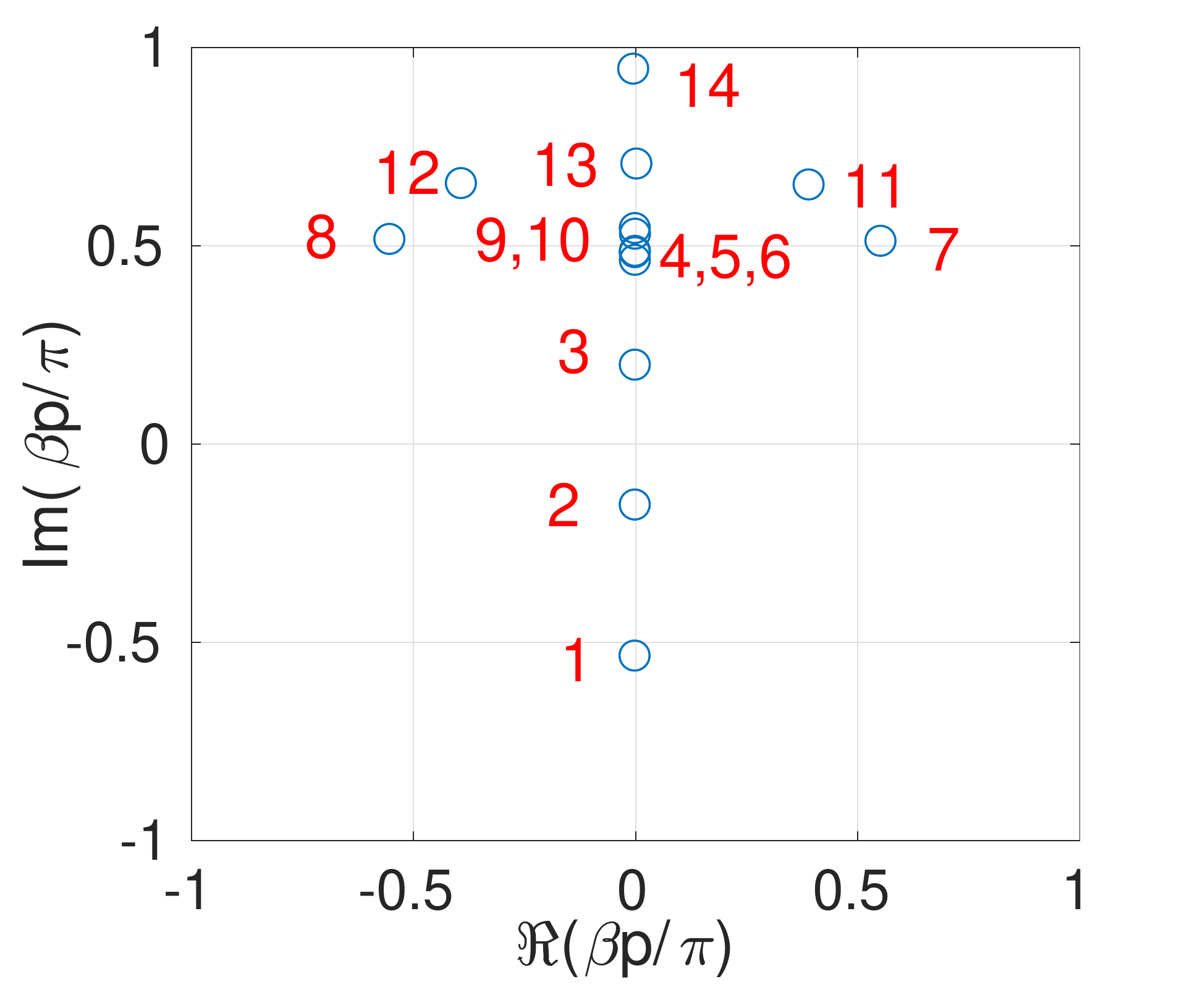}
\caption{}
\end{subfigure}
\begin{subfigure}{0.49\linewidth}
\includegraphics[width=\linewidth]{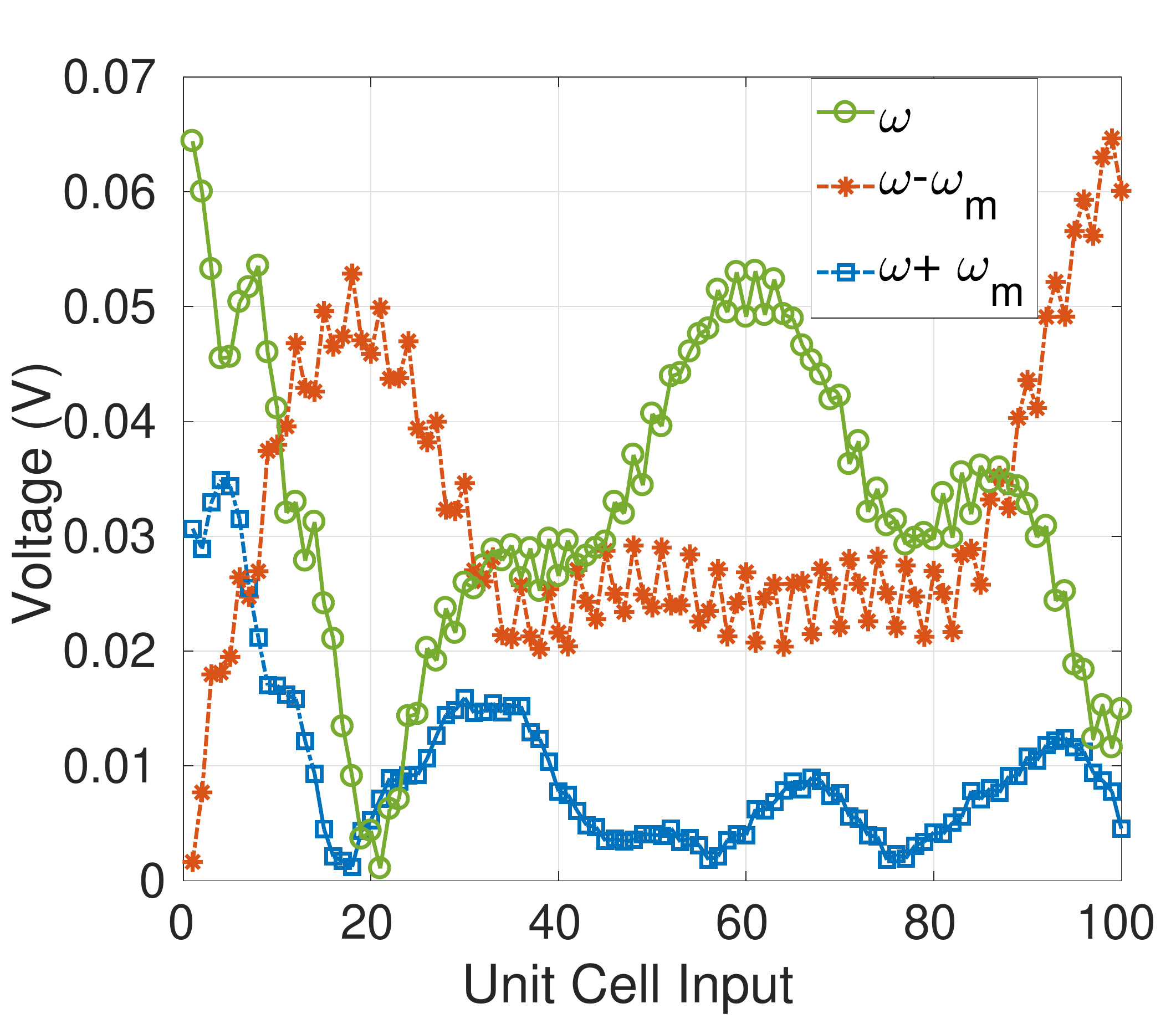}
\caption{}
\end{subfigure}
\caption{(a) Eigenvalues of the LTP circuit of NL RH TL. (b) Amplitudes of frequency components $\omega,\omega-\omega_m$ and $\omega+\omega_m$ computed using SSM at the dip in $S_{21}$, when $f_m=1$ GHz and for $N=100$ stages.}
\label{fig:nltl_eigenvals_100}
\end{figure}

When the modulation and signal are contra-directional,as Fig. \ref{fig:ltp_eigvals_ak_dispersion_evec_BWD}(a) demonstrates, the eigenvalues are generally different from those calculated above. The dispersion relation shows an increase in the separation between the forward branches as in Fig. \ref{fig:ltp_eigvals_ak_dispersion_evec_BWD}(c). Hence, the incident wave is expected to strongly couple to the main branch, labelled by the mode number 14. Note that other higher modes, for instance mode 15, are wrapped back to the negative side once $\beta p$ exceeds $\pi$. It is worth noting from the computed eigenvectors (Fig.\ref{fig:ltp_eigvals_ak_dispersion_evec_BWD}(d) that the 9\textsuperscript{th} and 10\textsuperscript{th} modes have a significant component at the 0\textsuperscript{th} harmonic. However due to the increased separation between the branches in the forward direction such modes are not excited. Therefore, one may conclude that when the signal and modulation are contra-directional the propagation is bascially that of the LTI system). Indeed, the calculated $a_k$ coefficients (Fig. \ref{fig:ltp_eigvals_ak_dispersion_evec_BWD}(b)) shows that coupling is mainly with the 14\textsuperscript{th} mode. Therefore, the mode couples with the forward main branch and the structure appears to be transparent in this mode of operation.  

\begin{figure}[!htb]
\centering
\begin{subfigure}{0.49\linewidth}
\includegraphics[width=\linewidth]{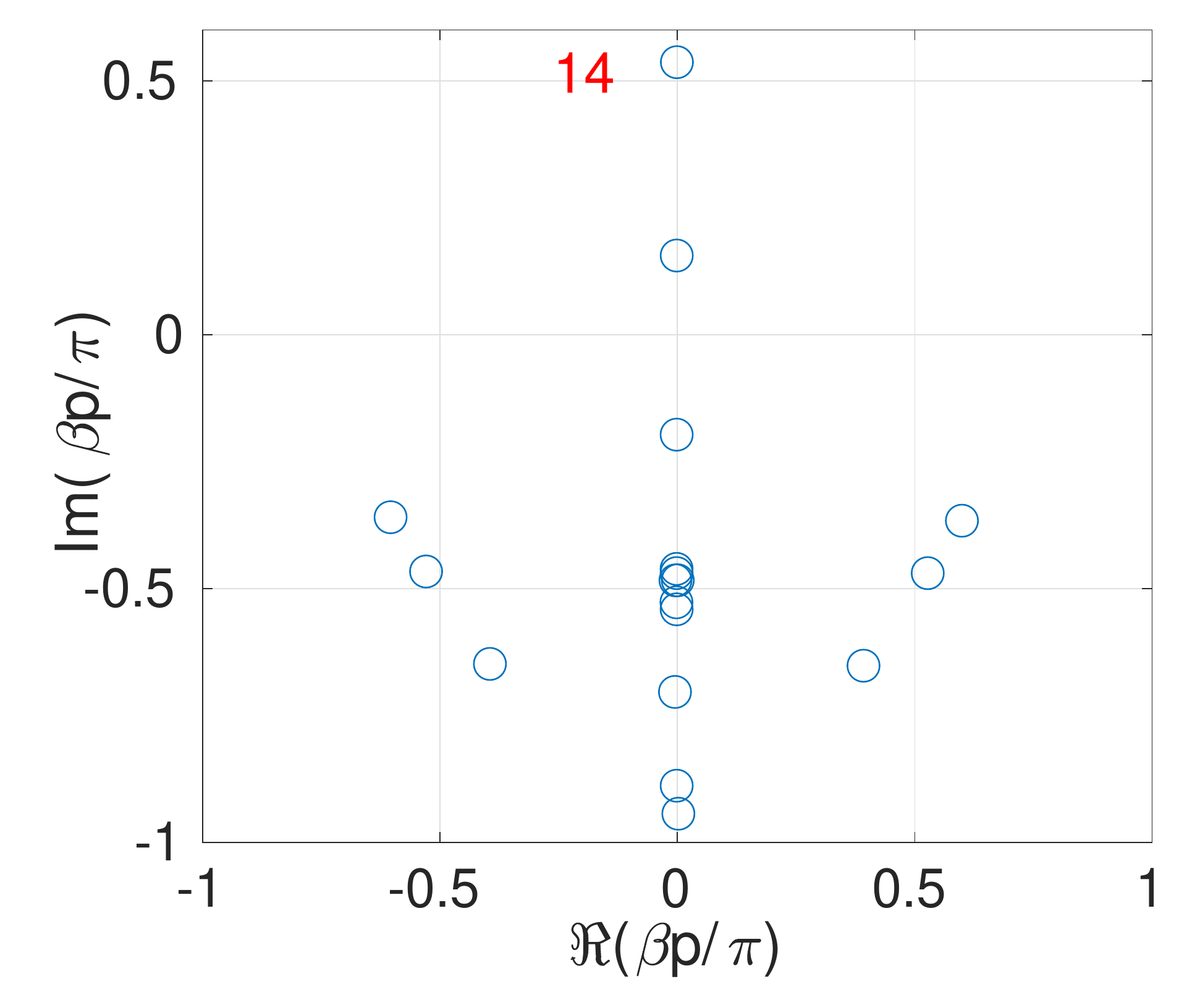}
\caption{}
\end{subfigure}
\begin{subfigure}{0.49\linewidth}
\includegraphics[width=\linewidth]{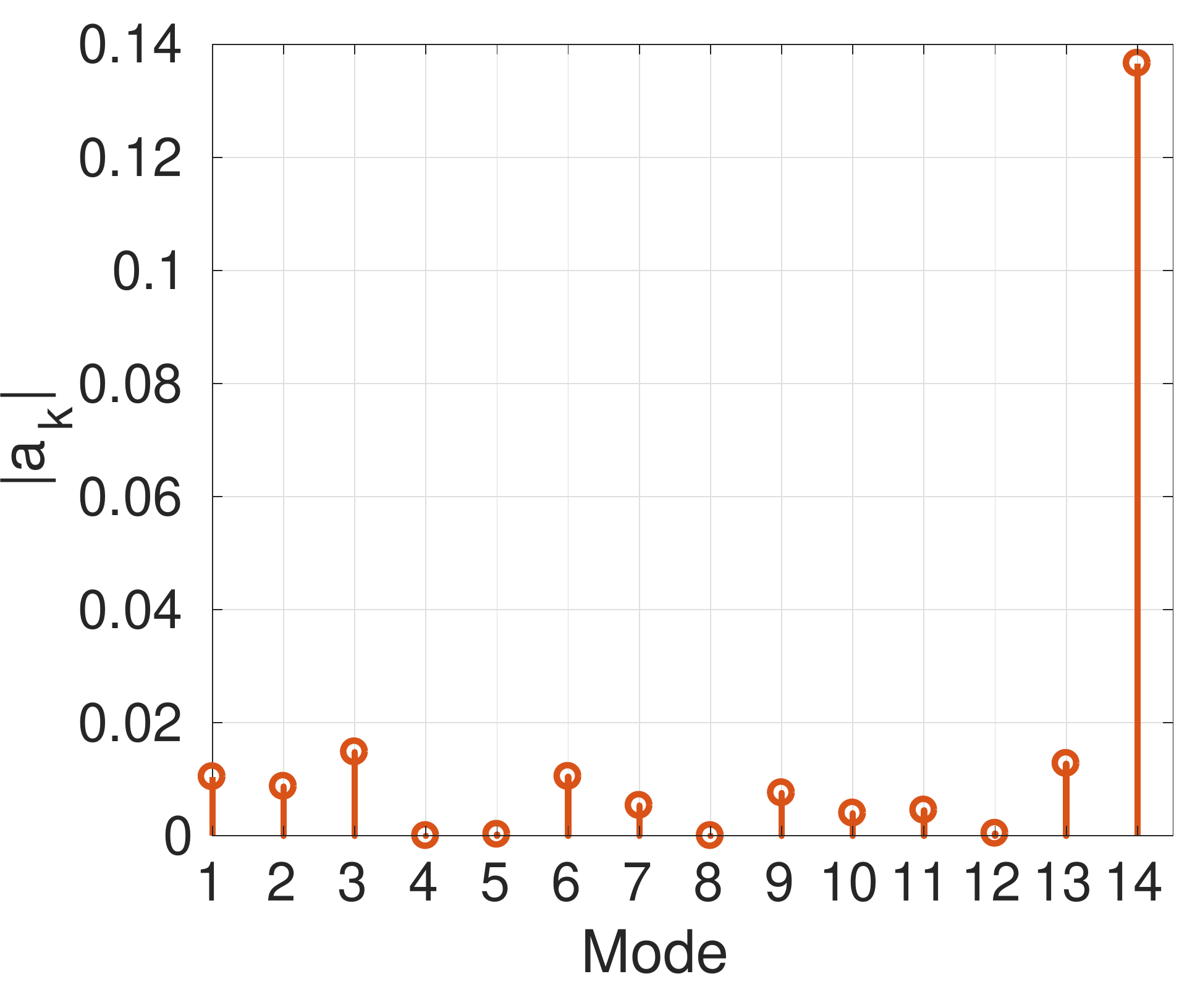}
\caption{}
\end{subfigure}
\begin{subfigure}{0.49\linewidth}
\includegraphics[width=\linewidth]{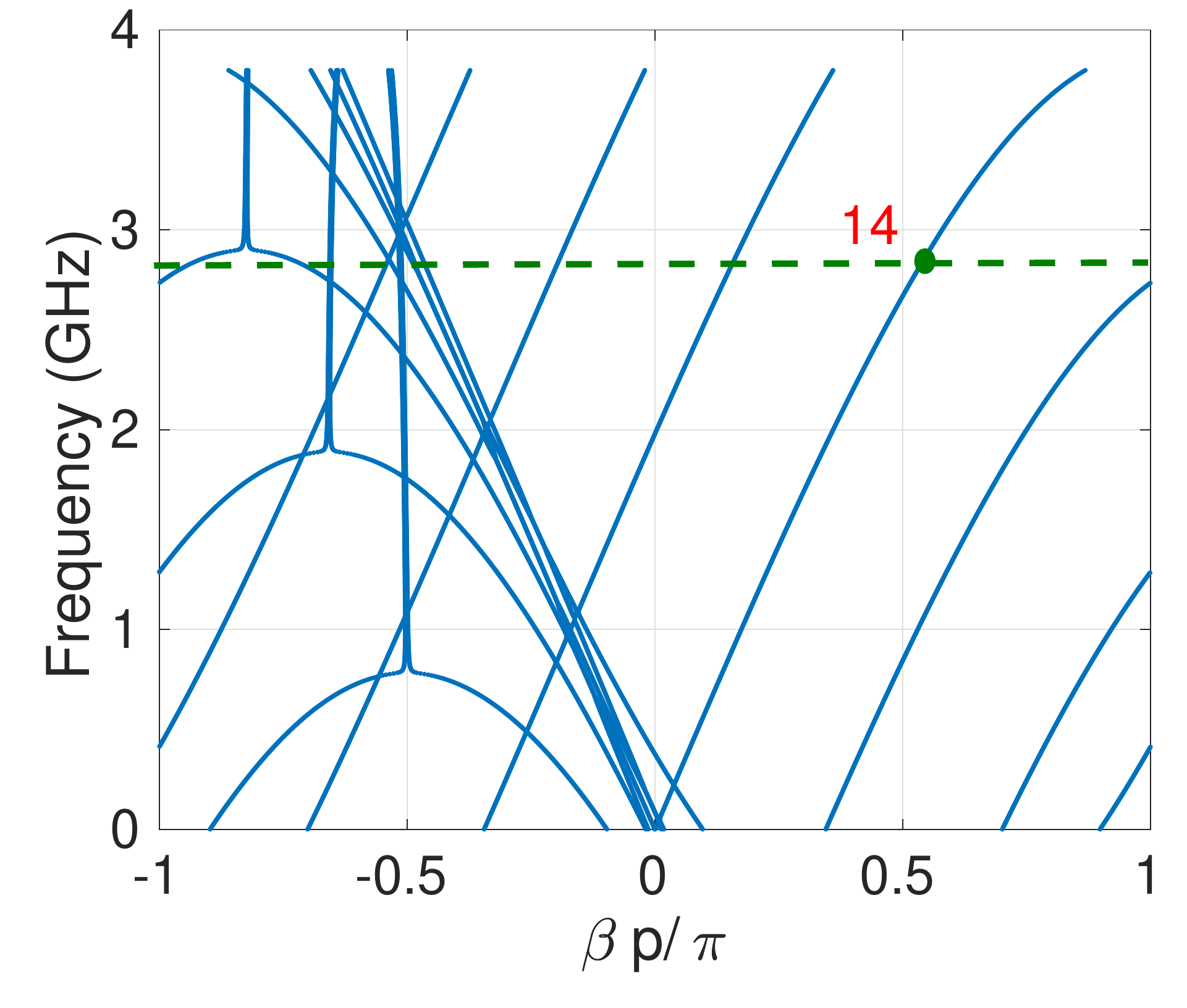}
\caption{}
\end{subfigure}
\begin{subfigure}{0.49\linewidth}
\includegraphics[width=\linewidth]{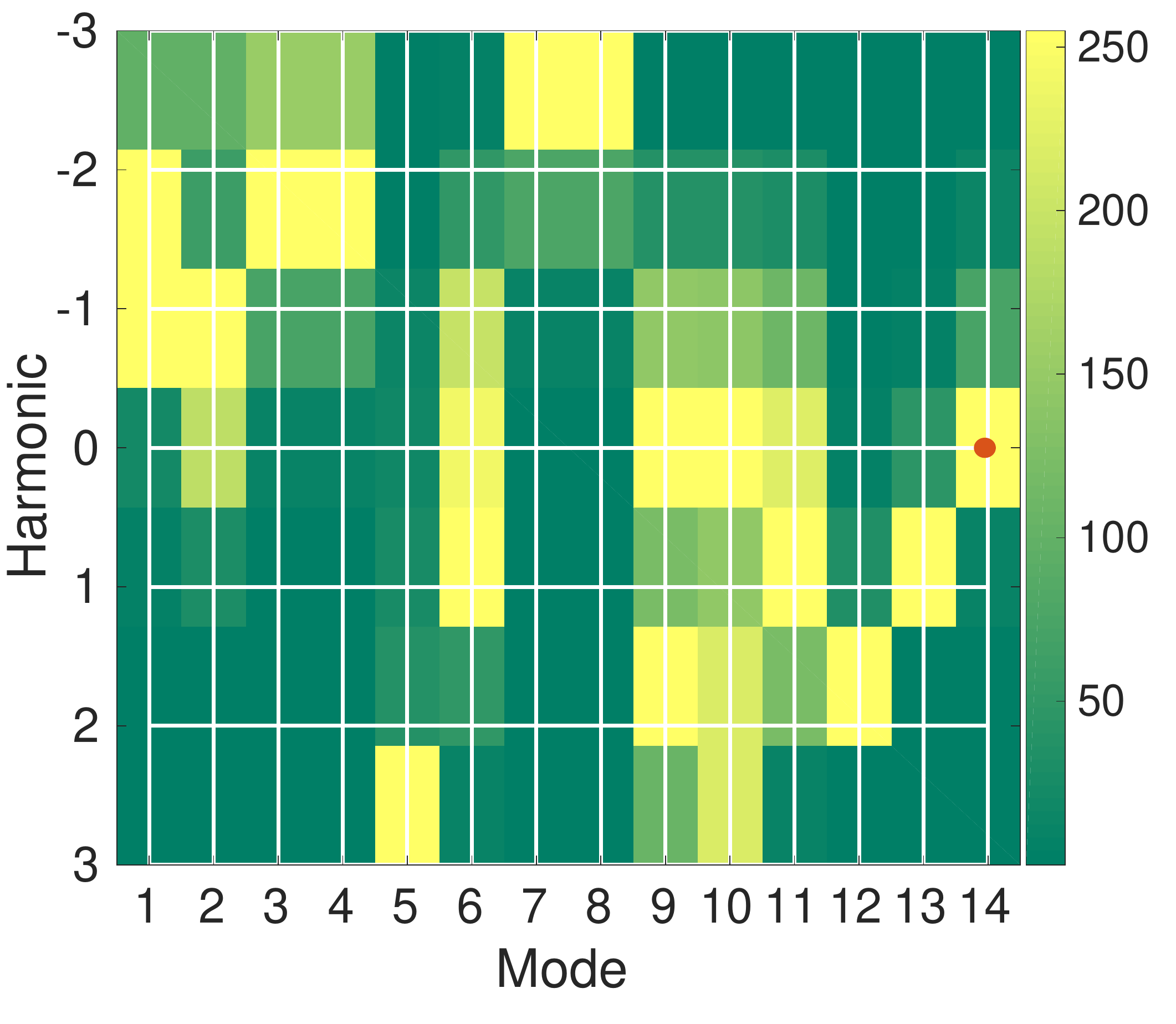}
\caption{}
\end{subfigure}
\caption{Modulation and signal are contra-directional. (a) Eigenvalues. (b) The magnitude of the expansion coefficients $a_k$. (c) Dispersion Relation. (d) Magnitude of the components of the eigenvectors.}
\label{fig:ltp_eigvals_ak_dispersion_evec_BWD}
\end{figure}

\begin{figure}[!htb]
\centering
\begin{subfigure}[h]{0.33\linewidth}
\includegraphics[width=\linewidth]{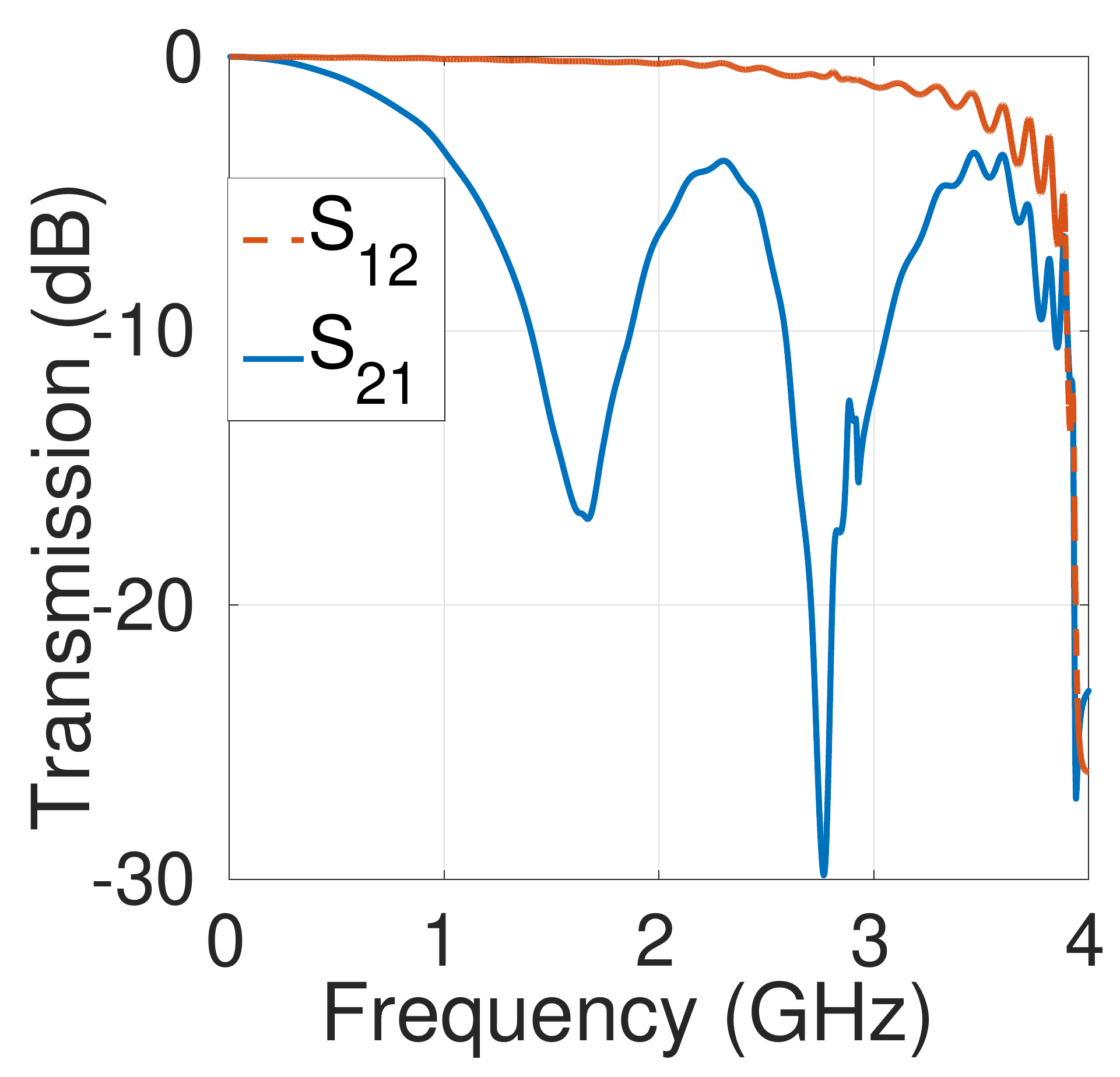}
\caption{}
\end{subfigure}
\begin{subfigure}[h]{0.31\linewidth}
\includegraphics[width=\linewidth]{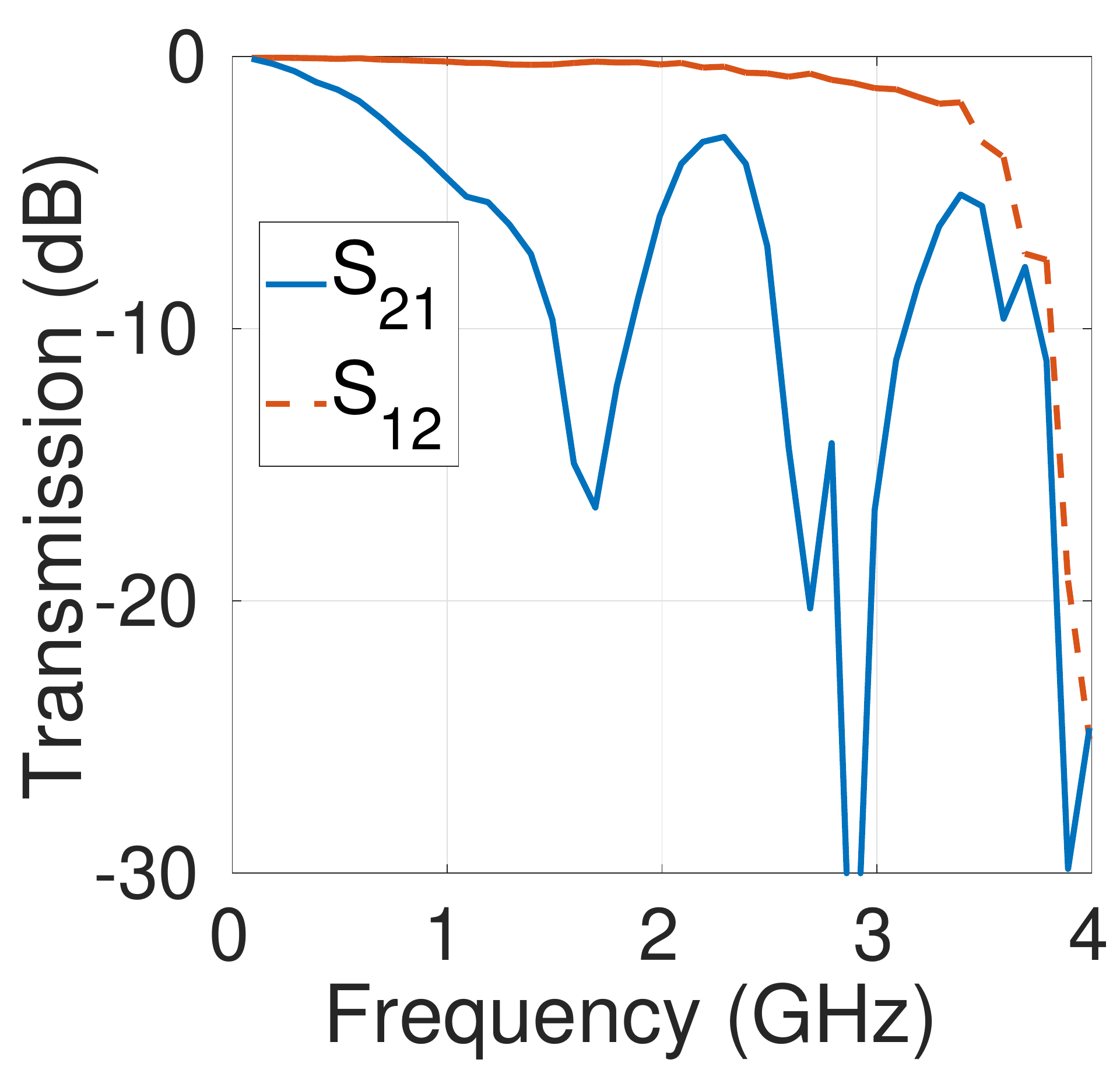}
\caption{}
\end{subfigure}
\begin{subfigure}{0.31\linewidth}
\includegraphics[width=\linewidth]{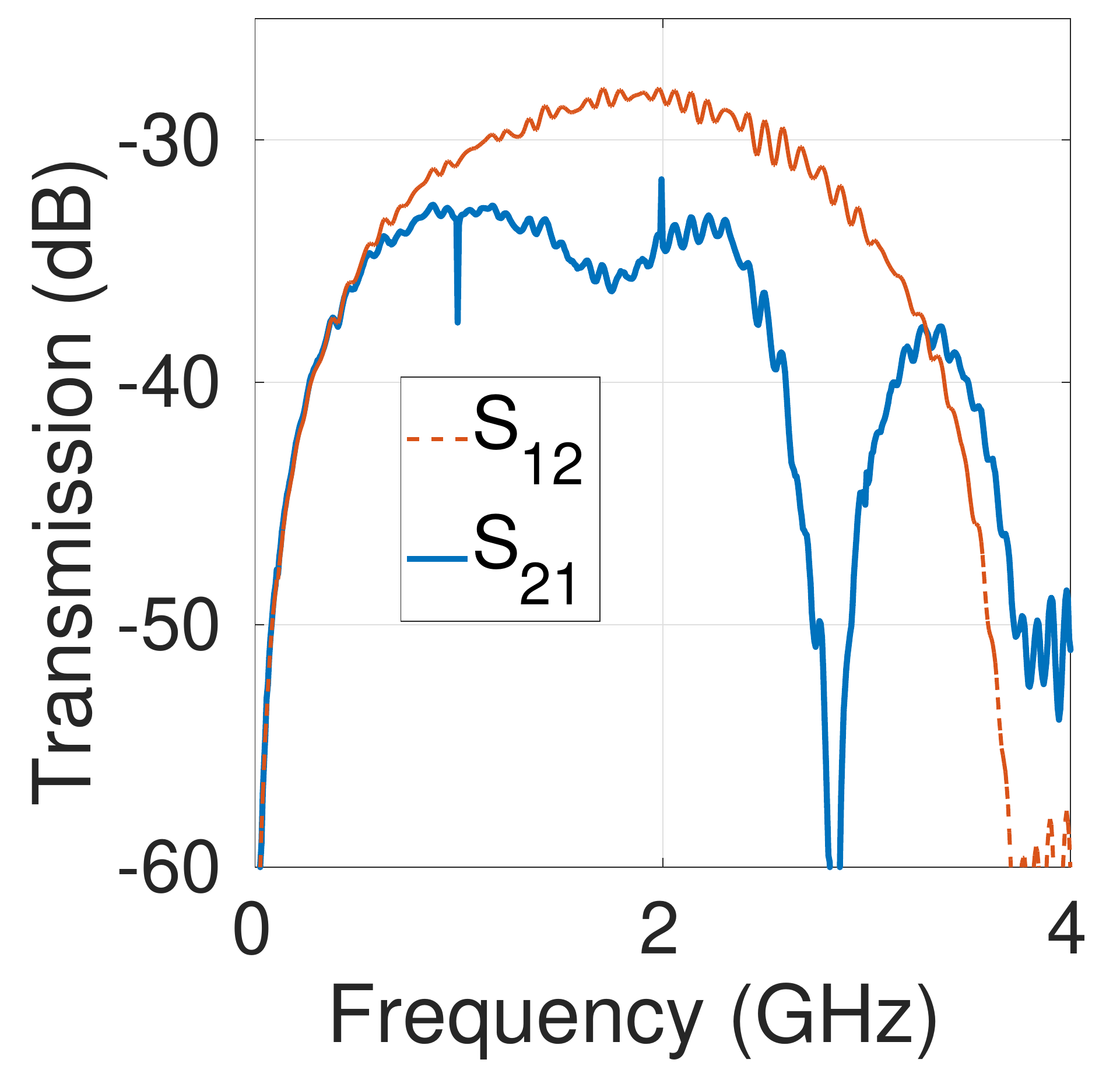}
\caption{}
\end{subfigure}

\begin{subfigure}{0.31\linewidth}
\includegraphics[width=\linewidth]{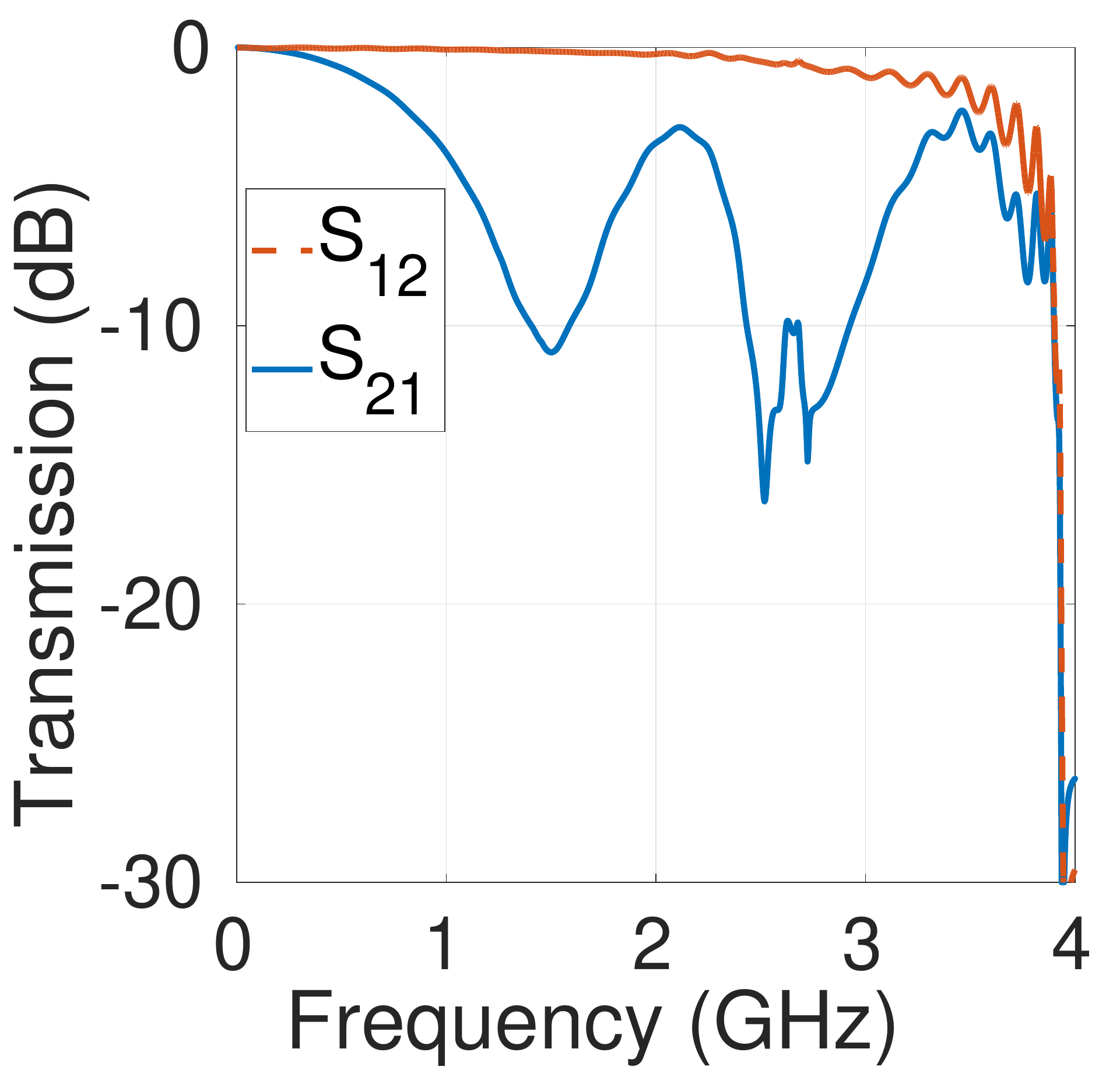}
\caption{}
\end{subfigure}
\begin{subfigure}{0.31\linewidth}
\includegraphics[width=\linewidth]{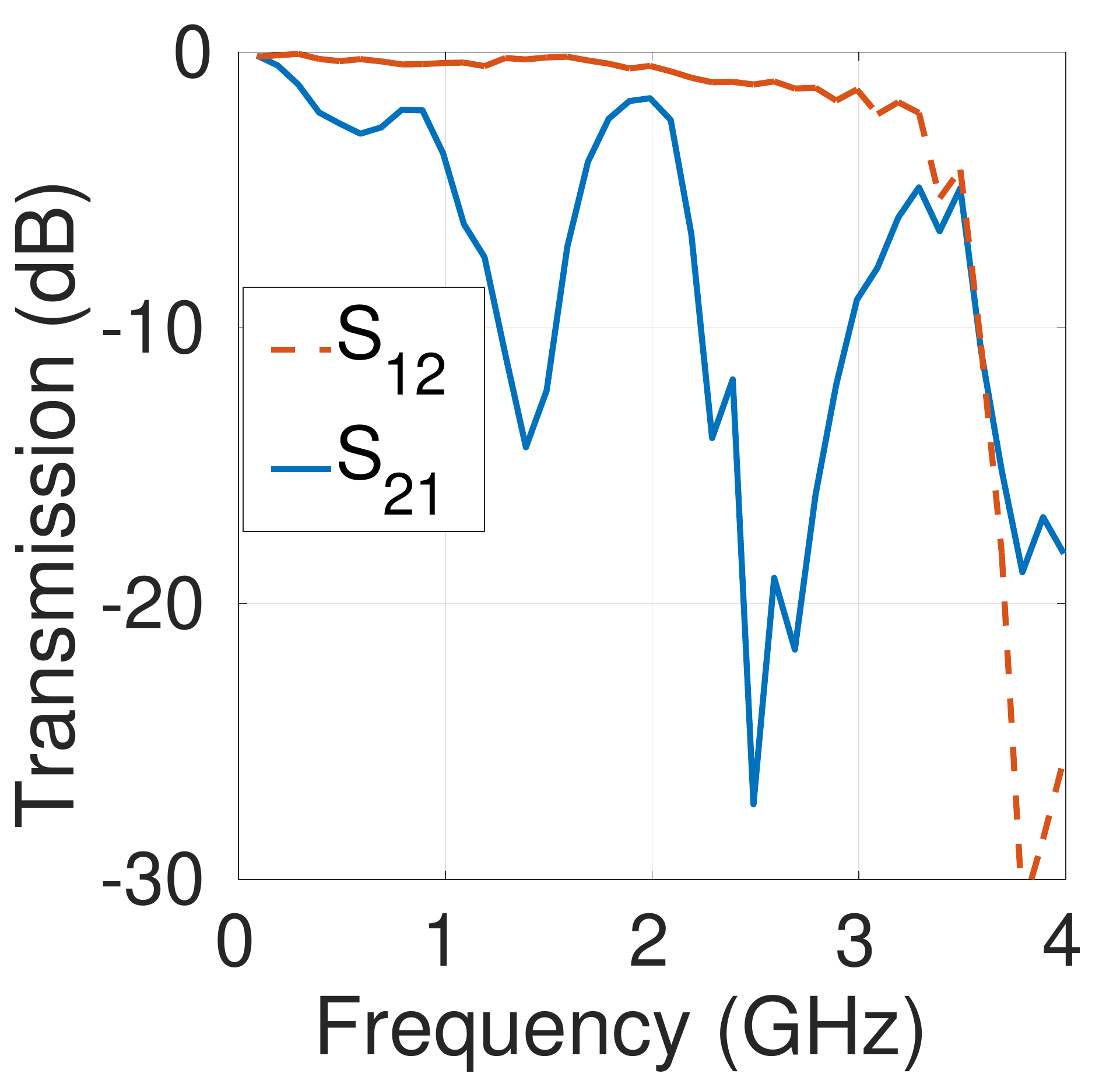}
\caption{}
\end{subfigure}
\begin{subfigure}{0.31\linewidth}
\includegraphics[width=\linewidth]{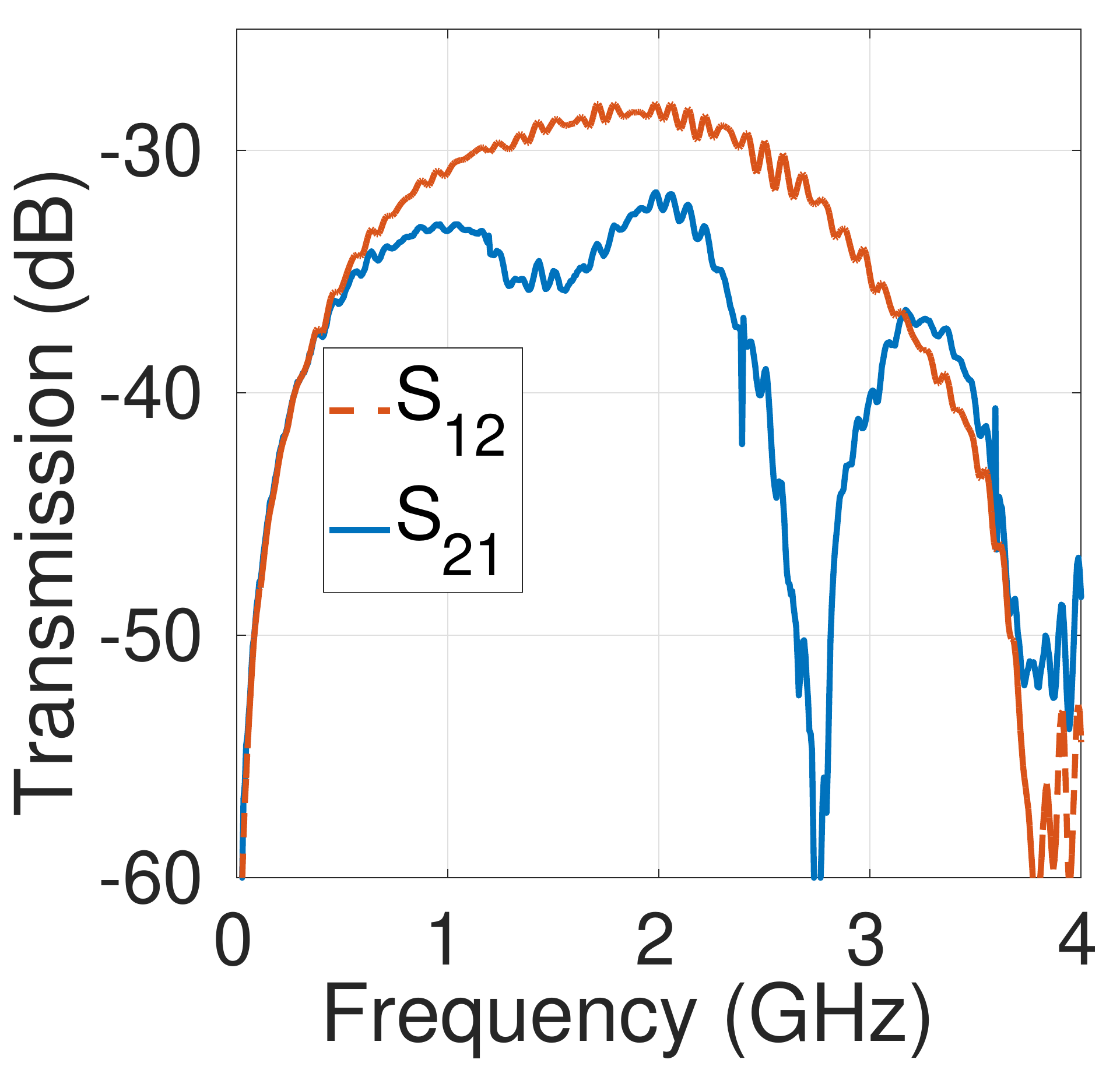}
\caption{}
\end{subfigure}
\caption{Transmission in the forward and backward directions. (a) LTP: $f_m=1$ GHz. (b) SSM: $f_m=1$ GHz. (c) Measurement: $f_m=1$ GHz. (d) LTP: $f_m=1.2$ GHz. (e) SSM: $f_m=1.2$ GHz. (f) Measurement: $f_m=1.2$ GHz.}
\label{fig:ltp_ssm_s21}
\end{figure}
\subsubsection{Transmission Coefficient}
Finally, the modes are superimposed according to (\ref{eq:s21}) to calculate the transmission for both the co- and contra-directional modes of operations. The transmission coefficient at the fundamental frequency is calculated using SSM and the process is repeated over the 0-4 GHz range. Fig. \ref{fig:ltp_ssm_s21} shows that the as a consequence of space time modulation and the asymmetric interaction between harmonics in the forward and backward directions, strong nonreciprocity between the forward and backward propagation arises. As has been shown, this is due to the \emph{passive} interaction between the fundamental mode and its lower harmonic at when the modulation and signal are co-directional. For an input frequency of 2.82 GHz, the coherent length is 20 unit cells long. Hence, maximum energy is transferred to the $-1^\textnormal{th}$ harmonic, reducing the signal at the output port. In the opposite direction, however, the distances between the forward branches are widened and the effect of modulation is negligible. Fig. \ref{fig:ltp_ssm_s21}(c) and (f) reveal that such nonreciprocal behaviour demonstrates itself in the measured scattering parameters. The bottom line, however, is reduced by approximately 30 dB due to the presence of the directional coupler.

\section{Conclusion}
The time periodic circuit theory was exploited in order to show some of the properties of the infinite dimensional spatial translation operator of space time modulated circuits. The modal behaviour of a generic space-time periodic structure can be explained after the solution of the system eigenvalue problem. Additionally, we showed that the translation operator as defined guarantees that solutions do not depend on the unit cell used. Furthermore it was shown that all points in the ($\beta$,$\omega$) plane parallel to the modulation velocity $\nu_m$ are equivalent in the sense that the eigenvectors are related by a shift operator. The wave waveforms inside the space time periodic circuit and the time periodic scattering parameters were determined through the expansion of the total solution in terms of the eigenmodes, and after imposing the suitable boundary conditions. Two examples were discussed. In the first, a space time modulated CRLH TL was studied using the developed approach and compared with time domain simulation. In the second example, the non-reciprocal behaviour observed on a nonlinear TL was explained. This was made possible via the extraction of circuit parameters from measurements that were then used to predict the wave behaviour inside the TL and its effect on the terminal properties. It was shown that the passive interaction between different harmonics resulteds in an observed non-reciprocal behaviour, where the difference between forward and backward transmission coefficients $S_{21}-S_{12}$ can be significant. The frequencies at which non-reciprocity occured and its strength agree with time domain simulation and measurements.
\appendix
\section{Transformation of Eigenvalues and Eigenvectors upon the change unit cell}
 \label{append:translationevp}
\begin{proof}
Consider the $k^\textnormal{th}$ equation of (\ref{eq:evb})
$$\left(\Gamma_{k}A_k^k-e^{i\beta p}\right)V_k+\Gamma_{k}B_k^kI_k+ \sum_{l\neq 0}\Gamma_{k+l}A_k^{k+l}V_{k+l}+\Gamma_{k+l}B_k^{k+l}I_{k+l}=0,$$
obtained from the use of the ABCD parameters of the $n^\textnormal{th}$ unit cell. Based on (\ref{eq:paramtranslated}), the $k^\textnormal{th}$ equation at $n+1$, after the direct substitution using (\ref{eq:betadash}) and (\ref{eq:n+1}), becomes
\begin{align}
&\left(\Gamma_{k}\underbrace{A_k^k}_{{\color{blue}{\textnormal{at }n+1}}}-e^{i\beta p}\right)V_k'+\Gamma_{k}\underbrace{B_k^k}_{{\color{blue}{\textnormal{at }n+1}}}I_k'+ \cdots\cdots \nonumber\\
&\sum_{l\neq 0}\Gamma_{k+l}\underbrace{A_k^{k+l}}_{{\color{blue}{\textnormal{at }n+1}}}V_{k+l}'+\Gamma_{k+l}\underbrace{B_k^{k+l}}_{{\color{blue}{\textnormal{at }n+1}}}I_{k+l}'=0. \label{eq:atn+1}
\end{align}
Noting that,
$$\underbrace{A_k^k}_{{\color{blue}{\textnormal{at }n+1}}}=\underbrace{A_k^k}_{{\color{red}{\textnormal{at }n}}},  \underbrace{B_k^k}_{{\color{blue}{\textnormal{at }n+1}}}=\underbrace{B_k^k}_{{\color{red}{\textnormal{at }n}}}$$
and
$$\underbrace{A_k^{k+l}}_{{\color{blue}{\textnormal{at }n+1}}}=\underbrace{A_k^{k+l}}_{{\color{red}{\textnormal{at }n}}}\Gamma_{-l}, \underbrace{B_k^{k+l}}_{{\color{blue}{\textnormal{at }n+1}}}=\underbrace{B_k^{k+l}}_{{\color{red}{\textnormal{at }n}}}\Gamma_{-l}$$ 
for $l\neq 0$.
Therefore, (\ref{eq:atn+1}) becomes
\begin{align*}
&\left(\Gamma_kA_k^k-e^{i\beta p}\right)V_k\Gamma_{k}+\Gamma_kB_k^kI_k\Gamma_{k}+\cdots\cdots\\
&\sum_{l\neq 0}\Gamma_{k+l}A_k^{k+l}\Gamma_{-l}V_{k+l}\Gamma_{k+l}+\Gamma_{k+l}B_k^{k+l}\Gamma_{-l}I_{k+l}\Gamma_{k+l}=\\
&\left(\Gamma_kA_k^k-e^{i\beta p}\right)V_k\Gamma_{k}+B_k^kI_k\Gamma_{2k}+\sum_{l\neq 0}\Gamma_{2k+l}A_k^{k+l}V_{k+l}+\Gamma_{2k+l}B_k^{k+l}I_{k+l},
\end{align*}
which after multiplying by $\Gamma_{-k}$ reduces to the LHS expression of the corresponding equation at the $n^\textnormal{th}$ unit cell.
\end{proof}

\section{Relation between Eigenvalues and Eigenvectors at $\omega$ and $\omega+l\omega_m$}
\label{append:relationbetomega}
\begin{proof}
Consider the $k^\textnormal{th}$ equation of (\ref{eq:evb}), where $\beta$ is a solution at $\omega$,
$$ \sum_{r=-\infty}^{+\infty}\left(e^{-i[r+k]\beta_mp}A_k^{k+r}(\omega)-\delta_0^re^{i\beta p}\right)V_{k+r}+e^{-i[k+r]\beta_mp}B_k^{k+r}(\omega)I_{k+r}=0$$

Since $A_k^{k+r}(\omega)=A_0^r(\omega+k\omega_m)$ and $B_k^{k+r}(\omega)=B_0^r(\omega+k\omega_m)$, the above equation becomes
\begin{align}
 &\sum_{r=-\infty}^{+\infty}\left(e^{-i[r+k]\beta_mp}A_0^{r}(\omega+k\omega_m)-\delta_0^re^{i\beta p}\right)V_{k+r}+\cdots\cdots \nonumber\\
 & e^{-i[k+r]\beta_mp}B_0^{r}(\omega+k\omega_m)I_{k+r}=0.\label{eq:ktheqn}
\end{align}
Let $\beta'$ be the solution at $\omega+l\omega_m$ and the vectors $\mathbf{V}'=[\cdots, V_{-1}',V_0',V_{+1}',\cdots]^t$ and $\mathbf{I}'=[\cdots, I_{-1}',I_0',I_{+1}',\cdots]^t$ form the corresponding eigenvector. We claim that $\beta'=\beta+l\beta_m$. To show this we revert to the $(k-l)^\textnormal{th}$ equation when $\omega \rightarrow \omega+l\omega_m$. Hence,
\begin{align}
 &\sum_{r=-\infty}^{+\infty}\left(e^{-i[r+k-l]\beta_mp}A_0^{r}(\omega+k\omega_m)-\delta_0^re^{i\beta' p}\right)V_{k-l+r}'+\cdots\cdots\nonumber\\
 & e^{-i[k+r-l]\beta_mp}B_0^{r}(\omega+k\omega_m)I_{k-l+r}'=0.\label{eq:kpleqn}
\end{align}
Note that (\ref{eq:kpleqn}) reduces to (\ref{eq:ktheqn}) when $\beta'=\beta+l\beta_m$ and $V_{k-l+r}'=V_{k-r}$. Consequently, the $k^\textnormal{th}$ component of the eigenvector at $\omega$ turns to be at the $(k-l)^\textnormal{th}$ location, when $\omega$ is changed to $\omega+l\omega_m$, i.e, $\mathbf{\Psi}_n(\omega+l\omega_m)=\mathcal{S}_\mathcal{U}^{l}\mathbf{\Psi}_n(\omega)$ is an eigenvector at $(\beta+l\beta_m,\omega+l\omega_m)$.
\end{proof}
\bibliography{NonReci}

\end{document}